\documentclass[AMA, STIX2COL]{styles/mrm/MRM}
\articletype{preprint - submitted to magnetic resonance in medicine}%

\usepackage{styles/defs}

\received{xx xxxxx xxxx}
\revised{xx xxxxx xxxx}
\accepted{xx xxxxx xxxx}

\begin{document}
\raggedbottom

\title{Strategies to Minimize Out-of-Distribution Effects in Data-Driven MRS Quantification}

\author[1]{Julian P. Merkofer}{\orcid{0000-0003-2924-5055}}
\author[2]{Antonia Kaiser}{\orcid{0000-0001-7805-6766}}
\author[3]{Anouk Schrantee}{\orcid{0000-0002-4035-4845}}
\author[3]{Oliver J. Gurney-Champion}{\textsuperscript{,*}\orcid{0000-0003-1750-6617}}
\author[1]{Ruud J. G. van Sloun}{\textsuperscript{,*}\orcid{0000-0003-2845-0495}}

\authormark{MERKOFER \textsc{et al}}
\titlemark{\MakeUppercase{Strategies to Minimize Out-of-Distribution Effects in Data-Driven MRS Quantification}}

\address[1]{\orgdiv{Department of Electrical Engineering}, \orgname{Eindhoven University of Technology}, \orgaddress{\state{Eindhoven}, \country{The Netherlands}}}

\address[2]{\orgdiv{CIBM Center for Biomedical Imaging}, \orgname{École Polytechnique Fédérale de Lausanne, EPFL}, \orgaddress{\state{Lausanne}, \country{Switzerland}}}

\address[3]{\orgdiv{Department of Radiology and Nuclear Medicine},  \orgname{Amsterdam University Medical Center}, \orgaddress{\state{Amsterdam}, \country{The Netherlands}}}

\address[4]{\llap{\textcolor{white}{\textsuperscript{4}}}\llap{\textsuperscript{*}}These authors share last authorship.}

\corres{Julian P. Merkofer, Eindhoven University of Technology, PO Box 513 5600 MB Eindhoven, The Netherlands. \email{\textcolor{blue}{j.p.merkofer@tue.nl}}}

\presentaddress{Eindhoven University of Technology
{\hfill\break}Electrical Engineering Department
{\hfill\break}Groene Loper 19, 5612 AP Eindhoven, The Netherlands}

\finfo{This work was in part funded by Spectralligence (EUREKA IA Call, ITEA4 project 20209) and received support from the NVIDIA Academic Hardware Grant Program.}

\abstract[Summary]{
\section{Purpose}
This study systematically compared data-driven and model-based strategies for metabolite quantification in \ac{mrs}, focusing on resilience to \ac{ood} effects and the balance between accuracy, robustness, and generalizability.

\section{Methods}
A neural network designed for \ac{mrs} quantification was trained using three distinct strategies: supervised regression, self-supervised learning, and test-time adaptation. These were compared against model-based fitting tools. Experiments combined large-scale simulated data, designed to probe metabolite concentration extrapolation and signal variability, with 1H single-voxel 7T in-vivo human brain spectra.

\section{Results}
In simulations, supervised learning achieved high accuracy for spectra similar to those in the training distribution, but showed marked degradation when extrapolated beyond the training distribution. Test-time adaptation proved more resilient to \ac{ood} effects, while self-supervised learning achieved intermediate performance. In-vivo experiments showed larger variance across the methods (data-driven and model-based) due to domain shift. Across all strategies, overlapping metabolites and baseline variability remained persistent challenges.

\section{Conclusion}
While strong performance can be achieved by data-driven methods for \ac{mrs} metabolite quantification, their reliability is contingent on careful consideration of the training distribution and potential \ac{ood} effects. When such conditions in the target distribution cannot be anticipated, test-time adaptation strategies can ensure consistency between the quantification, the data, and the model, enabling reliable data-driven \ac{mrs} pipelines.
\vspace{-1mm}}

\keywords{machine learning, magnetic resonance spectroscopy, metabolite quantification, generalization, test-time adaptation, out-of-distribution effects\vspace{-8mm}}


\maketitle
\clearpage


\section{Introduction} \label{sec:intro}

\acresetall

\Ac{mrs} is a non-invasive technique for measuring the metabolic composition of tissues, providing valuable insights into neurological disorders and serving as a tool to characterise tumours for treatment stratification and monitoring. \cite{faghihi_magnetic_2017, Maudsley2020AdvancedMR, Horska2023MRSClincalA} However, its clinical utility is constrained by several challenges, including an inherently low \ac{snr}, spectral overlap of metabolites, unparameterized baseline effects, and various experimental artifacts. \cite{Kreis2004IssuesOS, Hurd2009ArtifactsAPI} Together, these factors make accurate quantification of metabolite concentrations a challenging task, complicating the analysis and interpretation of \ac{mrs} data. 

Traditionally, metabolite quantification in \ac{mrs} has relied on model-based fitting approaches such as \ac{lcm} and peak fitting. \cite{near_preprocessing_2021} These methods construct a theoretical representation of the expected signal and optimize its parameters to match the measured spectrum, typically using nonlinear least-squares algorithms like the Levenberg–Marquardt method~\cite{Levenberg1944AMF, Marquardt1963AnAF}. Frequency-domain \ac{lcm} is widely adopted~\cite{provencher_estimation_1993, Soher2023Vespa, Gajdok2021INSPECTOR, Oeltzschner2020OspreyOP, clarke_fslmrs_2021}, fitting a linear combination of basis spectra (the idealized signal contributions of individual metabolites) to the data while accounting for baseline distortions, frequency and phase shifts, and lineshape variations~\cite{poullet_mrs_2008}. 
These purely model-based methods are generally computationally intensive, may require user expertise for proper setup and interpretation, and the inherently ill-posed nature yields non-unique solutions, requiring methods to employ specific regularizers, constraints, or other priors. \cite{near_preprocessing_2021, Landheer2021AreCRLBs, Marjanska2021InfuenceOfBaseline, Zoellner2021ComparisonLCMshortTE, provencher_estimation_1993} To overcome these challenges, increasing attention has been given to \ac{ml} methods as their ability to learn from data offer a promising alternative. \cite{Sande2023AReviewOfML, Luo2025DLReviewNMRS}

Early data-driven approaches to \ac{mrs} quantification explored direct regression from spectra to metabolite concentrations using supervised learning techniques, including random forests~\cite{das_quantification_2017} and \acp{cnn}~\cite{hatami_magnetic_2018, chandler_mrsnet_2019, shamaei_wavelet_2021, lee_deep_2020, iqbal_deep_2021}. These models are trained to predict metabolite amplitudes directly from the input spectrum by minimizing a loss function between predicted and ground truth concentrations. This requires access to reference concentrations during training, which are typically only available for synthetic data. More recently, self-supervised strategies have emerged that integrate a physics-based signal model into the training process~\cite{gurbani_incorporation_2019, shamaei_physics-informed_2023, Chen2024MRSQuantAided}. In these methods, the \ac{nn} estimates all relevant signal parameters, then reconstructs the spectrum using a forward signal model as done in traditional \ac{lcm}. The distinction lies in the optimization strategy: instead of explicitly solving a least-squares problem, a network learns to reduce the reconstruction error via stochastic gradient descent amortized over the training dataset. This brings several advantages including the use of learned priors as a regularizer that lowers prediction variance, reduces sensitivity to noise, and helps the optimizer avoid suboptimal local minima.

Nevertheless, the effectiveness of \ac{ml} models depends strongly on the quality and diversity of the training data. \cite{bishop_pattern_2006, Gudmundson2023AGNOSTIC, Mohammed2025EffectsDataQuality} 
%
Most of the previous work has relied heavily on simulated data with minimal in-vivo testing. \cite{Sande2023AReviewOfML} While investigations have explored diverse \ac{nn} architectures, spectroscopic input types, the use of ensemble learning~\cite{rizzo_quantification_2023}, and methods for uncertainty estimation~\cite{lee_deep_2020, lee_bayesian_2022, rizzo_reliability_2022}, a systematic analysis of the critical aspects of robustness and generalization is lacking. In particular, the influence of training paradigms on a model’s susceptibility to bias, its ability to maintain performance under challenging conditions, and its capacity to generalize to unseen, potentially \ac{ood}, in-vivo \ac{mrs} data has not been thoroughly studied and documented. Furthermore, \ac{tta}~\cite{Sun2020TestTimeTraining, Wilson2020UnsupDomainAdapt, Kouw2021DomainAdaptWOTargets, Fang2024SFDomainAdaptSurvey, Li2024SFDomainAdaptSurvey, Liang2025TestTimeAdaptSurvey}, where \acp{nn} are updated during inference, has received little attention in the context of \ac{mrs}, despite its potential to mitigate domain shift.

This study contributes to this ongoing effort by systematically comparing different data-driven strategies for \ac{mrs} metabolite quantification, explicitly focusing on their inherent data biases and their resilience to \ac{ood} samples. We evaluate a supervised regression approach, a self-supervised learning method that incorporates a signal model during training, and \ac{tta} techniques, as well as  compare these to purely model-based fitting. By assessing the performance of these strategies on both carefully controlled synthetic data and 7T in-vivo human brain proton spectra, we aim to provide valuable insights into the trade-offs between accuracy, robustness, and generalizability of different strategies for \ac{mrs} quantification.

\section{Methods} \label{sec:methods}

This section outlines the simulation framework, in-vivo data acquisition and processing, quantification strategies, evaluation metrics, and performed experiments.

\subsection{Simulated Data} \label{ssec:simulated_data}

Simulated spectra offer access to ground truth metabolite concentrations and acquisition parameters, providing a controlled setting for both model optimization and analysis.

\subsubsection{Signal Model}

To simulate proton \ac{mrs} spectra, we define a parametric signal model in the frequency domain, denoted by $X(f \mid \boldsymbol{\theta})$, where $f$ is frequency and $\boldsymbol{\theta}$ represents the set of signal model parameters. The modeled spectrum is expressed as:
\begin{equation} \label{eq:signal_model}
    X(f \mid \boldsymbol{\theta}) = e^{i (\phi_0 + f \phi_1)} \sum^{M}_{m=1} a_{m} \ S_{m}(f) + B(f),
\end{equation}
where $\phi_0$ and $\phi_1$ are zeroth- and first-order phases, $a_m$ is the amplitude of the $m$-th metabolite, and $B(f)$ denotes a spectral baseline, modeled as a complex-valued $K$-order polynomial. Each metabolite basis function is defined as:
\begin{equation}
    S_{m}(f) = \mathcal{F}\{s_{m}(t) \ e^{- (\gamma + \varsigma^2 t + i\epsilon) t}\},
\end{equation}
where $\gamma$ and $\varsigma$ denote global (for all metabolites) Lorentzian and Gaussian linewidth broadening parameters, respectively, and $\epsilon$ represents a global frequency shift. The operator $\mathcal{F}\{\cdot\}$ denotes the Fourier transform, and ${s_m(t)}_{m=1}^{M}$ are the time-domain basis functions representing the idealized signal contributions of individual metabolites. \Acp{mm} are included, experiencing the same broadening, phasing, and shifting as the metabolites. To simulate realistic measurement conditions, the observed spectrum $Y(f)$ is defined as:
\begin{equation} \label{eq:observation}
    Y(f) = X(f \mid \boldsymbol{\theta}) + N,
\end{equation} 
where $N$ denotes complex Gaussian noise.

\subsubsection{Parameter Ranges}

The simulation ranges for the signal model parameters 
\begin{equation} 
    \boldsymbol{\theta} = \left \{a_1, ..., a_M, \gamma, \varsigma, \epsilon, \phi_0, \phi_1, b_1, ..., b_{2(K+1)} \right \},
\end{equation}
were designed to capture the full variability observed in the in-vivo data. Therefore, metabolite concentration bounds were derived from a combination of literature values reported by De Graaf~\cite{de_graaf_vivo_2019}, and the empirical distributions obtained by fitting all in-vivo spectra using both LCModel~\cite{provencher_estimation_1993} and FSL-MRS~\cite{clarke_fslmrs_2021} (details are reported in Section~\ref{ssec:in-vivo_data}). For each metabolite, the lower and upper bounds were defined as the minimum and maximum observed values across these three sources. This ensured that the entire dynamic range of concentrations present in our in-vivo dataset was represented, while avoiding unrealistically narrow ranges. 

For the remaining signal parameters $\gamma$, $\varsigma$, $\epsilon$, $\phi_0$, $\phi_1$, $b_1$, ..., $b_{2(K+1)}$, and $N$, literature guidance was limited, so we used our own in-vivo data as the initial reference. Since this 7T in-vivo data, acquired with high \ac{snr}, good shimming, and consistent processing, resulted in relatively narrow parameter distributions, we deliberately selected wider simulation ranges. This ensured the simulated data reflected the greater variability that may be encountered in broader clinical or research settings. 

An overview of all simulation parameter distributions is provided in Table~\ref{tab:sim_params}. The individual metabolite range reported in De Graaf 2019 \cite{de_graaf_vivo_2019} and the obtained ranges of LCModel and FSL-MRS are listed in Tables \ref{tab:sim_params_deGraaf}, \ref{tab:sim_params_lcmodel}, and \ref{tab:sim_params_fsl} in Appendix~\ref{app:impl_details}.

\begin{table*}
\caption{\enspace Overview of the notations and distribution ranges of the simulation parameters. Metabolite bounds were set using the minimum and maximum values from De Graaf~\cite{de_graaf_vivo_2019} and fits to all in-vivo spectra using LCModel~\cite{provencher_estimation_1993} and FSL-MRS~\cite{clarke_fslmrs_2021}. $\mathcal{U}[p_{min}, p_{max}]$ and $\mathcal{CN}(p_{mean}, p_{var})$ denote continuous and complex Gaussian distributions, respectively, and curly braces $\{ \cdot \}$ indicate discrete sets of values.} 
\label{tab:sim_params}
\vspace{2mm}
\begin{adjustbox}{width=0.5\textwidth}
\begin{tabular*}{0.6\textwidth}{@{\extracolsep\fill\hspace{1mm}}lccc@{\extracolsep\fill\hspace{2mm}}}
\toprule
\textbf{Parameter} & \textbf{Notation} & \textbf{Range} & \textbf{Unit} \\
\midrule
Alanine (Ala) & \( a_1 \) & \( \mathcal{U}[0.0, 1.6] \) & \si{\mM} \\
Ascorbate (Asc) & \( a_2 \) & \( \mathcal{U}[0.0, 4.9] \) & \si{\mM} \\
Aspartate (Asp) & \( a_3 \) & \( \mathcal{U}[0.0, 4.8] \) & \si{\mM} \\
Creatine (Cr) & \( a_4 \) & \( \mathcal{U}[3.9, 12.3] \) & \si{\mM} \\
Gamma-Aminobutyric Acid (GABA) & \( a_5 \) & \( \mathcal{U}[0.0, 4.0] \) & \si{\mM} \\
Glutamine (Gln) & \( a_{6} \) & \( \mathcal{U}[0.0, 6.8] \) & \si{\mM} \\
Glutamate (Glu) & \( a_{7} \) & \( \mathcal{U}[6.0, 17.9] \) & \si{\mM} \\
Glycine (Gly) & \( a_{8} \) & \( \mathcal{U}[0.0, 1.0] \) & \si{\mM} \\
Glycerophosphocholine (GPC) & \( a_{9} \) & \( \mathcal{U}[0.0, 3.6] \) & \si{\mM} \\
Glutathione (GSH) & \( a_{10} \) & \( \mathcal{U}[0.0, 3.6] \) & \si{\mM} \\
Myo-Inositol (mIns) & \( a_{11} \) & \( \mathcal{U}[4.0, 12.1] \) & \si{\mM} \\
Lactate (Lac) & \( a_{12} \) & \( \mathcal{U}[0.0, 3.1] \) & \si{\mM} \\
N-Acetylaspartylglutamate (NAAG) & \( a_{13} \) & \( \mathcal{U}[0.0, 2.5] \) & \si{\mM} \\
N-Acetylaspartate (NAA) & \( a_{14} \) & \( \mathcal{U}[7.5, 16.3] \) & \si{\mM} \\
Phosphocholine (PCh) & \( a_{15} \) & \( \mathcal{U}[0.0, 2.4] \) & \si{\mM} \\
Phosphocreatine (PCr) & \( a_{16} \) & \( \mathcal{U}[0.0, 5.5] \) & \si{\mM} \\
Phosphoethanolamine (PE) & \( a_{17} \) & \( \mathcal{U}[0.0, 5.2] \) & \si{\mM} \\
Scyllo-Inositol (Scyllo) & \( a_{18} \) & \( \mathcal{U}[0.0, 0.6] \) & \si{\mM} \\
Serine (Ser) & \( a_{19} \) & \( \mathcal{U}[0.0, 7.3] \) & \si{\mM} \\
Taurine (Tau) & \( a_{20} \) & \( \mathcal{U}[1.2, 6.0] \) & \si{\mM} \\
Macromolecules (MMs) & \( a_{21} \) & \( \mathcal{U}[0.0, 400.0] \) & \si{\mM} \\
\bottomrule
\end{tabular*}
\end{adjustbox}
\begin{adjustbox}{width=0.5\textwidth}
\begin{tabular*}{0.6\textwidth}{@{\extracolsep\fill\hspace{1mm}}lccc@{\extracolsep\fill\hspace{2mm}}}
\toprule
\textbf{Parameter} & \textbf{Notation} & \textbf{Range} & \textbf{Unit} \\
\midrule
Number of Metabolites (+MMs) & \( M \) & \( \{ 21 \} \) & -- \\
Frequency Shifts & \( \epsilon \) & \( \mathcal{U}[-10, 10] \) & \si{\radian\per\second} \\
Lorentzian Broadening & \( \gamma \) & \( \mathcal{U}[2, 25] \) & \si{\per\second} \\
Gaussian Broadening & \( \varsigma \) & \( \mathcal{U}[2, 25] \) & \si{\per\second} \\
Zeroth-Order Phase & \( \phi_0 \) & \( \mathcal{U}[-0.5, 0.5] \) & \si{\radian} \\
First-Order Phase & \( \phi_1 \) & \( \mathcal{U}[-10^{-5}, 10^{-5}] \) & \si{\radian\per\hertz} \\
Baseline Polynomial Order & \( K \) & \( \{2\} \) & -- \\
Baseline Coefficients & \( b_1 \) & \( \mathcal{U}[-600, 200] \) & \si{\au} \\
                      & \( b_2 \) & \( \mathcal{U}[-800, 300] \) & \si{\au} \\
                      & \( b_3 \) & \( \mathcal{U}[-1000, 600] \) & \si{\au} \\
                      & \( b_4 \) & \( \mathcal{U}[-600, 1000] \) & \si{\au} \\
                      & \( b_5 \) & \( \mathcal{U}[-1600, 200] \) & \si{\au} \\
                      & \( b_6 \) & \( \mathcal{U}[-400, 1000] \) & \si{\au} \\
Random Walk Step Size & -- & \( \mathcal{U}[0, 10^5] \) & \si{\au} \\
Random Walk Smoothing & -- & \( \mathcal{U}[1, 10^5] \) & \si{\au} \\
Random Walk Min. Bound & -- & \( \mathcal{U}[-10^6, 0] \) & \si{\au} \\
Random Walk Max. Bound & -- & \( \mathcal{U}[0, 10^6] \) & \si{\au} \\
Complex Gaussian Noise$^{*}$ & \( N \) & \( \mathcal{CN}(0, \sigma^2) \) & \si{\au} \\
Noise Variance & \( \sigma^2 \) & \( \mathcal{U}[10, \sqrt{2} \cdot 5000] \) & \si{\au} \\
Chemical Shift Limits & \( f_{min} \) & \( \{ 0.5 \} \) & \si{\ppm} \\
                      & \( f_{max} \) & \( \{ 4.0 \} \) & \si{\ppm} \\
\bottomrule
\end{tabular*}
\end{adjustbox}
\begin{tablenotes}
\item[$^{*}$] \Ac{snr} ranging from 0 - 40 dB as computed from the ground truth noise and metabolite-only signals over 0.5 to 4.0 ppm.
\end{tablenotes}
\end{table*}

\subsubsection{Training Data Generation}


Synthetic examples were generated \textit{ad-hoc} at training time, with each batch composed of newly sampled signals. This setup allowed for virtually unlimited data during training and reduced the risk of overfitting to a discrete synthetic distribution. The signal parameters~$\boldsymbol{\theta}$ were drawn independently from their respective distributions and passed through Equation~\eqref{eq:observation} to generate spectra. Training was done with a batch size of 16 and every 256 batches, a set of 1024 new samples was used for validation (resulting in a 20/80\% validation/training split).

\subsection{In-Vivo Data} \label{ssec:in-vivo_data}
Data was obtained from 61 healthy volunteers as part of the BrainBeats study. The study adhered to the guidelines of the Institutional Review Board of the University of Amsterdam (the Netherlands). All participants provided written informed consent. Four subjects were excluded following visual inspection due to insufficient data quality, resulting in a final cohort of 57 participants. 

\subsubsection{Acquisition}
In-vivo single-voxel \ac{mrs} data were acquired in the anterior cingulate cortex using a semi-LASER sequence with TE/TR = 36/5000~ms on a 7T Philips scanner, as part of an interleaved fMRI/MRS protocol~\cite{Schrantee20237TfMRSandfMRI}. Acquisition parameters included a volume-of-interest of 25 x 18 x 18~mm$^3$, 1024 sample points, and a spectral bandwidth of 3000~Hz. Water suppression was performed using VAPOR and shimming was optimized using HOS-DLT~\cite{Boer2020HOS_DLT}. For each subject, spectra were obtained across three scan sessions. A total of 64 signal averages were used from the first session, 3×64 from the second, and 2×64 from the third, resulting in 342 spectra across all subjects. Further acquisition details are provided in Appendix~\ref{app:mrs_in_mrs}, Table~\ref{tab:mrsinmrs_invivo}, as part of the \ac{mrsinmrs}~\cite{lin_minimum_2021}.

\subsubsection{Processing}
Processing of the in-vivo \ac{mrs} data involved several steps. Coil combination was performed using a custom method that estimated phase correction and amplitude weighting parameters for each coil from an unsuppressed water reference scan acquired at the beginning of the time series. All subsequent processing was performed using FSL-MRS. Individual transients were frequency- and phase-aligned within the 0.2–4.2~ppm range and, subsequently, averaged. Eddy current correction was then applied using the unsuppressed water reference, followed by removal of nuisance peaks using \ac{hsvd}. Finally, the spectra were frequency- and phase-aligned to \ac{cr} at 3.027~ppm.


\subsubsection{Analysis}
Metabolite quantification of the processed in-vivo spectra was performed using both LCModel and FSL-MRS, using a basis set matched to the acquisition parameters (7T field strength, semi-LASER sequence, TE = 34 ms, 1024 points, 3000 Hz bandwidth), see Appendix~\ref{app:mrs_in_mrs}, Table~\ref{tab:mrsinmrs_invivo} for details on fit settings. The same basis set was employed for both synthetic spectrum generation and quantification to maintain consistency across training and evaluation. It consisted of 20 metabolites together with a single macromolecular baseline; detailed specifications are provided in Table~\ref{tab:sim_params}.

\subsection{Quantification Strategies} \label{sec:quantification_strategies}
\begin{figure*}
    \centering
    \includegraphics[width=2.0\columnwidth]{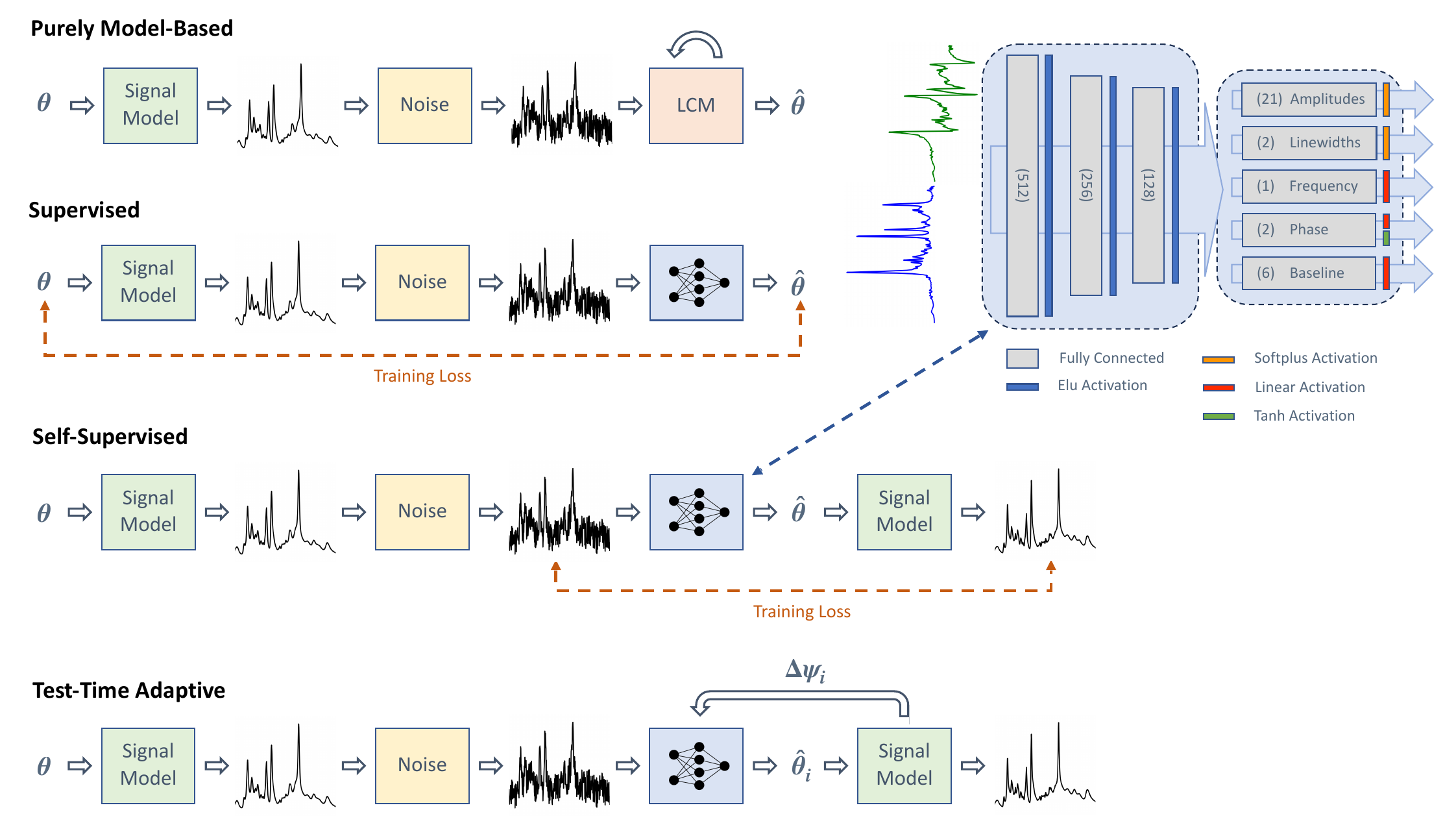}
    \caption{\enspace The schematic provides an overview of the relevant parameter estimation methods. From top to bottom: purely model-based fitting using \ac{lcm}, supervised regression trained to predict metabolite amplitudes directly, self-supervised training that utilizes the underlying signal model to map estimated parameters to their corresponding spectrum, and \ac{tta} to refine predictions via gradient-based optimization at inference. The \ac{nn} employed for the latter three is a simple \ac{mlp} with three fully connected layers, which takes the normalized real and imaginary parts of the frequency domain spectra as input and outputs the estimated parameters of the chosen signal model.} 
    \label{fig:training_strategies}
\end{figure*}

Quantification in \ac{mrs} aims to estimate metabolite concentrations ${a_m}_{m=1}^{M}$ from an observed spectrum. Standard tools such as LCModel fit a parametric signal model using the Levenberg–Marquardt algorithm, which combines gradient descent with Gauss-Newton updates \cite{Gavin2011LM}. We explicitly linked data-driven learning with traditional fitting by comparing strategies that all relied on gradient-based updates. We included a purely model-based baseline that replaced Levenberg-Marquardt with direct gradient descent, alongside \ac{nn}-based approaches. The strategies, summarized in Figure~\ref{fig:training_strategies}, provided a common framework to investigate how data-driven quantification could mitigate \ac{ood} effects and explore the bias--variance trade-offs between learned priors and adaptive fitting.

Let $\mathbf{y}, \mathbf{x}(\boldsymbol{\theta}) \in \mathbb{C}^L$ denote the observed and modeled spectra sampled at $L$ frequency points in the range $f_{\min} \le f \le f_{\max}$, corresponding to 0.5-4.0 ppm throughout this work. Unless stated otherwise, all methods were optimized using Adam~\cite{kingma2015adam} with a learning rate of $1\times10^{-4}$.

To ensure consistency across signal intensities, all methods applied normalization and scaling:
\begin{equation}
    \mathbf{y} \leftarrow \mathbf{y} \ / \ \|\mathbf{y}\|_2.
\end{equation}
The norm was then propagated forward to scale the predicted metabolite amplitudes and baseline parameters:
\begin{equation}
    \hat{a}_m \leftarrow \hat{a}_m \cdot \|\mathbf{y}\|, \quad \hat{b}_k \leftarrow \hat{b}_k \cdot \|\mathbf{y}\|.
\end{equation}
This procedure allowed the models to operate on normalized inputs while preserving the effective signal amplitude in the outputs. 

\subsubsection{Purely Model-Based Fitting}
Our purely model-based approach directly optimized the signal model parameters $\boldsymbol{\theta}$ using gradient descent (Adam with a learning rate of $1\times10^{-1}$, for 1000 epochs). No \ac{nn} was involved; instead, the parameters were treated as learnable tensors and refined iteratively to minimize the residual between the modeled ($\mathbf{x}(\boldsymbol{\theta})$) and observed spectra ($\mathbf{y}$):
\begin{equation}
    \hat{\boldsymbol{\theta}} = \arg\min_{\boldsymbol{\theta}} \| \mathbf{y} - \mathbf{x}(\boldsymbol{\theta}) \|_2^2.
\end{equation}
To enforce constraints, such as positivity for metabolite amplitudes and lineshape parameters, we applied the same activation functions used in the learning-based strategies (see Section~\ref{sssec:arch} for details). 

\subsubsection{Supervised Regression}
In the supervised setting, a \ac{nn} $g_{\boldsymbol{\psi}}$ with trainable weights~$\boldsymbol{\psi}$ maps each input spectrum $\mathbf{y}$ to signal model parameters $\boldsymbol{\hat{\theta}} = g_{\boldsymbol{\psi}}(\mathbf{y})$. The training objective minimizes the \ac{mae} between predicted and reference parameters, amortized over the training dataset $\mathcal{D}_{train}$:
\begin{equation} 
    \boldsymbol{\psi}^* = \arg\min_{\boldsymbol{\psi}} \frac{1}{|\mathcal{D}_{train}|} \sum_{(\mathbf{y}, \boldsymbol{\theta}) \in \mathcal{D}_{train}} \mathcal{L}_{\text{MAE*}}(\boldsymbol{\theta}, g_{\boldsymbol{\psi}}(\mathbf{y})). 
\end{equation}
The scaled \ac{mae} loss is defined as
\begin{equation}
    \mathcal{L}_{\text{MAE*}}(\boldsymbol{\theta}, \boldsymbol{\hat{\theta}}) = \left| \frac{\boldsymbol{\theta} - p_{\min}}{p_{\max} - p_{\min}} - \frac{\boldsymbol{\hat{\theta}} - p_{\min}}{p_{\max} - p_{\min}} \right|,
\end{equation}
where $p_{\min}$ and $p_{\max}$ denote the lower and upper bounds of each parameter, derived from the simulation priors. This scaling ensures balanced optimization across all components of $\boldsymbol{\theta}$. Although strictly only the metabolite concentrations need to be estimated, we chose to predict all signal parameters to allow direct comparison with alternative methods and to ensure that the resulting fits are structurally analogous. After optimization, the trained network $g_{\boldsymbol{\psi}^*}$ is fixed and used for rapid prediction (inference) on unseen test data.

\subsubsection{Self-Supervised Regression}
The self-supervised strategy integrates the physics-based signal model directly into the training loop. Unlike supervised learning, this method trains the network~$g_{\boldsymbol{\psi}}$ without requiring reference parameters $\boldsymbol{\theta}$. Instead, the optimization objective minimizes the reconstruction error between the modeled spectrum $\mathbf{x}(\boldsymbol{\hat{\theta}}) = \mathbf{x}(g_{\boldsymbol{\psi}}(\mathbf{y}))$ and the observed spectrum $\mathbf{y}$: 
\begin{equation} 
    \boldsymbol{\psi}^* = \arg\min_{\boldsymbol{\psi}} \frac{1}{|\mathcal{D}_{train}|} \sum_{\mathbf{y} \in \mathcal{D}_{train}} \| \mathbf{y} - \mathbf{x}(g_{\boldsymbol{\psi}}(\mathbf{y})) \|_2^2. 
\end{equation}
The key distinction from purely model-based least-squares fitting is the optimization target: minimizing the loss by updating the network weights $\boldsymbol{\psi}$ enables the network to learn priors from the dataset $\mathcal{D}_{train}$. Furthermore, the trained network $g_{\boldsymbol{\psi}^*}$ can instantly output parameter predictions during inference, rather than requiring iterative parameter fitting for each new spectrum.

\subsubsection{Test-Time Adaptation}
\Ac{tta} refers to refining model predictions during inference, allowing the network to adjust dynamically to previously unseen data $\mathcal{D}_{test}$. 
This is done by fine-tuning the pretrained network $g_{\boldsymbol{\psi}^*}$ using least-squares. Unless stated otherwise, all \ac{tta} procedures in this work initialize the network from the supervised pretrained model, with alternative initializations reported in Appendix~\ref{app:add_materials}. 

\subsubsubsection{Test-Time Instance Adaptation}
Adapts the pretrained network $g_{\boldsymbol{\psi}^*}$ to a single spectrum $\mathbf{y} \in \mathcal{D}_{test}$. The approach is particularly relevant for clinical deployment, where predictions must be robust to single-subject variability. It fine-tunes the network weights $\boldsymbol{\psi}$ for $j \in \{1, ..., J = 50\}$ steps using least-squares:
\begin{equation} 
    \boldsymbol{\psi}^{(j+1)} = \arg\min_{\boldsymbol{\psi}^{(j)}} \| \mathbf{y} - \mathbf{x}(g_{\boldsymbol{\psi}^{(j)}}(\mathbf{y})) \|_2^2. 
\end{equation}
After adaptation, the network $g_{\boldsymbol{\psi}^{(J)}}$ produces the parameter prediction $\boldsymbol{\hat{\theta}} = g_{\boldsymbol{\psi}^{(J)}}(\mathbf{y})$. This procedure is repeated independently for each spectrum, allowing the network to refine predictions in response to distribution shifts while retaining priors learned from the training dataset.

\subsubsubsection{Test-Time Online Adaptation}
Updates the network $g_{\boldsymbol{\psi}^*}$ continuously as new batches $\mathcal{B}_i \subset \mathcal{D}_{test}$ arrive. This strategy allows the model to adjust continually to evolving data characteristics, making it suitable for streaming or high-throughput acquisition settings. For each batch, the network weights $\boldsymbol{\psi}$ are adapted,
\begin{equation} 
    \boldsymbol{\psi}_{i+1} = \arg\min_{\boldsymbol{\psi}_i} \frac{1}{|\mathcal{B}_{i}|} \sum_{\mathbf{y} \in \mathcal{B}_{i}} \| \mathbf{y} - \mathbf{x}(g_{\boldsymbol{\psi}_i}(\mathbf{y})) \|_2^2,
\end{equation}
so that the updated model $g_{\boldsymbol{\psi}_{i+1}}$ and the corresponding predictions $\boldsymbol{\hat{\theta}}_i = g_{\boldsymbol{\psi}}(\mathbf{y})$ for all $\mathbf{y} \in \mathcal{B}_{i}$ are obtained as part of the same update process. The adapted weights $\boldsymbol{\psi}_{i+1}$ then initialize the model for the next incoming batch~$\mathcal{B}_{i+1}$ (default batch size $|\mathcal{B}_{i}| = 16$).

\subsubsubsection{Test-Time Domain Adaptation}
Refines the pretrained network $g_{\boldsymbol{\psi}^*}$ using the entire test dataset $\mathcal{D}_{test}$ to account for distribution shifts. Domain adaptation is particularly beneficial in research settings where the full test dataset is available before inference, but could also be used to recalibrate a network to a new institute/scanner. The network weights are adapted by minimizing:
\begin{equation} 
    \boldsymbol{\psi}^* = \arg\min_{\boldsymbol{\psi}} \frac{1}{|\mathcal{D}_{test}|} \sum_{\mathbf{y} \in \mathcal{D}_{test}} \| \mathbf{y} - \mathbf{x}(g_{\boldsymbol{\psi}_i}(\mathbf{y})) \|_2^2.
\end{equation}
The adapted network $g_{\boldsymbol{\psi}^*}$ is then used to produce predictions $\boldsymbol{\hat{\theta}} = g_{\boldsymbol{\psi}^*}(\mathbf{y})$ for all spectra in $\mathcal{D}_{test}$. A mini batch size of 16 is used and optimization is run for 1000 epochs. 

\subsubsection{Neural Network Architecture} \label{sssec:arch}
All data-driven strategies used the same \ac{nn} architecture, illustrated in Figure~\ref{fig:training_strategies}: a three-layer \ac{mlp} with ELU~\cite{clevert2016elu} activations. Outputs were constrained via activation functions to enforce physical plausibility: metabolite amplitudes are passed through a softplus function to ensure non-negativity, linewidths were softplus-transformed and shifted by 1, frequency offsets and other linear parameters remained unconstrained, and the first-order phase was restricted to a narrow range using a scaled hyperbolic tangent activation.

The choice of an \ac{mlp} over other architectures such as \acp{cnn} was intentional: our analysis focused on the impact of training strategy rather than architectural design. By using a general-purpose function approximator, we maximized the \ac{nn}'s flexibility while minimizing architecture-specific biases. The \ac{mlp} was optimized using sweep runs for hyperparameter tuning (depth, width, activation, etc.). Exact architecture implementation details are provided in Appendix~\ref{app:impl_details}, Table~\ref{tab:arch_compare} along with results obtained with a \ac{cnn} in Appendix \ref{app:add_materials}.

\subsection{Evaluation}
The prediction accuracy was assessed using both absolute and relative error metrics. 
The \ac{mae} was computed directly between the predicted amplitudes $\hat{a}_m$ and the true concentrations $a_m$:
\begin{equation}
\text{MAE} = \frac{1}{M-1} \sum_{m=1}^{M-1} \left| \hat{a}_m - a_m \right|,
\end{equation}
where $M-1$ is the number of metabolites (excluding \acp{mm}). 

To enable fair comparison of relative estimates, we calculated an optimal scaling factor $w_{\mathrm{opt}}$ that minimized the absolute error between the scaled relative metabolite concentration estimates $\hat{a}_m$ and absolute ground truth values $a_m$:
\begin{equation} \label{eq:optimal_scaling}
    w_{\mathrm{opt}} = \arg\min_w \sum_{m=1}^{M-1} \left| w  \  \hat{a}_m - a_m \right|.
\end{equation}
Using this scaling, the \ac{mosae} is defined as
\begin{equation} \label{eq:mosae}
    \mathrm{MOSAE} = \frac{1}{M} \sum_{m=1}^{M-1} \left| w_{\mathrm{opt}} \ \hat{a}_m - a_m \right|.
\end{equation}
This metric allowed a fair comparison of concentration estimates independent of a water reference or other metabolite references such as \ac{tcr}.

We further assessed the agreement between predicted and true concentrations via linear regression
\begin{equation}
a_m = \alpha \ \hat{a}_m + \beta,
\end{equation}
yielding four interpretable metrics: slope $\alpha$ (proportional bias), intercept $\beta$ (constant bias), coefficient of determination $R^2$ (explained variance), and \ac{rmse}
\begin{equation}
\sigma = \sqrt{\frac{1}{|\mathcal{D}_{test}|} \sum_{u=1}^{|\mathcal{D}_{test}|} \left ((\hat{a}_m)_u - (a_m)_u \right )^2}.
\end{equation}

\subsection{Experiments}
The primary purpose of the experiments is to investigate the inherent data biases and resilience of these methods to \ac{ood} effects and domain shift, ultimately assessing trade-offs between accuracy, robustness, and generalizability.

\subsubsection{Controlled Simulation Experiments}
For each test scenario, 10,000 synthetic spectra were generated, focusing on two main categories of perturbations.

\subsubsubsection{Metabolite Concentration Effects}
\begin{itemize}
    \item ID (Mid-Range): Training and testing on the central 50\% of the concentration range (baseline).
    \item OoD (Full-Range): Training on mid-range, testing across the full range (extrapolation stress test).
    \item ID (Full-Trained): Training and testing on the full range (reference).
\end{itemize}

\subsubsubsection{Signal Parameter Perturbations}
Test scenarios with wider ranges of \ac{snr}, linewidth, frequency/phase shifts, and baseline variations than seen in training. 

\subsubsection{In-Vivo Data Testing}
From the 342 original 7T acquisitions with 64 transients each, we produced 1,710 spectra by forming subsets of 4, 8, 16, and 32 transients to control \ac{snr}. Since true in-vivo concentrations are unknown, FSL-MRS fits from the full 64 averages served as pseudo ground truth for signal parameter estimates and error estimation. LCModel and mixed references are described in Appendix \ref{app:add_materials}. The in-vivo spectra were then filtered to match the simulation conditions: ID (Mid Range), OoD (Full Range), and ID (Full Trained). This setup enabled a systematic study of two factors: the domain shift from synthetic to real data, and the impact of training on narrow versus broad synthetic ranges when applied to in-vivo spectra.

\section{Results} \label{sec:results}

\subsection{Results on Simulated Data} \label{ssec:results_sim}

\subsubsection{Overall Quantification Performance}

\begin{table*}[!t]
\centering
\caption{\enspace 
Comparison of quantification methods across test scenarios of 10,000 spectra.
\textbf{ID (Mid-Range)}: Models trained and tested on the central 50\% metabolite concentration range.
\textbf{OoD (Full-Range)}: Models trained on the central 50\% concentration range, but tested across the entire range of concentrations to assess extrapolation.
\textbf{ID (Full-Trained)}: Models trained and tested on the full concentration range. Lowest error values are highlighted in \textbf{bold}.}
\label{tab:sim_performance_sel} 
\vspace{2mm}
\begin{adjustbox}{width=1.0\textwidth}
\begin{tabular}{lccc c}
\toprule
\multirow{2}{*}{\textbf{Method}} 
  & \multicolumn{3}{c}{\textbf{MOSAE ↓ ($\pm$ SE)}} 
  & \multirow{2}{*}{\textbf{Time ↓ (ms/sample)}} \\
\cmidrule(lr){2-4}
  & \textbf{ID (Mid-Range)} & \textbf{OoD (Full-Range)} & \textbf{ID (Full-Trained)} & \\
\midrule
Supervised            & \bftab 0.2764 (± 0.0032) & 0.5537 (± 0.0058) & \bftab 0.3896 (± 0.0047) & \bftab 0.1565 \\
Self-Supervised       & 0.3734 (± 0.0044) & 0.5401 (± 0.0059) & 0.4548 (± 0.0059) & 0.1574 \\
Test-Time Instance Adaptive  & 0.3767 (± 0.0049) &  \bftab 0.4420 (± 0.0052) & 0.4378 (± 0.0060) & 30.4637 \\
Test-Time Online Adaptive    & 0.2788 (± 0.0032) & 0.5520 (± 0.0058) & 0.3923 (± 0.0048) & 0.2454 \\
Test-Time Domain Adaptive    & 0.3621 (± 0.0044) & 0.4717 (± 0.0053) & 0.4478 (± 0.0055) & 104.5304 \\
Purely Model-Based    & 0.5299 (± 0.0070) & 0.5238 (± 0.0071) & 0.5229 (± 0.0071) & 424.9928 \\
FSL-MRS               & 0.5369 (± 0.0073) & 0.5210 (± 0.0073) & 0.5201 (± 0.0074) & 962.7825 \\
LCModel               & 0.5704 (± 0.0067) & 0.5843 (± 0.0072) & 0.5832 (± 0.0072) & 78.0310 \\
\bottomrule
\end{tabular}
\end{adjustbox}
\begin{tablenotes}
\end{tablenotes}
\end{table*}

The performance, measured by \ac{mosae} across three metabolite concentration scenarios, diverged significantly between methods reliant on learned priors and adaptive strategies (Table~\ref{tab:sim_performance_sel}). In the ideal, restricted-range \ac{id} setting, supervised regression achieved the lowest errors (0.2764 ± 0.0032). However, when restricted-range models were evaluated on the full concentration range (\ac{ood} extrapolation), supervised regression showed the strongest decrease in accuracy, with errors nearly doubling (0.5537 ± 0.0058). Domain adaptation also improved over baseline regression. Training on the full concentration range removed the extrapolation gap, but all data-driven methods exhibited increased errors in this broader setting (e.g. supervised: 0.3896 ± 0.0047).

Classical fitting approaches (purely model-based gradient descent, FSL-MRS, LCModel) yielded similar error magnitudes across all three scenarios, but with runtimes orders of magnitude slower than the data-driven models.
Extended results, including equivalent \ac{mae} performance, \ac{cnn}-based baselines, self-supervised initialization of \ac{tta}, and alternative iteration counts for instance adaptation, are provided in Tables~\ref{tab:sim_performance_mae} and \ref{tab:sim_performance_mosae} in Appendix~\ref{app:add_materials}. A comparison of \ac{mae} and \ac{mosae} showed similar relative performance trends across quantification methods. Furthermore, similar observations regarding performance degradation from \ac{id} to \ac{ood} were found for the \ac{cnn}-based baselines. 

\subsubsection{Metabolite Distributions}

\begin{figure*}
    \vspace{2cm}
    \centering
    \includegraphics[width=2.0\columnwidth]{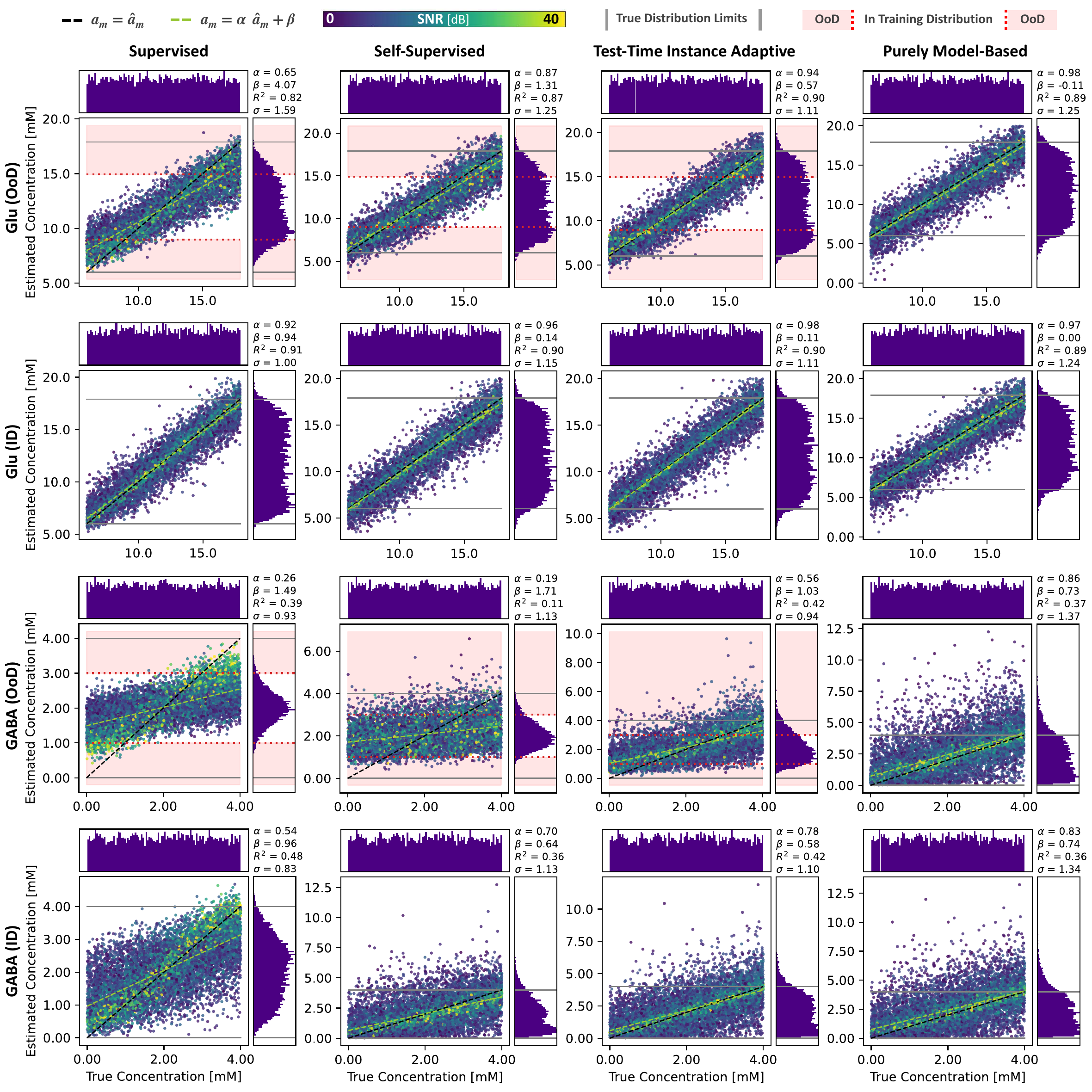}
    \caption{\enspace Scatter plots with marginal histograms comparing predicted versus true concentrations of \ac{glu} and \ac{gaba} across 10,000 simulated spectra. Models were evaluated under two scenarios for the full concentration range: trained on mid-range concentrations (\ac{ood}) or trained on the full range (\ac{id}). 
    Points are colored by \ac{snr}, and regression lines with corresponding statistics (slope $\alpha$, intercept $\beta$, R$^2$, and \ac{rmse} $\sigma$) are shown.
    \vspace{2cm}
    }
    \label{fig:sim_dist_comb_1_mae}
\end{figure*}

Analysis of metabolite distributions revealed that for \ac{glu}, the predicted values closely matched the uniform ground truth, with only minor deviations (Figure~\ref{fig:sim_dist_comb_1_mae}). However, for \ac{gaba} the predicted distributions were noticeably narrower. The predictions of the supervised method exhibited a confinement to its training distribution, with few estimates observed outside its trained range, particularly evident for lower \ac{snr} spectra. Self-supervised training similarly showed a bias towards its training distribution, with no significant trend observed in relation to \ac{snr} variations. In contrast, test-time instance adaptation demonstrated better coverage of the full range of concentrations. In the full-range (\ac{id}) scenario, its \ac{rmse} was slightly higher. The purely model-based approach showed constant performance and reasonable agreement with the ground truth distribution, though it consistently exhibits a high \ac{rmse}.
Overall, higher \acp{rmse} were observed for the \ac{ood} cases compared to the \ac{id} cases for both supervised and self-supervised methods, whereas test-time instance adaptive and purely model-based approaches maintained their performance across these scenarios.

The corresponding plots for the remaining quantification methods and the respective figures illustrating optimally scaled concentrations for relative metabolite quantification comparison are in Appendix~\ref{app:add_materials}, Figures \ref{fig:sim_dist_comb_2_mae}, \ref{fig:sim_dist_comb_1_mosae}, \ref{fig:sim_dist_comb_2_mosae}. Alternative metabolites showed similar effects and are provided in Appendix~\ref{app:add_materials}, Figure~\ref{fig:sim_dist_ood_1_sel_mae} and \ref{fig:sim_dist_ood_2_sel_mae} for \ac{ood} and Figure~\ref{fig:sim_dist_idft_1_sel_mae} and \ref{fig:sim_dist_idft_2_sel_mae} for \ac{id}.

\subsubsection{Signal Parameter Perturbations}
\begin{figure*}
    \includegraphics[width=2.0\columnwidth]{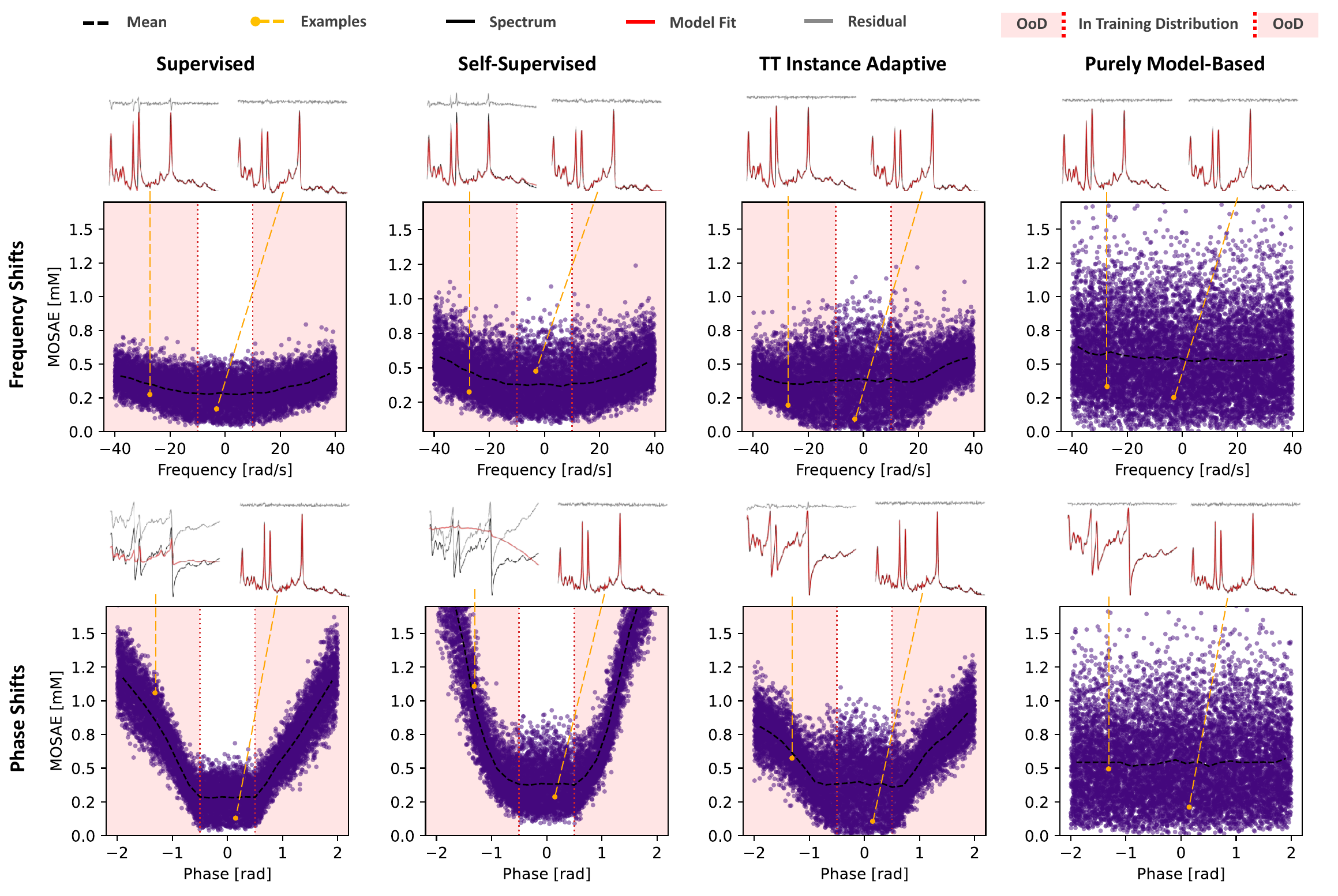}
    \caption{\enspace Scatter plots of quantification accuracy (\ac{mosae}) across 10,000 simulated spectra as a function of ground truth frequency and zeroth-order phase shifts. Data-driven methods (supervised, self-supervised, and test-time instance adaptive) were compared against purely model-based fitting. Each point represents a single spectrum, illustrating method-specific sensitivity to core signal parameter variations. Example spectra from \ac{id} and \ac{ood} regions are shown above the scatter plots, with lines connecting each spectrum to its corresponding point, including fitted signals and residuals for all methods.}
    \label{fig:sim_rang_sel_mosae}
\end{figure*}

Test-time instance adaptive had the least effect on \ac{mosae} in \ac{ood} data as compared to the other \ac{nn} methods (Figure~\ref{fig:sim_rang_sel_mosae}). 

For \ac{id} spectra, all methods maintain uniform performance and good visual fits. However, under \ac{ood} conditions, frequency shifts caused minor degradation, while \ac{ood} phase shifts led to more pronounced errors for data-driven methods. Other signal parameter variations (\ac{snr}, linewidth, \acp{mm}, baseline, random nuisance effects) showed only minor \ac{ood} degradation, as can be seen in Appendix~\ref{app:add_materials}, Figures \ref{fig:sim_rang_1_1_mosae}, \ref{fig:sim_rang_1_2_mosae}, \ref{fig:sim_rang_2_1_mosae}, and \ref{fig:sim_rang_2_2_mosae}.

\subsubsection{Performance Across Parameter Ranges}

\begin{figure*}
    \vspace{1cm}
    \centering
    \includegraphics[width=2.0\columnwidth]{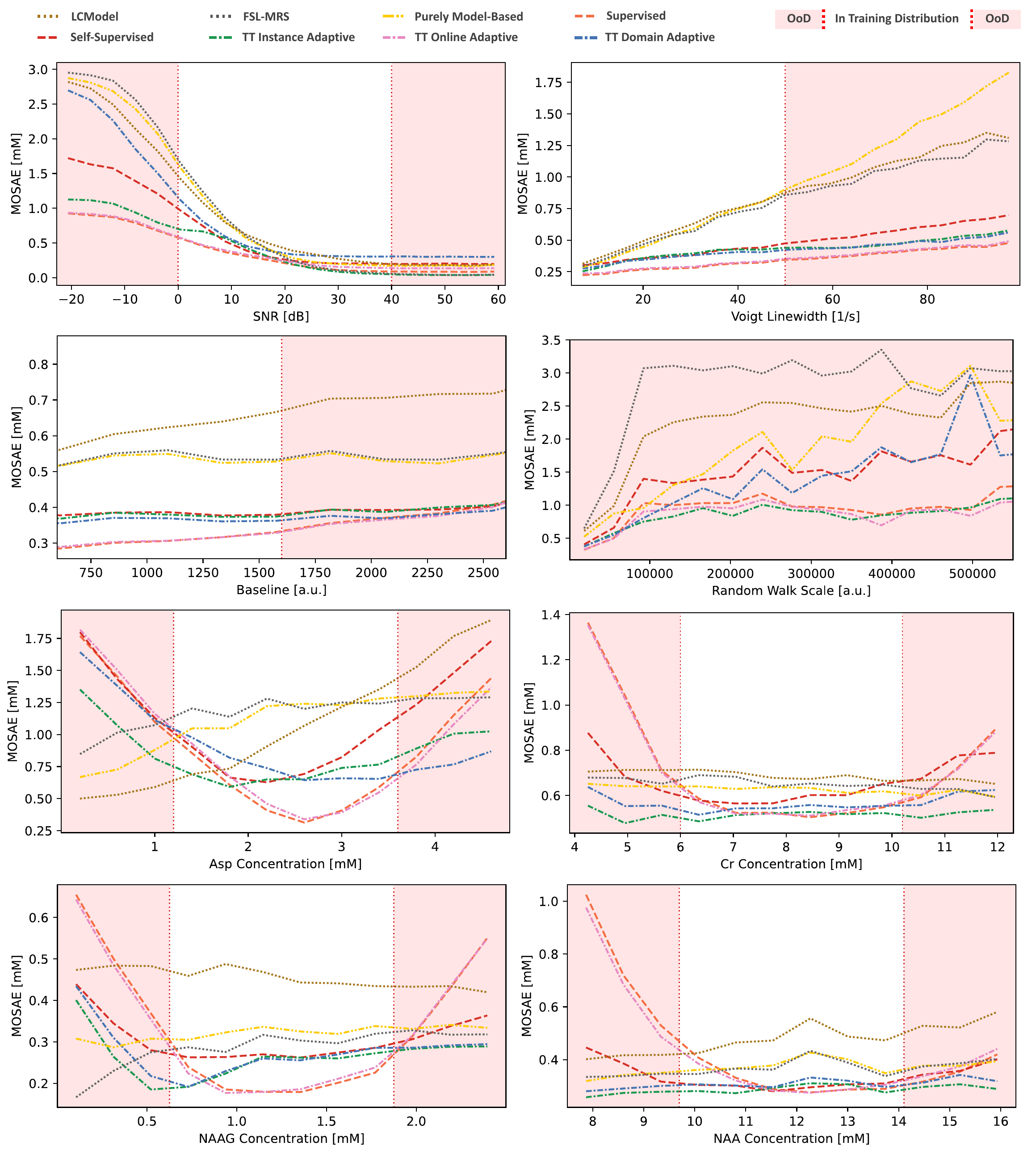}
    \caption{\enspace Mean quantification error (\ac{mosae}) across 10,000 simulated spectra for all methods under \ac{id} and \ac{ood} conditions. Results are shown for key signal parameters (\ac{snr}, linewidth, baseline, random walk) and metabolites (\ac{asp}, \ac{cr}, \ac{naag}, \ac{naa}). Each curve represents the mean error across spectra binned by the corresponding parameter value or metabolite concentration, summarizing method performance trends and sensitivity to challenging conditions.
    \vspace{1cm}
    }
    \label{fig:sim_mean_lines_all}
\end{figure*}

Figure~\ref{fig:sim_mean_lines_all} complements the single-spectrum scatter plots of Figure~\ref{fig:sim_rang_sel_mosae}, highlighting systematic trends across methods and scenarios. Data-driven approaches generally show increased errors under extreme \ac{ood} conditions, while \ac{tta} and model-based methods maintain more stable performance across all parameters and metabolites. The observed trends hold for additional metabolites and signal parameters, as can be seen in Appendix~\ref{app:add_materials}, Figures \ref{fig:sim_perf_metabs_1_1_mosae}, \ref{fig:sim_perf_metabs_1_2_mosae}, \ref{fig:sim_perf_metabs_1_3_mosae}, and \ref{fig:sim_perf_other_1_mosae}.

\subsection{Results on In-Vivo Data} \label{ssec:results_invivo}

\subsubsection{Overall Quantification Performance}
The \acp{mosae} reported for in-vivo data (Table~\ref{tab:invivo_fslmrs_sel_mosae}) were generally higher across most data-driven methods, indicating a domain shift between synthetic and in-vivo spectra. Supervised and self-supervised models exhibited the strongest increase in error. \Ac{tta} methods retained lower errors, with domain adaptation achieving the best overall performance among adaptive approaches (0.3425 ± 0.0102 \ac{mosae} in the \ac{ood} scenario). The purely model-based approach performed comparably well to the best adaptive methods (0.3431 ± 0.0106 \ac{mosae} in the \ac{ood} scenario).

Additional results utilizing alternative pseudo ground truths (including the mean of FSL-MRS and LCModel, and LCModel alone) confirm these general trends across the methods (Tables \ref{tab:invivo_mix_mae}, \ref{tab:invivo_mix_mosae}, \ref{tab:invivo_lcmodel_mae}, and \ref{tab:invivo_lcmodel_mosae} in Appendix~\ref{app:add_materials}). Furthermore, extended initialization experiments reveal that test-time instance adaptation initialized from scratch achieved the lowest deviations from FSL-MRS pseudo ground truths, outperforming models initialized from the pretrained network (Tables \ref{tab:invivo_fslmrs_mae} and \ref{tab:invivo_fslmrs_mosae} in Appendix~\ref{app:add_materials}).

\begin{table*}
\caption{\enspace 
Comparison of quantification methods on 1,710 in-vivo spectra using pseudo ground truth: \textbf{FSL-MRS}. The spectra were filtered to create equivalent scenarios to the simulated test scenarios: \textbf{ID (Mid-Range)}, \textbf{OoD (Full-Range)}, and \textbf{ID (Full-Trained)} Lowest error values are highlighted in \textbf{bold}. Comparisons where FSL-MRS is evaluated against its own estimates (using different numbers of averages: 4, 8, 16, 32, 64 vs. 64) are shown in \textit{italics}.
}
\label{tab:invivo_fslmrs_sel_mosae} 
\vspace{2mm}
\begin{adjustbox}{width=1.0\textwidth}
\begin{tabular}{lccc c}
\toprule
\multirow{2}{*}{\textbf{Method}} 
  & \multicolumn{3}{c}{\textbf{MOSAE ↓ ($\pm$ SE)}} 
  & \multirow{2}{*}{\textbf{Time ↓ (ms/sample)}} \\
\cmidrule(lr){2-4}
  & \textbf{ID (Mid-Range)} & \textbf{OoD (Full-Range)} & \textbf{ID (Full-Trained)} \\
\midrule
Supervised            & 0.5164 (± 0.0170) & 0.5533 (± 0.0136) & 0.5562 (± 0.0159) & \bftab 0.1145 \\
Self-Supervised       & 0.5773 (± 0.0217) & 0.6249 (± 0.0171) & 0.5037 (± 0.0138) & 0.1347 \\
Test-Time Instance Adaptive  & 0.4498 (± 0.0144) & 0.4818 (± 0.0115) & 0.4489 (± 0.0118) & 212.2983 \\
Test-Time Online Adaptive    & 0.4793 (± 0.0158) & 0.5156 (± 0.0127) & 0.5101 (± 0.0143) & 0.3400 \\
Test-Time Domain Adaptive    & \bftab 0.3052 (± 0.0120) & \bftab 0.3425 (± 0.0102) & 0.3490 (± 0.0104) & 312.6671 \\
Purely Model-Based    & 0.3168 (± 0.0130) & 0.3431 (± 0.0106) & \bftab 0.3431 (± 0.0106) & 3154.4558 \\
FSL-MRS               & \ittab 0.2104 (± 0.0119) & \ittab 0.2210 (± 0.0097) & \ittab 0.2210 (± 0.0097) & 475.4806 \\
LCModel               & 0.7027 (± 0.0262) & 0.7611 (± 0.0207) & 0.7611 (± 0.0207) & 274.6317 \\
\bottomrule
\end{tabular}
\end{adjustbox} 
\end{table*}

\subsubsection{Metabolite Distributions}

\begin{figure*}
    \vspace{2cm}
    \centering
    \includegraphics[width=2.0\columnwidth]{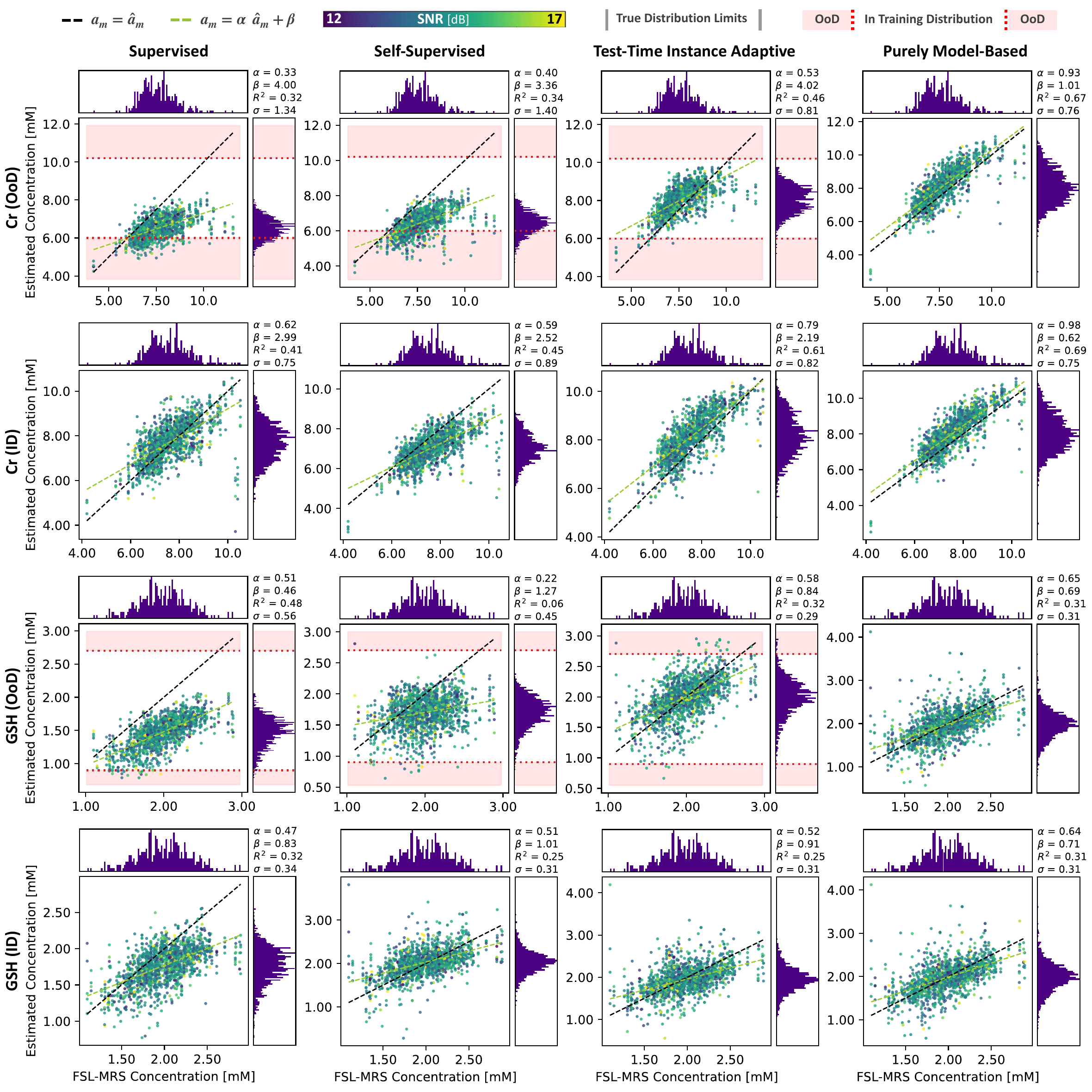}
    \caption{\enspace Scatter plots with marginal histograms comparing predicted versus pseudo-true (FSL-MRS estimates) concentrations of \ac{cr} and \ac{gsh} across 1,710 in-vivo spectra. Models were evaluated under two scenarios for the full concentration range: trained on mid-range concentrations (\ac{ood}) or trained on the full range (\ac{id}). 
    Data-driven methods include supervised, self-supervised, and test-time instance adaptive approaches, compared with purely model-based fitting.  
    \vspace{2cm}
    }
    \label{fig:invivo_fsl_64_dist_comb_1_mae}
\end{figure*}

\begin{figure*}
    \vspace{2cm}
    \centering
    \includegraphics[width=2.0\columnwidth]{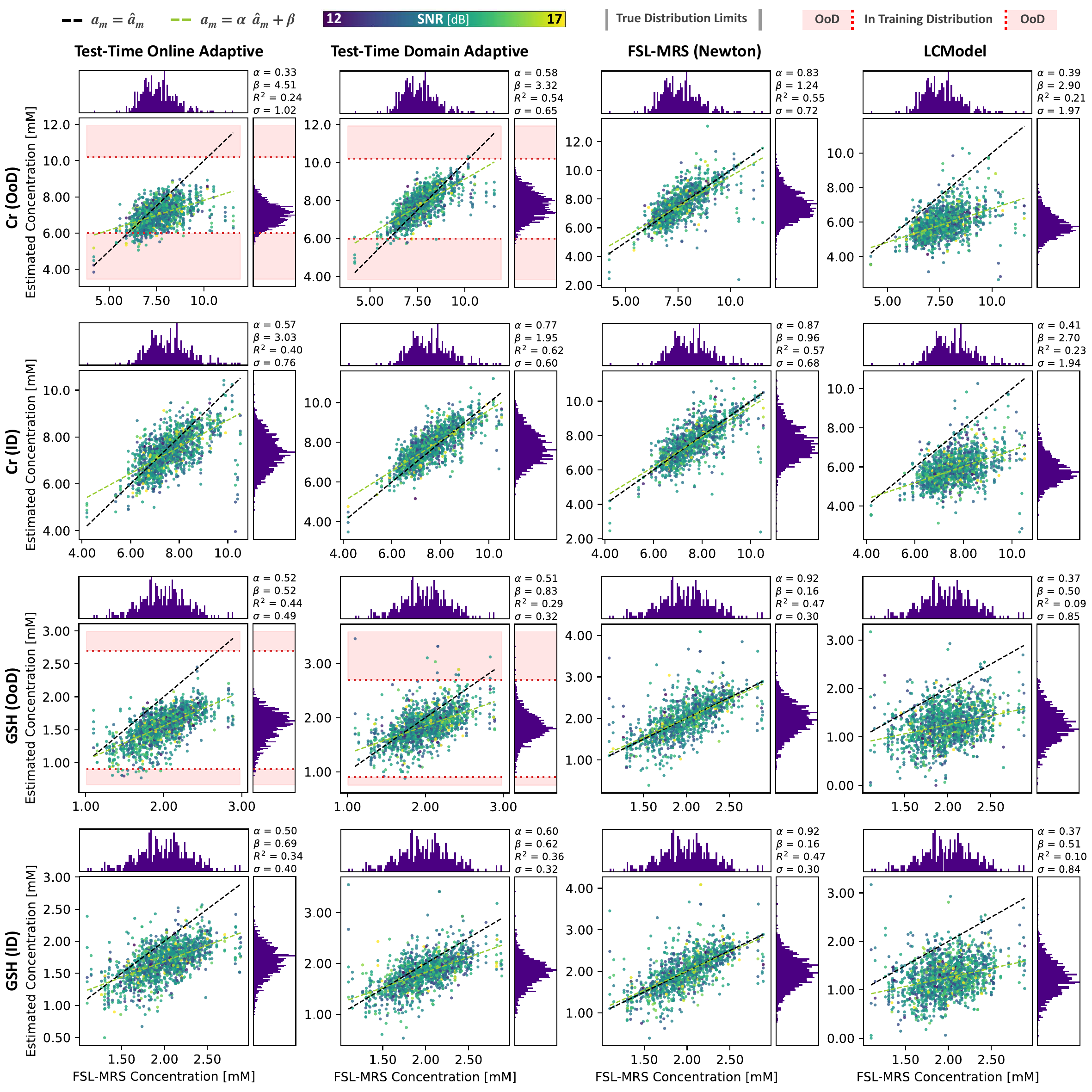}
    \caption{\enspace Scatter plots with marginal histograms comparing predicted versus pseudo-true (FSL-MRS estimates) concentrations of \ac{cr} and \ac{gsh} across 1,710 in-vivo spectra. Models were evaluated under two scenarios for the full concentration range: trained on mid-range concentrations (\ac{ood}) or trained on the full range (\ac{id}).
    Methods included test-time online adaptive approaches, test-time domain adaptive approaches compared with FSL-MRS (Newton) and LCModel.  
    \vspace{2cm}
    }
    \label{fig:invivo_fsl_64_dist_comb_2_mae}
\end{figure*}

Analyzing the metabolite distributions, supervised and self-supervised models exhibit narrow distributions with reduced slopes, consistent with regression toward the mean (Figures~\ref{fig:invivo_fsl_64_dist_comb_1_mae} and \ref{fig:invivo_fsl_64_dist_comb_2_mae}). This effect was less pronounced when considering relative concentrations (see Appendix, Figures \ref{fig:invivo_fsl_64_dist_comb_1_mosae} and \ref{fig:invivo_fsl_64_dist_comb_2_mosae}). Test-time online adaptation remained similar to the supervised baseline, while instance and domain adaptation produced broader distributions with slopes closer to unity. The purely model-based approach was closely aligned with FSL-MRS, whereas LCModel showed the largest deviation. 

As in the simulations, the transition from \ac{id} to \ac{ood} testing was reflected in the distributions, with adaptive models extrapolating more effectively. Additional metabolites showed similar effects and are shown in Appendix~\ref{app:add_materials}, Figures \ref{fig:invivo_fsl_64_dist_ood_1_sel_mae}, \ref{fig:invivo_fsl_64_dist_ood_2_sel_mae}, \ref{fig:invivo_fsl_64_dist_idft_1_sel_mae}, and \ref{fig:invivo_fsl_64_dist_idft_2_sel_mae}.

\subsubsection{Signal Parameter Perturbations}

\begin{figure*}
    \centering
    \includegraphics[width=2.0\columnwidth]{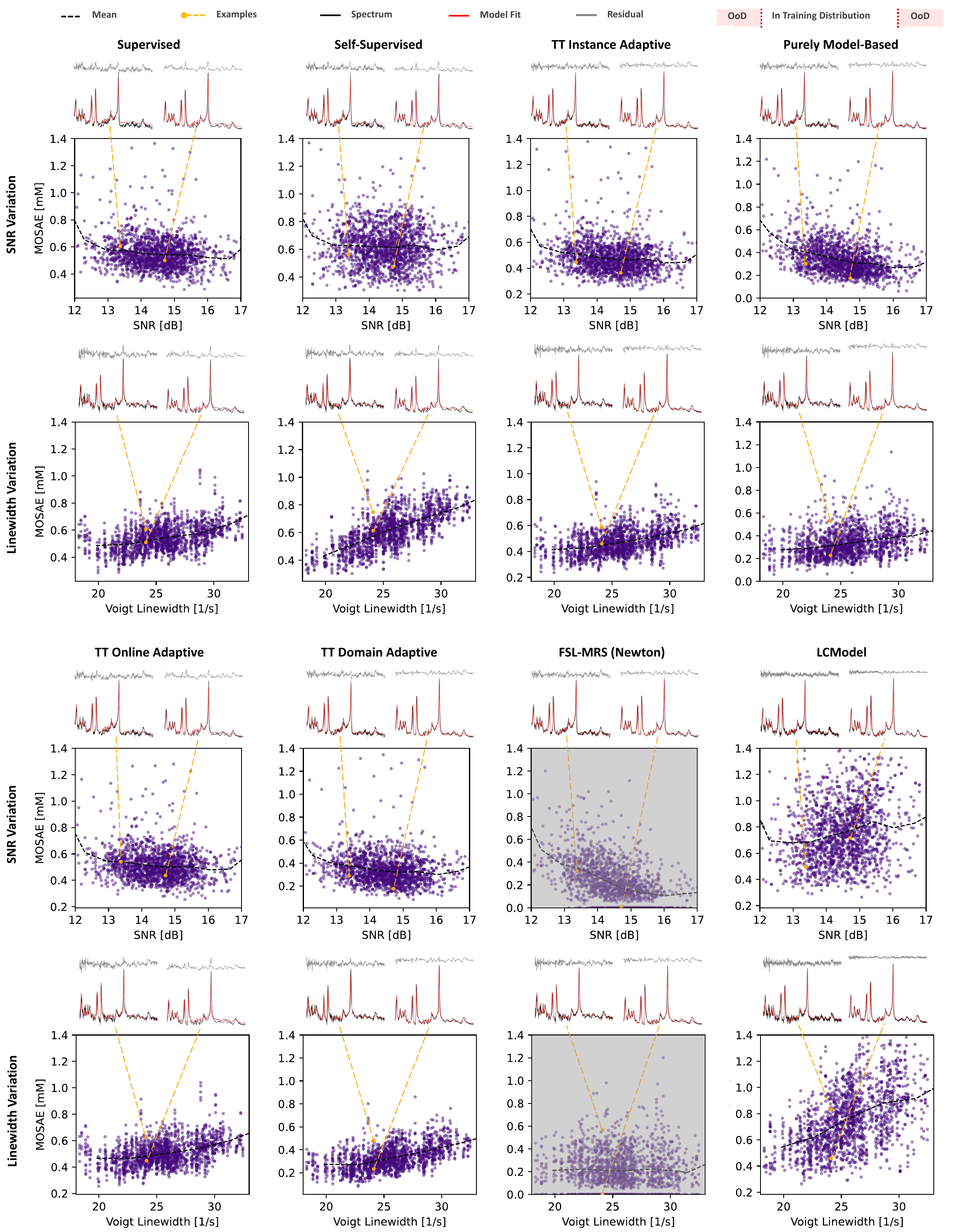}
    \caption{\enspace Scatter plots showing quantification accuracy (\ac{mosae}) across 1,710 in-vivo spectra as a function of estimated \ac{snr}, linewidth. All methods are compared, with FSL-MRS re-estimating for 4, 18, 16 and 32 \ac{nsa} then comparing against the 64 \ac{nsa} pseudo ground truth. Each point represents one spectrum, illustrating method-specific sensitivity to core signal parameter variations.}
    \label{fig:invivo_fsl_64_rang_sel_mosae}
\end{figure*}

Most data-driven models maintain stable deviations from the pseudo ground truth across different \ac{snr} and linewidth conditions (Figure~\ref{fig:invivo_fsl_64_rang_sel_mosae}). Fitting-based methods show stronger variability: LCModel produces larger errors at high \ac{snr} compared to FSL-MRS, and all fitting approaches show increasing errors with broader linewidths. Supervised and self-supervised regressions display some inaccuracies even under \ac{id} conditions, while \ac{tta} methods remain relatively stable. 

Extended sensitivity analyses, including frequency and phase shifts, baseline, and \ac{mm} variations, showed the same stable deviations from the pseudo ground truth, as seen in Appendix~\ref{app:add_materials}, Figures \ref{fig:invivo_fsl_64_rang_1_1_mosae}, \ref{fig:invivo_fsl_64_rang_1_2_mosae}, \ref{fig:invivo_fsl_64_rang_2_1_mosae}, and \ref{fig:invivo_fsl_64_rang_2_2_mosae}. As with the simulations, overall performance trends aggregated in binned means across the full parameter ranges are provided in Appendix~\ref{app:add_materials}, Figures \ref{fig:invivo_fsl_64_perf_metabs_1_1_mosae}, \ref{fig:invivo_fsl_64_perf_metabs_1_2_mosae}, \ref{fig:invivo_fsl_64_perf_metabs_1_3_mosae}, and \ref{fig:invivo_fsl_64_perf_other_1_mosae}.

\section{Discussion} \label{sec:discussion}

Our core findings revealed a coupled bias--variance and computational tradeoff. Supervised regression achieved the lowest errors under ideal \ac{id} simulated conditions (0.2764 ± 0.0032 \ac{mosae} in the restricted range case). This reflects a low variance model that relies heavily on its learned prior. Once forced to extrapolate beyond its trained concentration range, that same prior introduced systematic bias, nearly doubling the error (0.5537 ± 0.0058 \ac{mosae}) and constraining metabolite estimates such as \ac{gaba} to the training interval.
\Ac{tta} reduced this bias by relaxing the prior per sample, but at the cost of higher variance and additional computation. Instance adaptation proved substantially more resilient to concentration extrapolation in simulation (0.4420 ± 0.0052 \ac{mosae}), and domain adaptation achieved the best overall adaptive performance when tested on unseen in-vivo spectra (0.3425 ± 0.0102 \ac{mosae} \ac{ood} against the FSL-MRS reference).

\subsection{Performance of Data-Driven Methods}

The simulations, although simplified, incorporated central challenges of \ac{mrs} such as low \ac{snr}, peak overlap, broad linewidths, and baseline variability. Importantly, these factors make quantification inherently difficult even in the idealized case where the forward signal model used for fitting is identical to the generated spectra. The challenged performance of purely model-based methods in this setting suggests that the simulations reproduce meaningful aspects of the quantification problem and therefore provide a valuable evaluation for different strategies.

Across the simulated test scenarios, data-driven methods achieved low errors \ac{id} (Table~\ref{tab:sim_performance_sel}) and maintained stable performance under perturbations in \ac{snr}, linewidth, etc. (Figure~\ref{fig:sim_mean_lines_all} and Appendix~\ref{app:add_materials}). This indicates that learning-based approaches capture relevant spectral structure and yield consistent metabolite estimates when confronted with moderate variations in signal characteristics. 
However, the priors learned by these methods reflect the source distribution, and shifts in the target metabolite concentrations can introduce mismatches that degrade performance.

Interestingly, training networks from scratch on individual spectra outperformed direct model-based optimization using the same signal model (Appendix~\ref{app:add_materials}, Tables~\ref{tab:sim_performance_mae}, \ref{tab:sim_performance_mosae}, \ref{tab:invivo_fslmrs_mae}, \ref{tab:invivo_fslmrs_mosae}). Here, the improvement did not come from priors across multiple examples but from the architectural structure of the network and the properties of gradient-based training, which together act as a form of implicit regularization.

\subsection{Role of Physics-Informed Models}
Incorporating the signal model into training enabled self-supervised learning without ground truth metabolite concentrations, which is especially valuable for \ac{tta} methods. 
However, residual minimization can be ambiguous, as different parameter sets may yield similar fits. We observed that while the residual continues to decrease during training, the quantification error increases, indicating overfitting to the residual and limiting further improvements. 

By incorporating the physics model into the optimization, the problem is no longer a black-box mapping from spectra to concentrations. Instead, the optimization landscape becomes physics-informed: certain parameters collapse to constrained, meaningful subspaces. For example, phase, which was previously an arbitrary number, becomes cyclic and physically interpretable. This constraining improves plausibility, but the model has no incentive to produce estimates outside the training distribution, as doing so increases errors during training. Consequently, while extrapolations are slightly smoother, overall \ac{ood} performance for metabolite concentrations remains limited (Figure~\ref{fig:sim_mean_lines_all}).

\subsection{Regularization via Learned Priors}
For a fixed likelihood, the \ac{crb} sets a lower bound on the variance of an unbiased estimator. \cite{Landheer2021AreCRLBs} Reducing variance beyond this bound requires either additional information or the introduction of bias. Data-driven methods learn priors on metabolite parameters from the training data. These priors act as a form of regularization, lowering variance in the predictions, but they can also introduce systematic bias, particularly when the test/target distribution differs from the training/source distribution. \Ac{tta} addresses this by adapting the network parameters to incoming test spectra, retaining some of the variance reduction offered by the learned priors while mitigating bias caused by domain shift.

\subsection{Adaptation \& Initialization Effects}
Supervised regression provides an effective initialization for \ac{tta} methods, placing estimates near the global optimum rather than in local minima defined only by residual minimization. Residual-based updates then refine these estimates on test spectra, improving alignment with the new distribution while remaining anchored to accurate starting points.

While our main results focused on overall performance, supplementary iteration experiments (Appendix~\ref{app:add_materials}, Tables~\ref{tab:sim_performance_mae} and \ref{tab:sim_performance_mosae}) indicated that too many fine-tuning steps can lead to overfitting on individual spectra, whereas a moderate number of updates provided more stable improvements. Overall, \ac{tta} offered a practical means of adapting to distribution shifts, but its computational cost and sensitivity to the number of updates remain important considerations.

\subsection{From Simulations to In-Vivo: Domain Shift}

A key limitation was the lack of ground truth metabolite concentrations in-vivo, which prevented direct evaluation of absolute errors. We therefore assessed generalization only relative to alternative model-based quantification approaches. Consistent with prior work \cite{Bhogal2017ProcessingParamsAffect, Zoellner2021ComparisonLCMshortTE}, even established tools can disagree substantially. To maintain a controlled comparison, we adopted a single signal model as the reference pseudo ground truth, with results from alternative models provided in Appendix~\ref{app:add_materials} (Tables~\ref{tab:invivo_mix_mae}, \ref{tab:invivo_mix_mosae}, \ref{tab:invivo_lcmodel_mae}, and \ref{tab:invivo_lcmodel_mosae}).

Methods that shared the same signal model, including FSL-MRS, deviated less from each other than from models using a different parametrization such as LCModel. This indicates that differences in the signal model contributed more to variability than generalization or domain shift itself. Across metabolites, deviations were primarily systematic offsets rather than random errors. Supervised and self-supervised methods generalize most poorly relative to the FSL-MRS reference, but they exhibit similar trends for \ac{id} and \ac{ood} spectra as observed in simulations, with reasonable performance for \ac{id} cases. 
\Ac{tta}, however, performed well under both conditions, with domain adaptation in particular showing strong performance by converging reliably. Instance adaptation remained more sensitive to the number of iterations, mirroring the iteration-dependent effects seen in simulations.

\subsection{Training on Target Data}
Supervised approaches are constrained to simulated training due to their reliance on ground-truth metabolite concentrations, and are consequently more affected by domain shift. In contrast, physics-informed methods, including self-supervised and \ac{tta}, can be trained directly on unlabeled in-vivo data. Data leakage is not an issue because these methods train solely on the measured spectrum and the forward model, never accessing ground truths.

When network initialization is ignored, training a self-supervised model on target data is equivalent to test-time domain adaptation, which refines the network using the entire test dataset. Domain adaptation effectively bridges systematic domain shifts, achieving strong overall adaptive performance (Table~\ref{tab:invivo_fslmrs_sel_mosae}).

For instance adaptation, the impact of domain shift from supervised initialization is evident: more iterations progressively reduce deviation from all tested pseudo ground truths (Appendix~\ref{app:add_materials}, Tables \ref{tab:invivo_fslmrs_mae}, \ref{tab:invivo_fslmrs_mosae}, \ref{tab:invivo_mix_mae}, \ref{tab:invivo_mix_mosae}, \ref{tab:invivo_lcmodel_mae}, and \ref{tab:invivo_lcmodel_mosae}). When initialized with self-supervised priors, instance adaptation starts from a better baseline, achieving the best overall performance (Appendix~\ref{app:add_materials}, Tables \ref{tab:invivo_mix_mae}, \ref{tab:invivo_mix_mosae}, \ref{tab:invivo_lcmodel_mae}, and \ref{tab:invivo_lcmodel_mosae}). Training from scratch, which removes any domain shift entirely, produces the lowest overall deviation from the pseudo ground truth FSL-MRS (Appendix~\ref{app:add_materials}, Tables \ref{tab:invivo_fslmrs_mae} and \ref{tab:invivo_fslmrs_mosae}).

Online adaptation provides a practical trade-off between performance and speed, consistently outperforming the supervised baseline on in-vivo spectra (Table~\ref{tab:invivo_fslmrs_sel_mosae}). 

\subsection{Limitations}


One of the main limitations of this study is the simplified signal model used for both the simulations and \ac{lcm}. The model included only global linewidths and frequency shifts, a single \ac{mm} component, and a second-order polynomial baseline. It did not account for local variations, metabolite-specific lineshapes, or distortions such as eddy currents. Incorporating additional parameters or using more flexible models could better capture the complexity of real spectra, but this comes at the cost of reduced interpretability, as it becomes harder to disentangle the influence of individual effects.

Another limitation concerns the training strategy. The networks were optimized for gradual and steady convergence on simulated data with known ground truths, but these settings may not translate optimally to in-vivo spectra, where variability is higher and ground truth is unavailable.

A further challenge observed across all data-driven methods is phase handling. These approaches are particularly sensitive to \ac{ood} phase variations. Model architecture plays a significant role: some architectures have demonstrated more robust phase handling in \ac{id} settings \cite{tapper_frequency_2021, ma_mr_2022, shamaei_modelinformed_2022}, but it remains unclear whether such robustness extends to \ac{ood} cases. Alternative input representations may also help address this limitation. \cite{rizzo_quantification_2023}

Finally, other limitations must be considered, including potential biases from preprocessing, sensitivity to specific artifacts, and generalization across acquisition protocols. Addressing these factors is essential for developing robust \ac{mrs} quantification methods that perform reliably across different datasets and experimental conditions.

\subsection{Broader Implications}
Despite these limitations, the results offer several insights for the development of data-driven methods in \ac{mrs}. Learned priors and adaptive strategies can stabilize parameter estimation and handle moderate variability, suggesting that networks trained on realistic simulations can provide robust quantification even in challenging conditions. Furthermore, training with in-vivo data with representative metabolite and signal perturbation parameters can lead to well performing methods \ac{id} and with proper adaptation strategies in place also \ac{ood}. These findings extend beyond metabolite quantification, emphasizing the value of incorporating prior knowledge, physics-informed modeling, and adaptive updates when designing data-driven pipelines in \ac{mrs}.

\section{Conclusion} \label{sec:conclusion}
Data-driven strategies for MRS metabolite quantification, including supervised, self-supervised, and \ac{tta}, were systematically compared with purely model-based approaches. While data-driven methods showed strong performance in simulated \ac{id} scenarios, inherent biases and sensitivity to \ac{ood} conditions were observed. \Ac{tta} techniques were found to significantly enhance robustness and generalizability in \ac{ood} settings for both simulated and in-vivo data, mitigating the impact of domain shift. The study highlighted the importance of integrating prior knowledge, physics-informed modeling, and adaptive updates for developing robust data-driven \ac{mrs} pipelines.


\section*{Acknowledgments}
This work was in part funded by Spectralligence (EUREKA IA Call, ITEA4 project 20209). We further acknowledge support from the NVIDIA Academic Hardware Grant Program, and acknowledge the CIBM Center for Biomedical Imaging for providing expertise and resources to conduct this study.




\section*{Data Availability Statement}
The source code used in our experiments for the methods, data simulation, and analysis can be found at \url{https://github.com/julianmer/OoD-Robust-MRS-Quantification}. The in-vivo data are available from the corresponding author upon reasonable request.

\bibliography{refs/refs, refs/lr_refs, refs/my_refs}

\appendix

\section{Additional Materials} \label{app:add_materials}
This section provides supplementary analyses to support the findings and discussions of this work.

\subsection{Additional Results on Simulated Data}

\begin{figure*}
    \vspace{2cm}
    \centering
    \includegraphics[width=2.0\columnwidth]{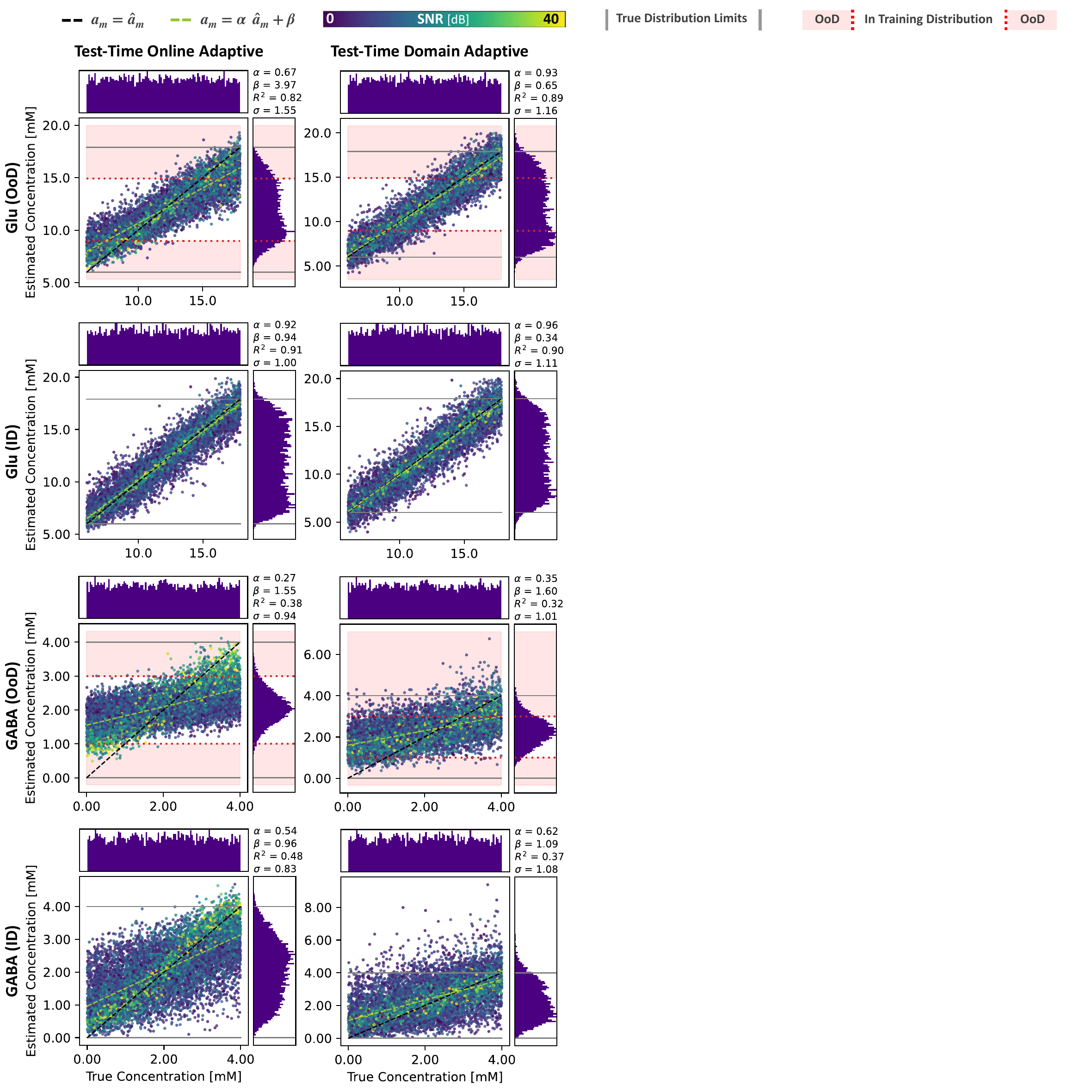}
    \caption{\enspace Scatter plots with marginal histograms comparing predicted versus true concentrations of \ac{glu} and \ac{gaba} across 10,000 simulated spectra. Models are evaluated under two scenarios for the full concentration range: trained on mid-range concentrations (\ac{ood}) or trained on the full range (\ac{id}). Methods include test-time online adaptive approaches and test-time domain adaptive approaches. Points are colored by \ac{snr}, and regression lines with corresponding statistics (slope $\alpha$, intercept $\beta$, R$^2$, and \ac{rmse} $\sigma$) are shown.
    \vspace{2cm}
    }
    \label{fig:sim_dist_comb_2_mae}
\end{figure*}

\begin{figure*}
    \vspace{2cm}
    \centering
    \includegraphics[width=2.0\columnwidth]{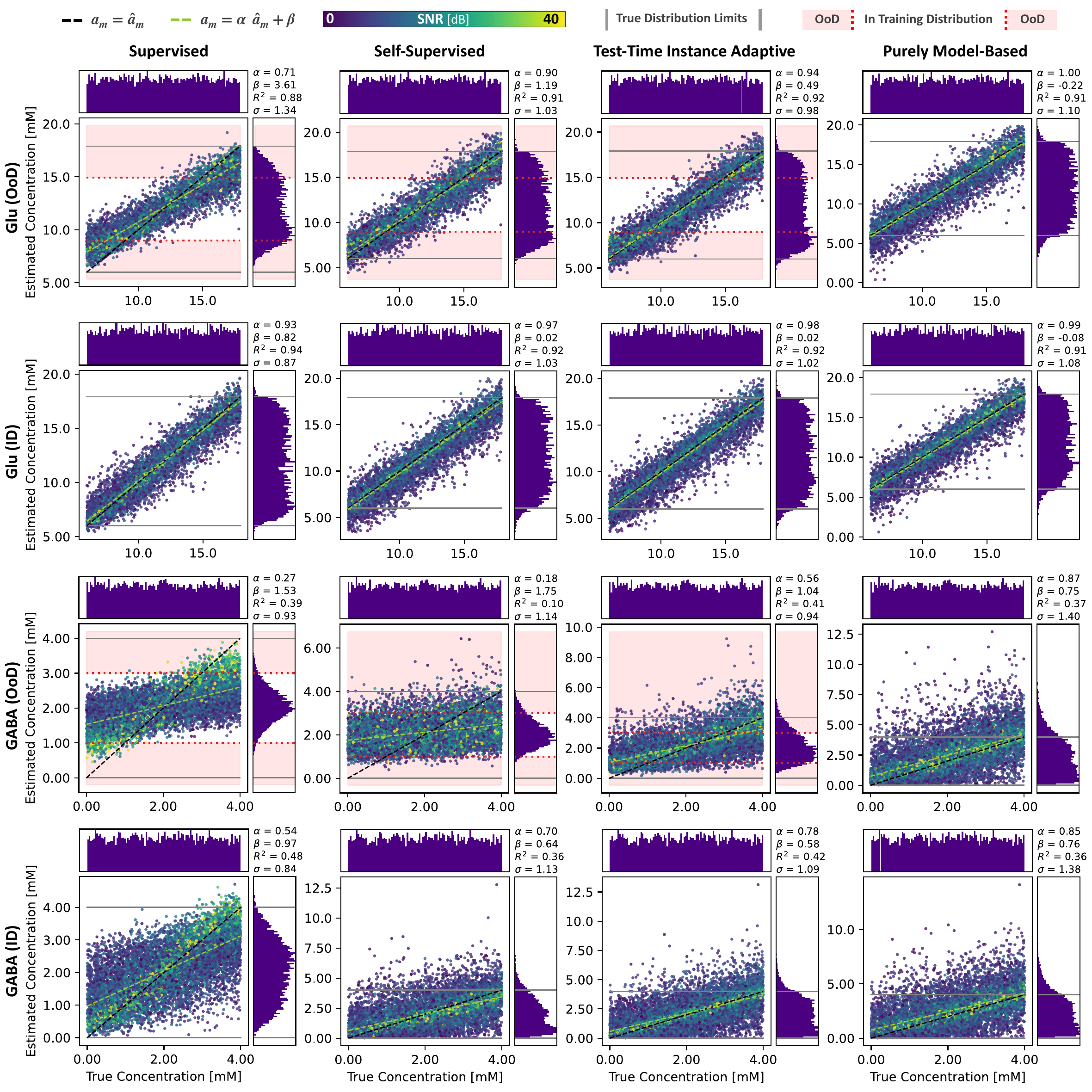}
    \caption{\enspace Scatter plots with marginal histograms comparing optimally scaled predicted versus true concentrations of \ac{glu} and \ac{gaba} across 10,000 simulated spectra. Models are evaluated under two scenarios for the full concentration range: trained on mid-range concentrations (\ac{ood}) or trained on the full range (\ac{id}). Data-driven methods include supervised, self-supervised, and test-time instance adaptive approaches, compared with purely model-based fitting. Points are colored by \ac{snr}, and regression lines with corresponding statistics (slope $\alpha$, intercept $\beta$, R$^2$, and \ac{rmse} $\sigma$) are shown.
    \vspace{2cm}
    }
    \label{fig:sim_dist_comb_1_mosae}
\end{figure*}

\begin{figure*}
    \vspace{2cm}
    \centering
    \includegraphics[width=2.0\columnwidth]{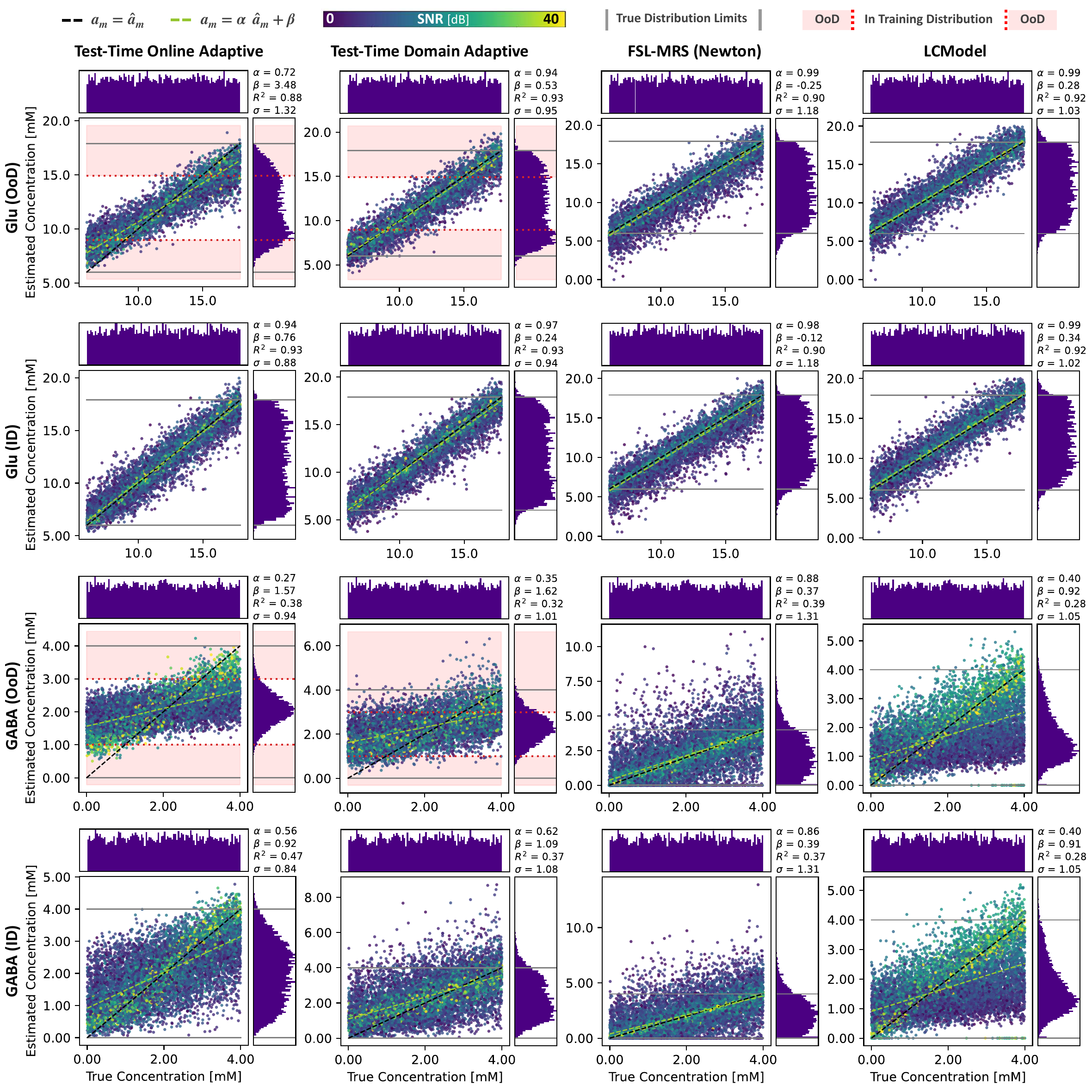}
    \caption{\enspace Scatter plots with marginal histograms comparing optimally scaled predicted versus true concentrations of \ac{glu} and \ac{gaba} across 10,000 simulated spectra. Models are evaluated under two scenarios for the full concentration range: trained on mid-range concentrations (\ac{ood}) or trained on the full range (\ac{id}). Methods include test-time online adaptive approaches, test-time domain adaptive approaches compared with FSL-MRS (Newton) and LCModel. Points are colored by \ac{snr}, and regression lines with corresponding statistics (slope $\alpha$, intercept $\beta$, R$^2$, and \ac{rmse} $\sigma$) are shown.
    \vspace{2cm}
    }
    \label{fig:sim_dist_comb_2_mosae}
\end{figure*}

\begin{figure*}
    \centering
    \includegraphics[width=2.0\columnwidth]{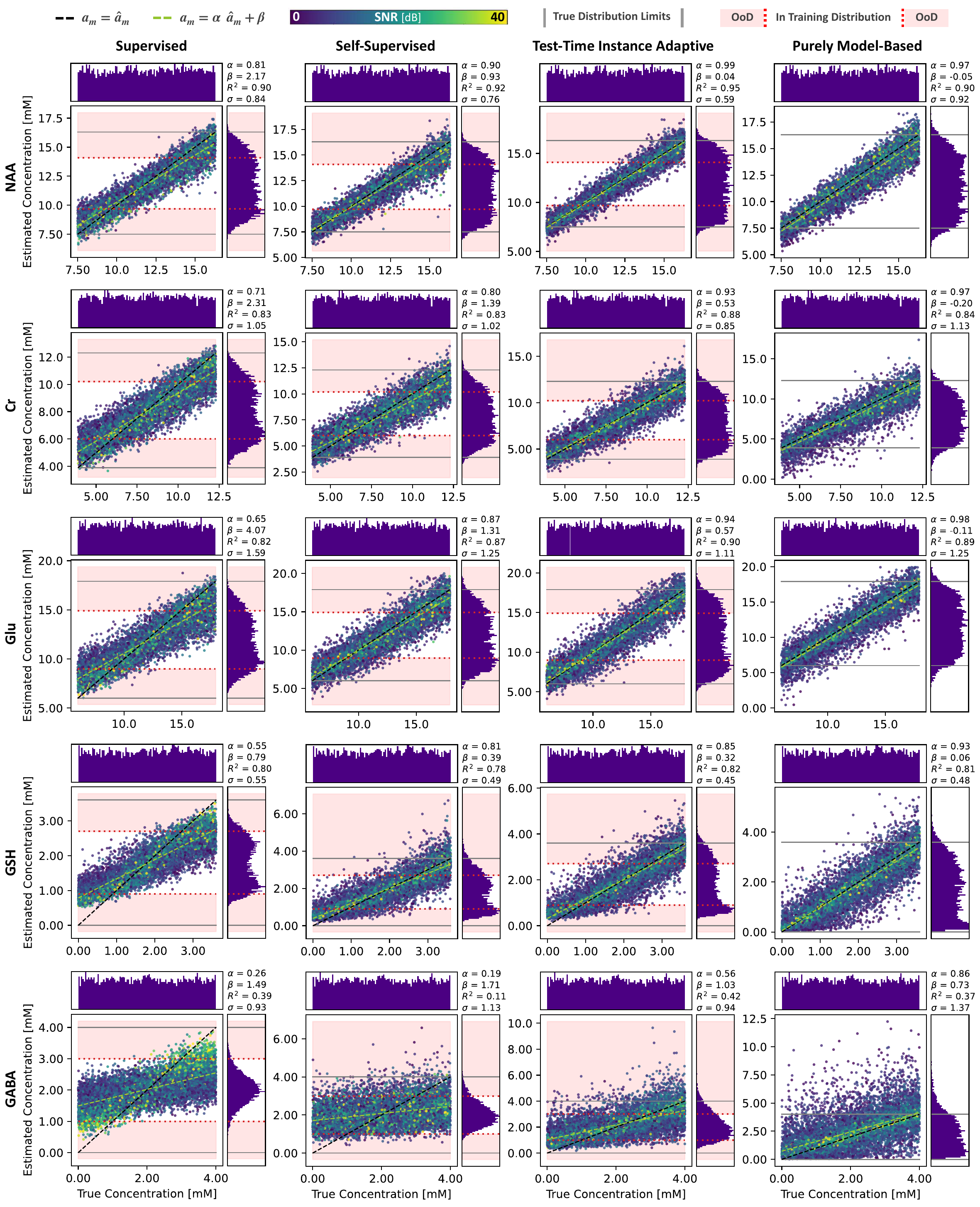}
    \caption{\enspace Scatter plots with marginal histograms comparing predicted versus true concentrations for \ac{naa}, \ac{cr}, \ac{glu}, \ac{gsh}, and \ac{gaba} across 10,000 simulated spectra. Models trained on mid-range concentrations are evaluated across the full concentration range (\ac{ood}). Data-driven methods include supervised, self-supervised, and test-time instance adaptive compared against purely model-based fitting. Points are colored by \ac{snr}, and regression lines with corresponding statistics (R$^2$, slope, intercept, \ac{rmse}) are included.
    }
    \label{fig:sim_dist_ood_1_sel_mae}
\end{figure*}

\begin{figure*}
    \centering
    \includegraphics[width=2.0\columnwidth]{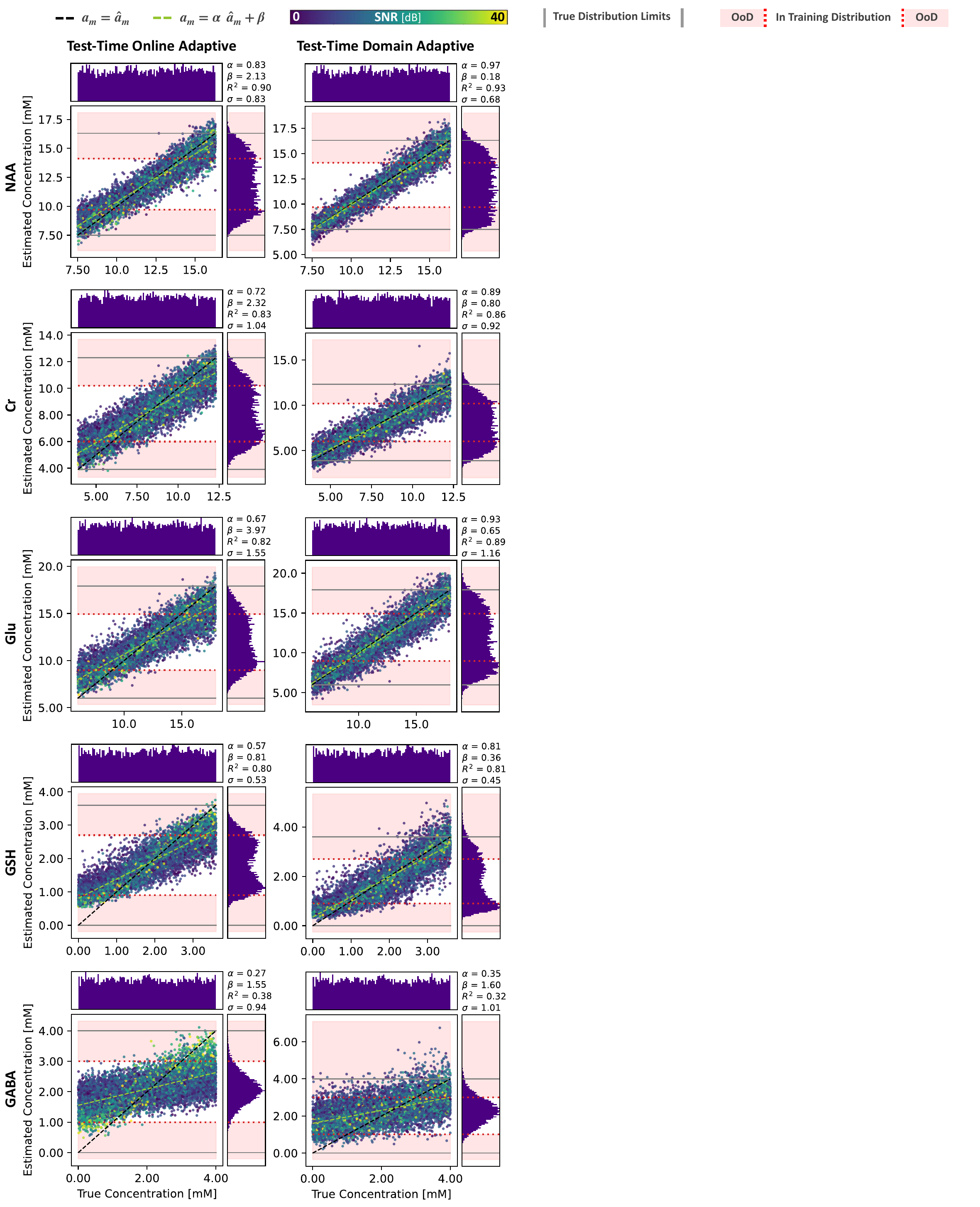}
    \caption{\enspace Scatter plots with marginal histograms comparing predicted versus true concentrations for \ac{naa}, \ac{cr}, \ac{glu}, \ac{gsh}, and \ac{gaba} across 10,000 simulated spectra. Models trained on mid-range concentrations are evaluated across the full concentration range (\ac{ood}). Adaptive methods include test-time online adaptive and test-time domain adaptive. Points are colored by \ac{snr}, and regression lines with corresponding statistics are included.}
    \label{fig:sim_dist_ood_2_sel_mae}
\end{figure*}

\begin{figure*}
    \centering
    \includegraphics[width=2.0\columnwidth]{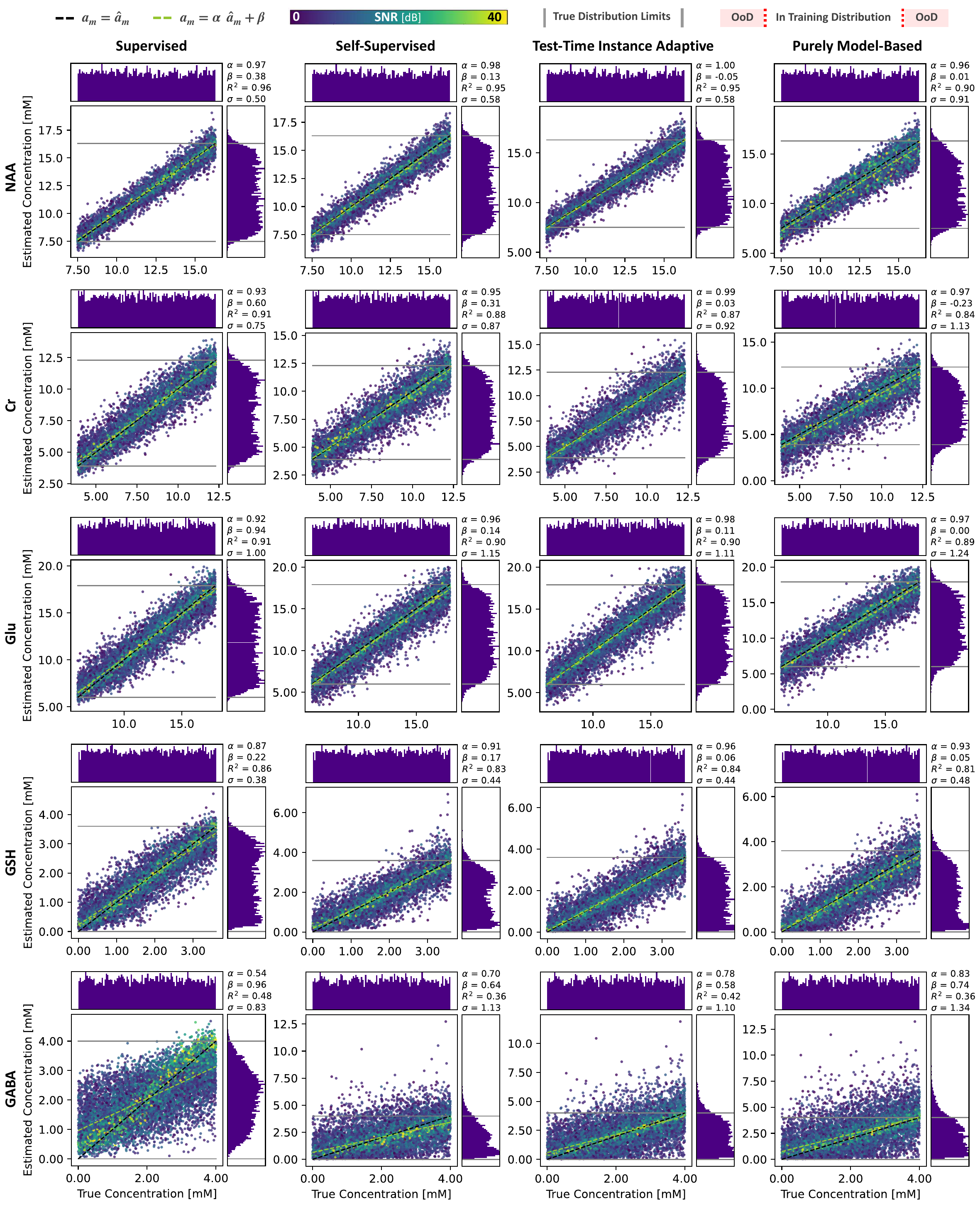}
    \caption{\enspace Scatter plots with marginal histograms comparing predicted versus true concentrations for \ac{naa}, \ac{cr}, \ac{glu}, \ac{gsh}, and \ac{gaba} under the \ac{id} full-range scenario, where models are trained and tested on the full concentration range. This figure shows data-driven methods: supervised, self-supervised, and test-time instance adaptive against purely model-based. Points are colored by \ac{snr}, and regression lines with corresponding statistics are included.}
    \label{fig:sim_dist_idft_1_sel_mae}
\end{figure*}

\begin{figure*}
    \vspace{0.5cm}
    \centering
    \includegraphics[width=2.0\columnwidth]{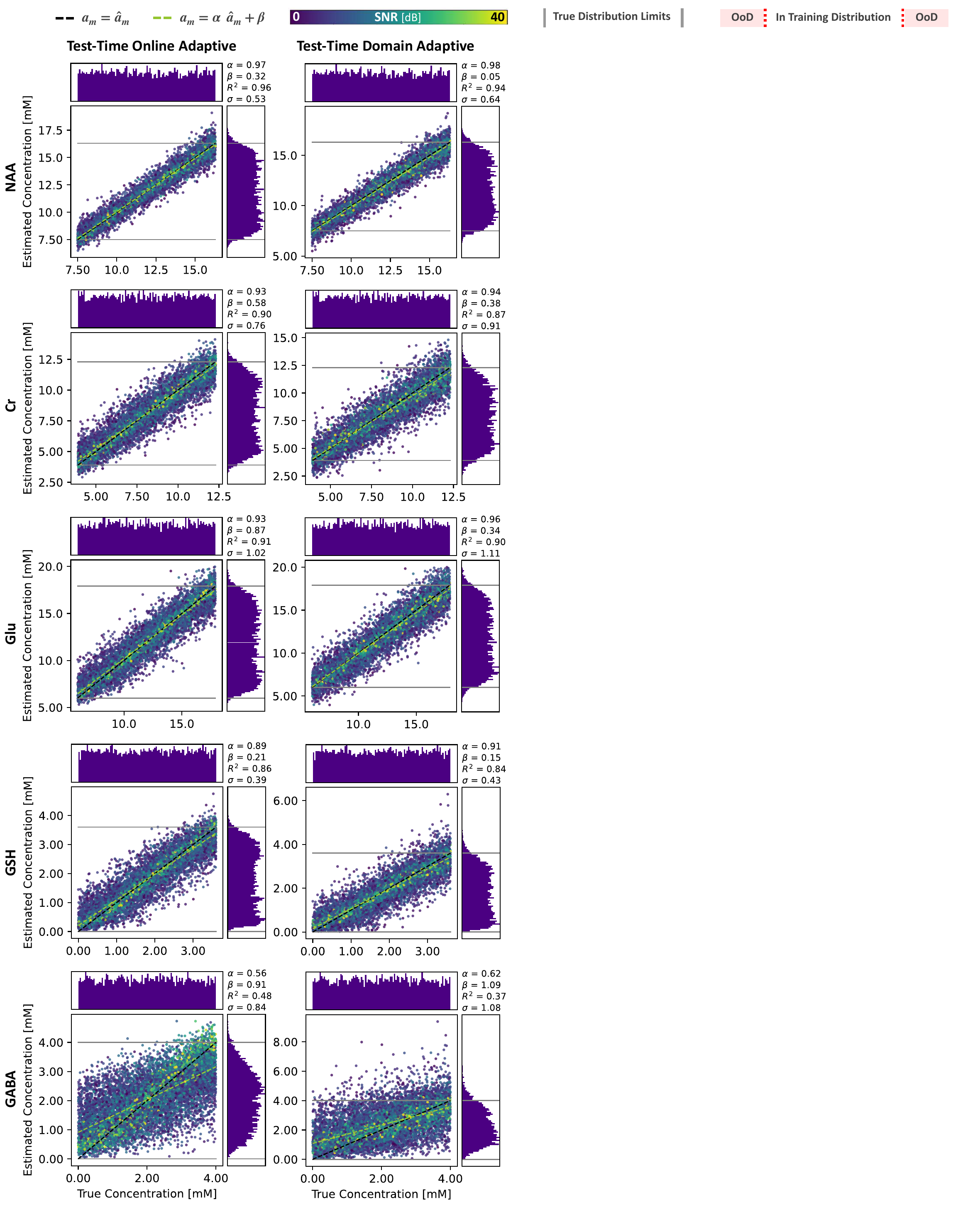}
    \caption{\enspace Scatter plots with marginal histograms comparing predicted versus true concentrations for \ac{naa}, \ac{cr}, \ac{glu}, \ac{gsh}, and \ac{gaba} under the \ac{id} full-range scenario. This figure shows adaptive methods: test-time online adaptive and test-time domain adaptive. Points are colored by \ac{snr}, and regression lines with corresponding statistics are included.
    \vspace{0.5cm}
    }
    \label{fig:sim_dist_idft_2_sel_mae}
\end{figure*}

\begin{figure*}
    \centering
    \includegraphics[width=2.0\columnwidth]{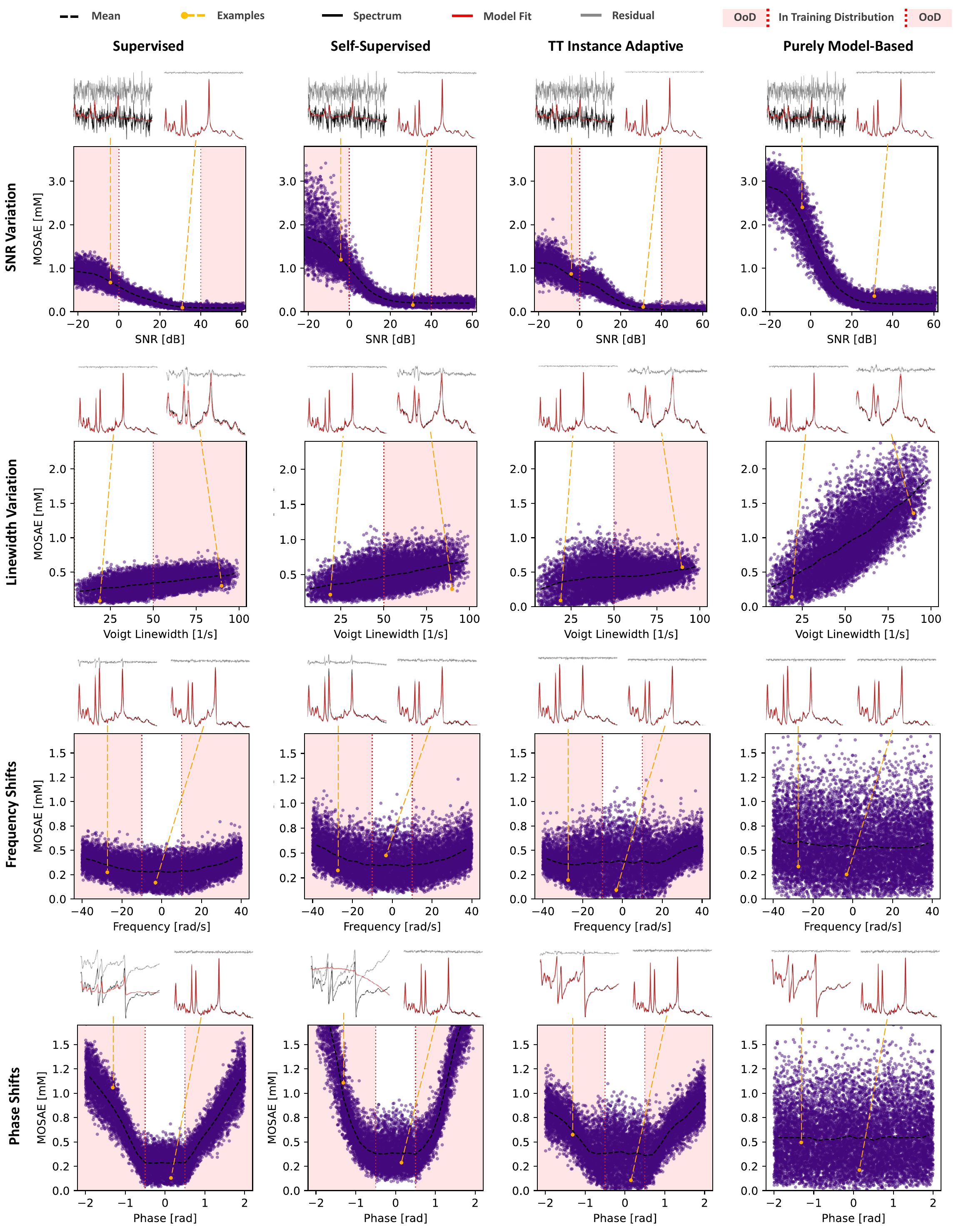}
    \caption{\enspace Scatter plots showing quantification accuracy (\ac{mosae}) across 10,000 simulated spectra as a function of ground truth \ac{snr}, linewidth, zeroth-order phase shift, and frequency offset. Data-driven methods include supervised, self-supervised, and test-time instance adaptive compared against purely model-based fitting. Each point represents one spectrum, illustrating method-specific sensitivity to core signal parameter variations.}
    \label{fig:sim_rang_1_1_mosae}
\end{figure*}

\begin{figure*}
    \vspace{3cm}
    \centering
    \includegraphics[width=2.0\columnwidth]{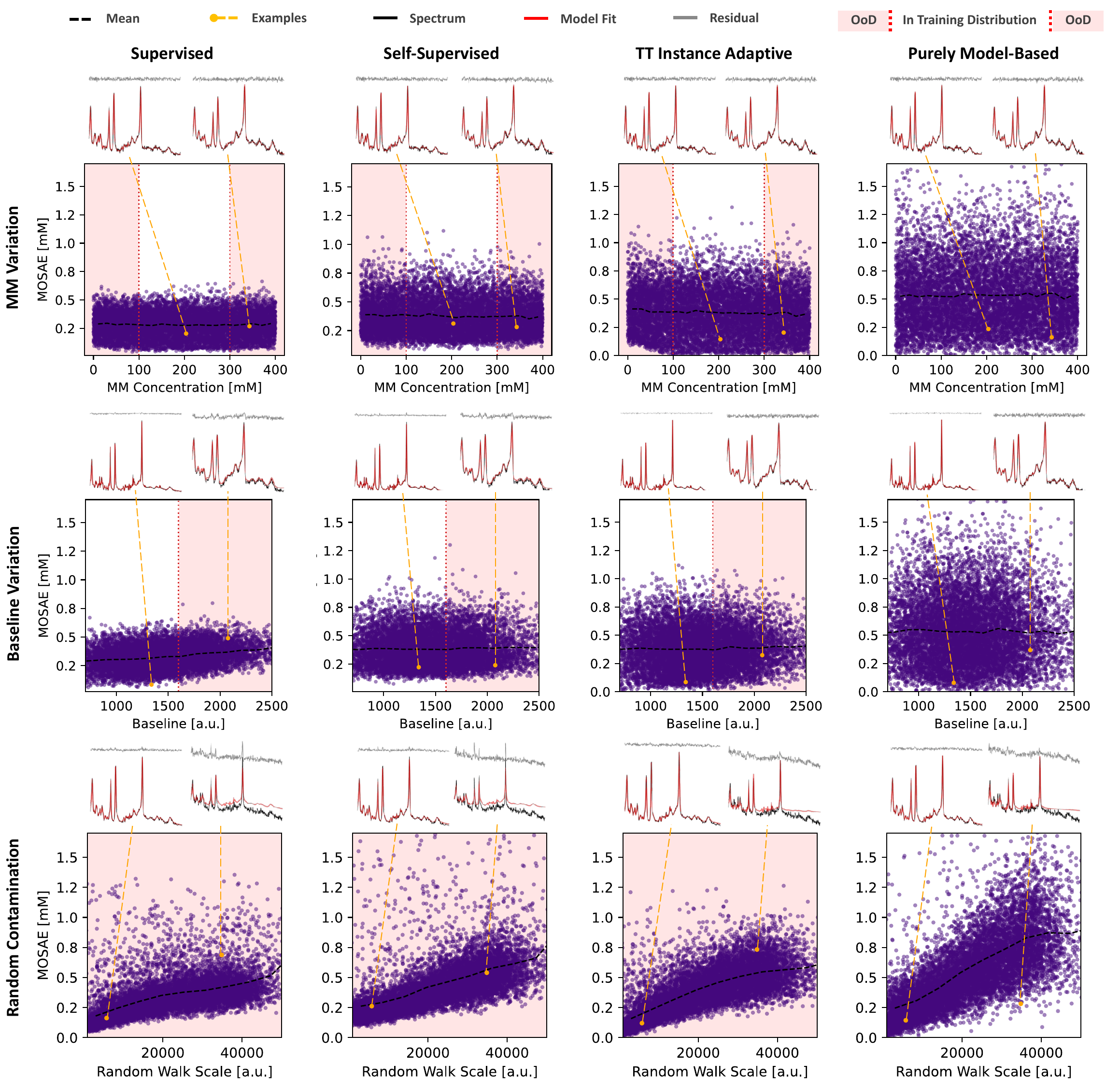}
    \caption{\enspace Scatter plots showing quantification accuracy (\ac{mosae}) across 10,000 simulated spectra under \ac{mm}, baseline, and random signal corruptions. Data-driven methods include supervised, self-supervised, and test-time instance adaptive compared against purely model-based fitting. Each point represents one spectrum, illustrating method-specific robustness to unmodeled spectral deviations.
    \vspace{3cm}
    }
    \label{fig:sim_rang_1_2_mosae}
\end{figure*}

\begin{figure*}
    \centering
    \includegraphics[width=2.0\columnwidth]{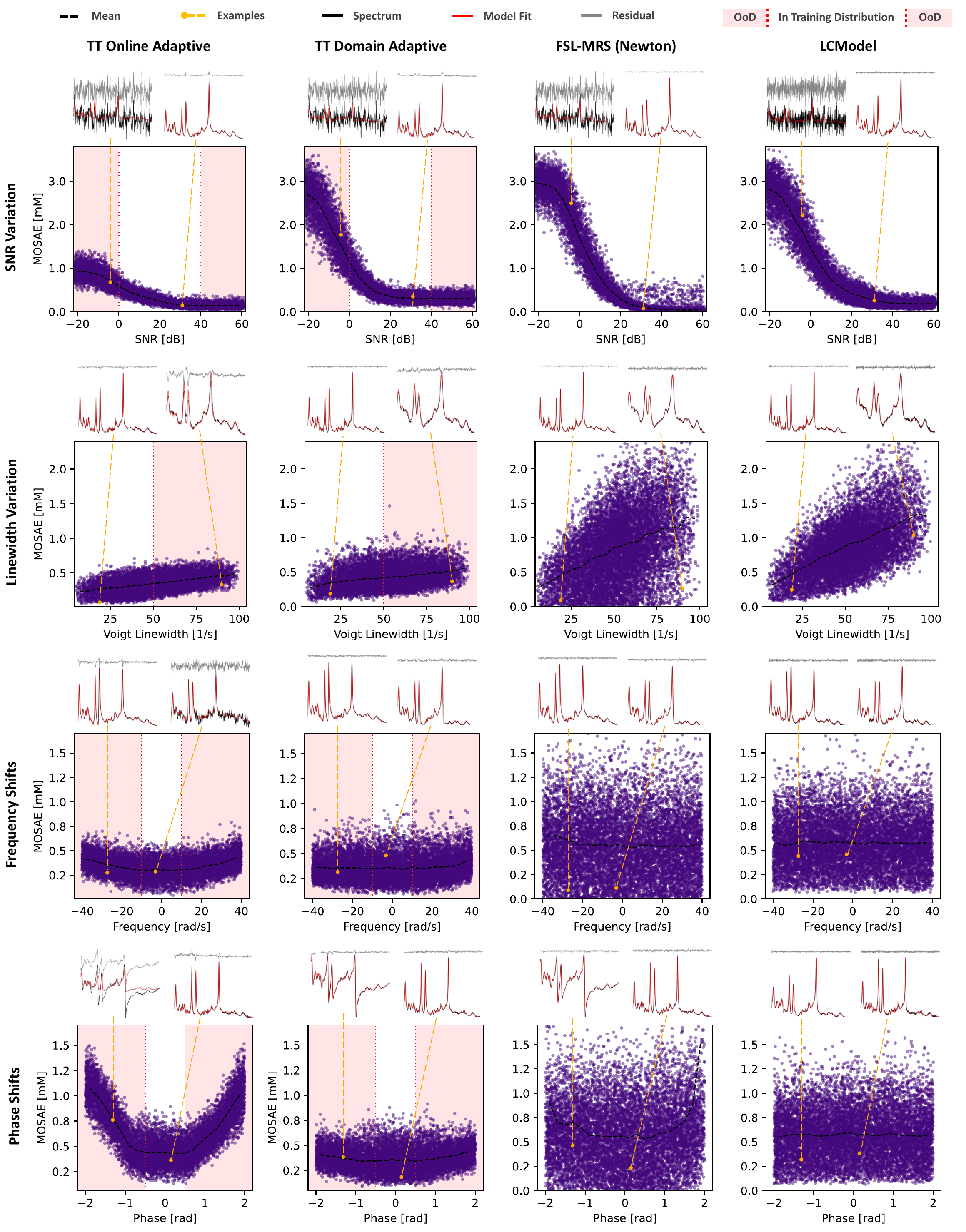}
    \caption{\enspace Scatter plots showing quantification accuracy (\ac{mosae}) across 10,000 simulated spectra as a function of ground truth \ac{snr}, linewidth, zeroth-order phase shift, and frequency offset. Adaptive and classical methods include test-time online adaptive, test-time domain adaptive, FSL-MRS (Newton), and LCModel. Each point represents one spectrum, illustrating method-specific robustness to unmodeled spectral deviations.}
    \label{fig:sim_rang_2_1_mosae}
\end{figure*}

\begin{figure*}
    \vspace{3cm}
    \centering
    \includegraphics[width=2.0\columnwidth]{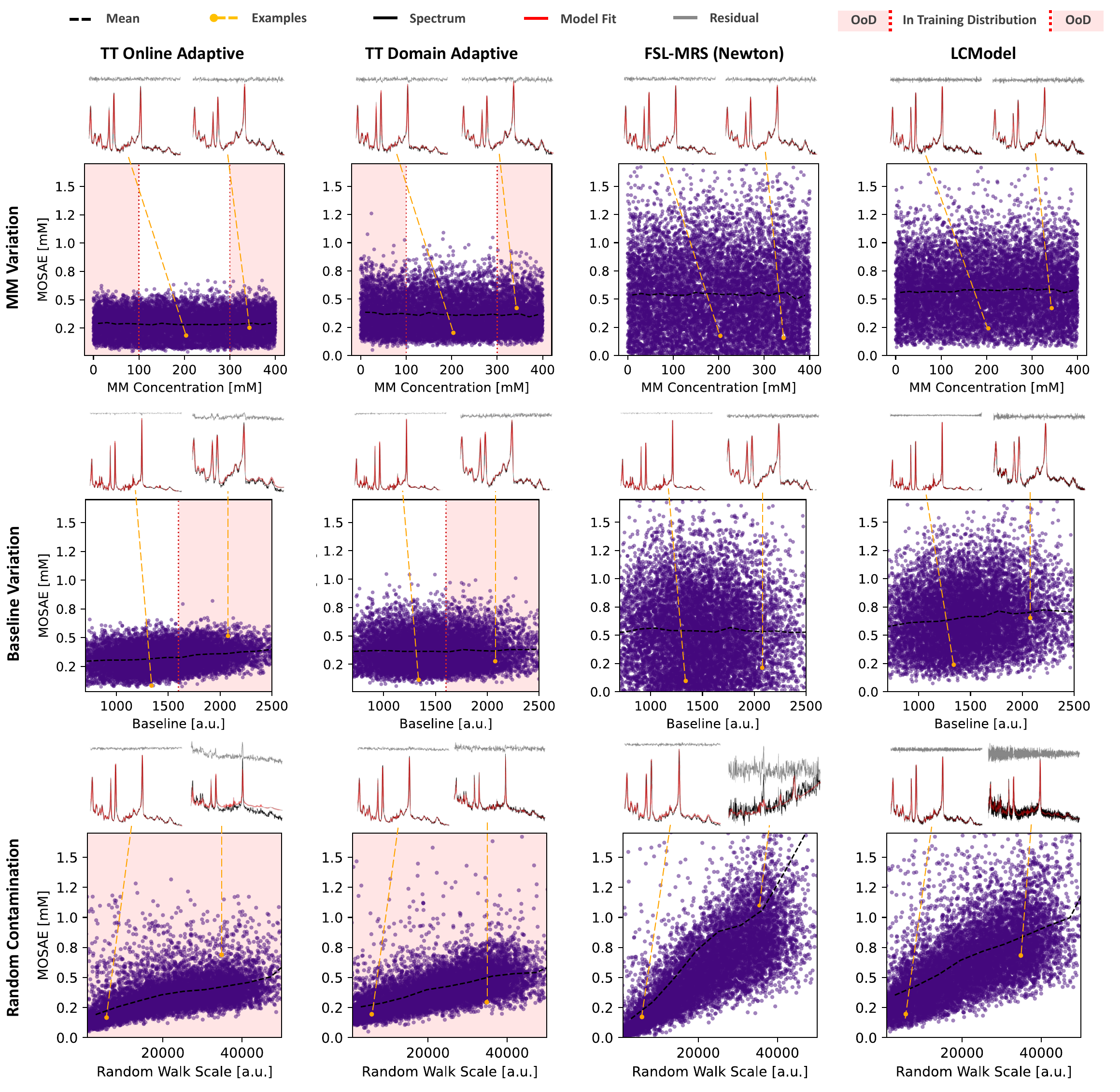}
    \caption{\enspace Scatter plots showing quantification accuracy (\ac{mosae}) across 10,000 simulated spectra under \ac{mm}, baseline, and random signal corruptions. Adaptive and classical methods include test-time online adaptive, test-time domain adaptive, FSL-MRS (Newton), and LCModel. Each point represents one spectrum, illustrating method-specific robustness to unmodeled spectral deviations.
    \vspace{3cm}
    }
    \label{fig:sim_rang_2_2_mosae}
\end{figure*}

\begin{figure*}
    \vspace{1cm}
    \centering
    \includegraphics[width=2.0\columnwidth]{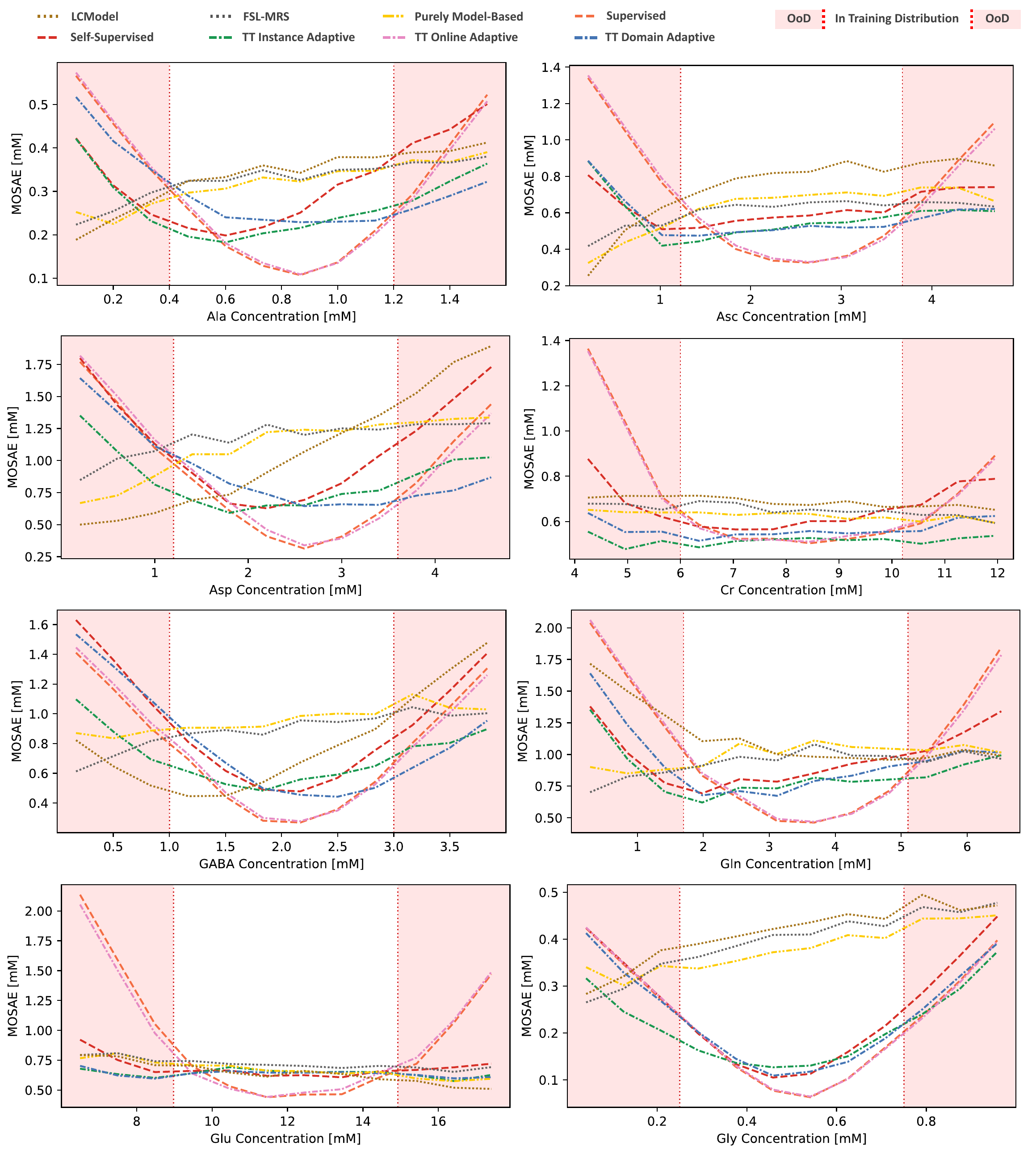}
    \caption{\enspace Summary of quantification performance across 10,000 simulated spectra for eight metabolites (Ala, Asc, Asp, Cr, GABA, Gln, Glu, Gly). Each subplot corresponds to one metabolite, showing the \ac{mosae} for all methods: purely model-based gradient descent, FSL-MRS (Newton), LCModel, supervised, self-supervised, and \ac{tta} strategies. This visualization allows comparison of method performance across metabolites under \ac{ood} conditions.
    \vspace{1cm}
    }
    \label{fig:sim_perf_metabs_1_1_mosae}
\end{figure*}

\begin{figure*}
    \vspace{1cm}
    \centering
    \includegraphics[width=2.0\columnwidth]{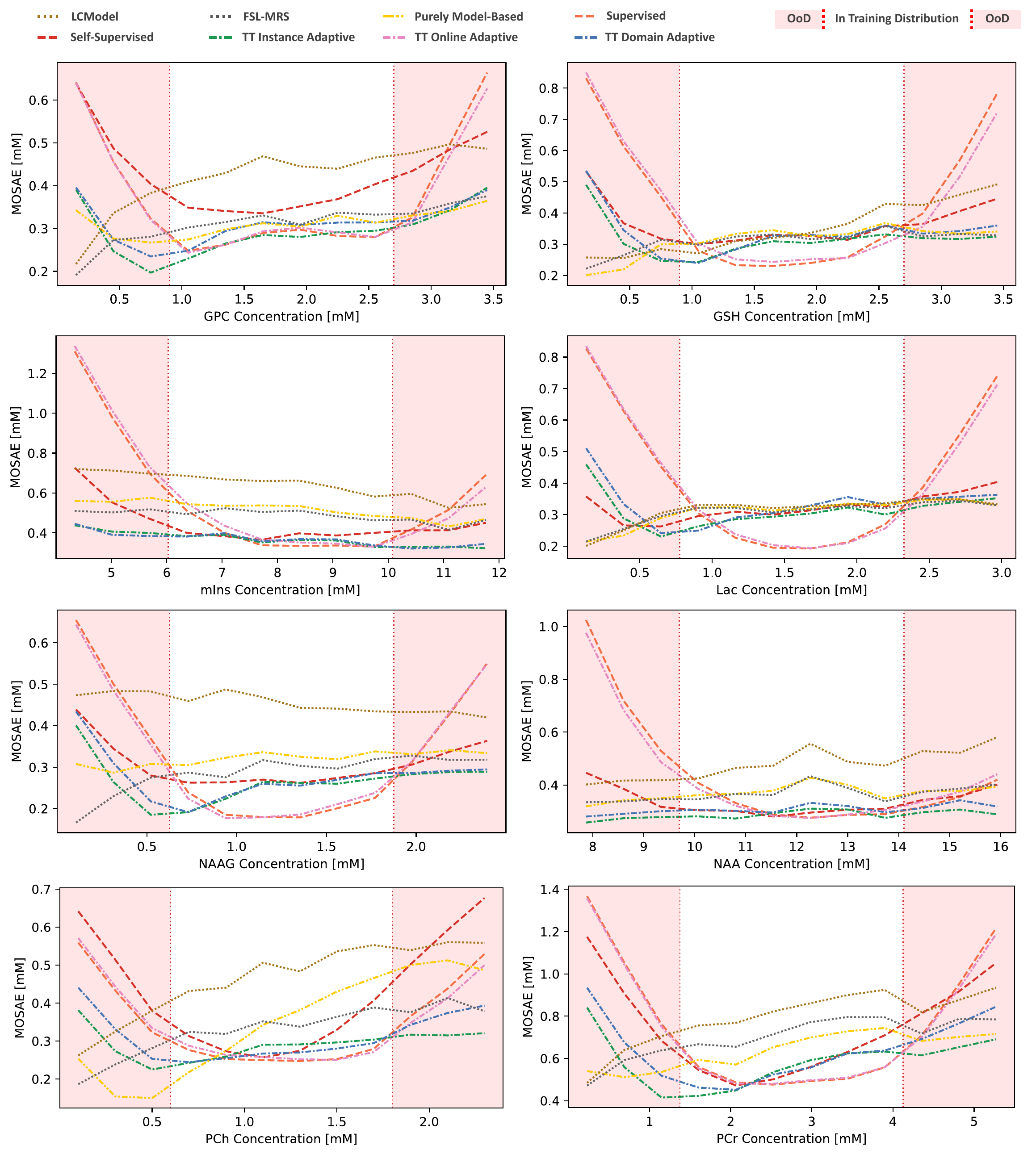}
    \caption{\enspace Summary of quantification performance across 10,000 simulated spectra for eight metabolites (GPC, GSH, mIns, Lac, NAAG, NAA, PCh, PCr). Each subplot corresponds to one metabolite, showing the \ac{mosae} for all methods: purely model-based gradient descent, FSL-MRS (Newton), LCModel, supervised, self-supervised, and \ac{tta} strategies. This visualization allows comparison of method performance across metabolites under \ac{ood} conditions.
    \vspace{1cm}
    }
    \label{fig:sim_perf_metabs_1_2_mosae}
\end{figure*}

\begin{figure*}
    \vspace{7cm}
    \centering
    \includegraphics[width=2.0\columnwidth]{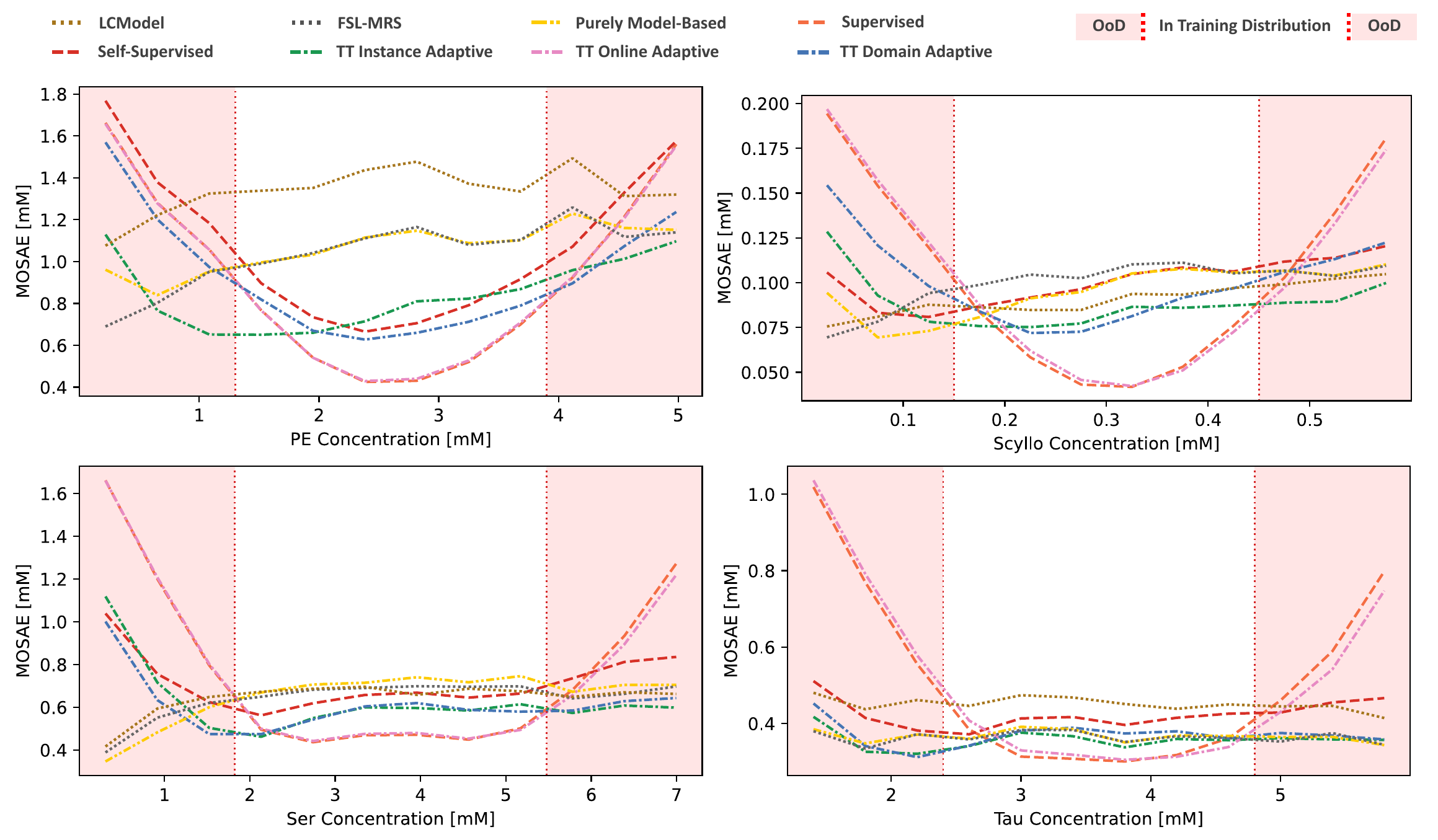}
    \caption{\enspace Summary of quantification performance across 10,000 simulated spectra for four metabolites (PE, Scyllo, Ser, Tau). Each subplot corresponds to one metabolite, showing the \ac{mosae} for all methods: purely model-based gradient descent, FSL-MRS (Newton), LCModel, supervised, self-supervised, and \ac{tta} strategies. This visualization allows comparison of method performance across metabolites under \ac{ood} conditions.
    \vspace{7cm}
    }
    \label{fig:sim_perf_metabs_1_3_mosae}
\end{figure*}

\begin{figure*}
    \vspace{1cm}
    \centering
    \includegraphics[width=2.0\columnwidth]{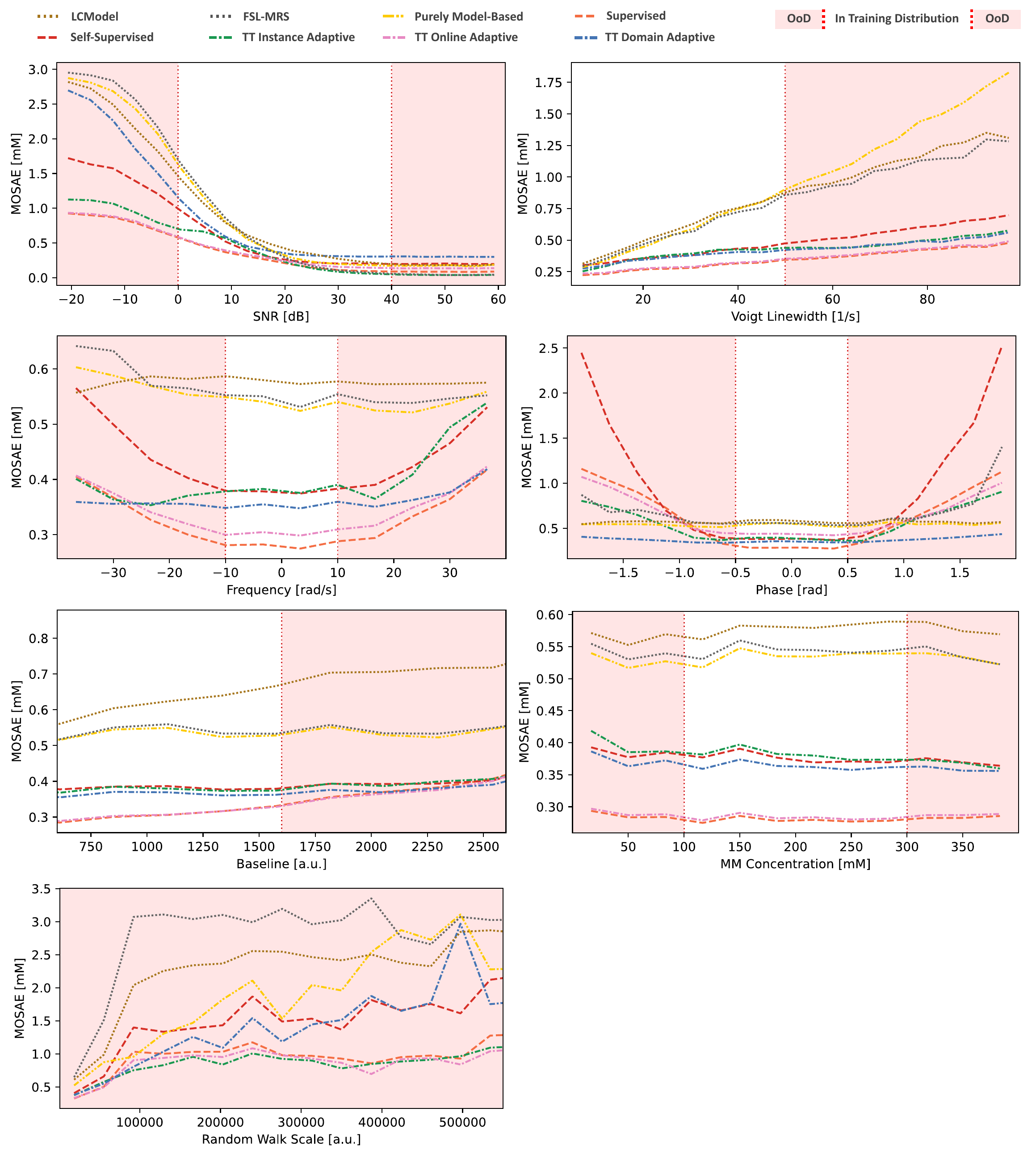}
    \caption{\enspace Summary of quantification performance under \ac{ood} signal perturbations. For each signal parameter, \ac{snr}, linewidth, frequency offset, phase shift, \ac{mm} baseline, and polynomial baseline, the \ac{mosae} is averaged within parameter bins. All methods are overlaid in each subplot, including model fitting (purely model-based gradient decent, FSL-MRS (Newton), and LCModel), supervised, self-supervised, along with the \ac{tta} strategies.
    \vspace{1cm}
    }
    \label{fig:sim_perf_other_1_mosae}
\end{figure*}

This part offers extended results from the simulated \ac{mrs} data, including:
\begin{itemize}
    \item \textbf{Tables \ref{tab:sim_performance_mae} and \ref{tab:sim_performance_mosae}}: These tables provide comprehensive comparisons of quantification methods across 10,000 simulated spectra, detailing both \ac{mae} and \ac{mosae}, respectively. Three primary scenarios for metabolite concentration ranges are covered:
    \begin{itemize}
        \item \textbf{ID (Mid-Range)}: Models trained and tested on the central 50\% metabolite concentration range.
        \item \textbf{OoD (Full-Range)}: Models trained on the central 50\% concentration range but tested across the entire range of concentrations to assess extrapolation.
        \item \textbf{ID (Full-Trained)}: Models trained and tested on the full concentration range.
    \end{itemize}
    These tables also include performance metrics for \ac{cnn}-based baselines, self-supervised initialization of \ac{tta} models, and various iteration counts for instance adaptation.

    \item \textbf{Figures \ref{fig:sim_dist_comb_2_mae}, \ref{fig:sim_dist_comb_1_mosae}, \ref{fig:sim_dist_comb_2_mosae}}: These scatter plots with marginal histograms compare predicted versus true concentrations for \ac{glu} and \ac{gaba} across 10,000 simulated spectra. They are evaluated under \ac{ood} and \ac{id} scenarios, including optimally scaled concentrations for relative metabolite quantification comparison.

    \item \textbf{Figures \ref{fig:sim_dist_ood_1_sel_mae}, \ref{fig:sim_dist_ood_2_sel_mae}, \ref{fig:sim_dist_idft_1_sel_mae}, \ref{fig:sim_dist_idft_2_sel_mae}}: These figures show scatter plots with marginal histograms comparing predicted versus true concentrations for other metabolites such as \ac{naa}, \ac{cr}, \ac{glu}, \ac{gsh}, and \ac{gaba}, under both \ac{ood} (mid-range trained, full-range tested) and \ac{id} (full-range trained) conditions.

    \item \textbf{Figures \ref{fig:sim_rang_1_1_mosae}, \ref{fig:sim_rang_1_2_mosae}, \ref{fig:sim_rang_2_1_mosae}, \ref{fig:sim_rang_2_2_mosae}}: These scatter plots illustrate quantification accuracy (\ac{mosae}) for various signal parameter \ac{snr}, linewidth, zeroth-order phase shift, frequency offset, \ac{mm}, baseline, and random signal corruptions.
    \begin{itemize}
        \item The random signal corruption is generated by $R(f)$, a complex-valued random walk that introduces arbitrary spectral distortions. This process is defined as a bounded, smoothed stochastic process with independent real and imaginary components \cite{Rayleigh1905ThePO}. This random walk component $R(f)$ is exclusively added during evaluation to assess robustness to random spectral artifacts and is excluded during the training phase.
    \end{itemize}

    \item \textbf{Figures \ref{fig:sim_perf_metabs_1_1_mosae}, \ref{fig:sim_perf_metabs_1_2_mosae}, \ref{fig:sim_perf_metabs_1_3_mosae}, \ref{fig:sim_perf_other_1_mosae}}: These figures summarize quantification performance across various metabolites and other signal parameters, displaying the mean \ac{mosae} within parameter bins.
\end{itemize}

\subsection{Additional Results on In-Vivo Data}
\begin{figure*}
    \vspace{3cm}
    \centering
    \includegraphics[width=2.0\columnwidth]{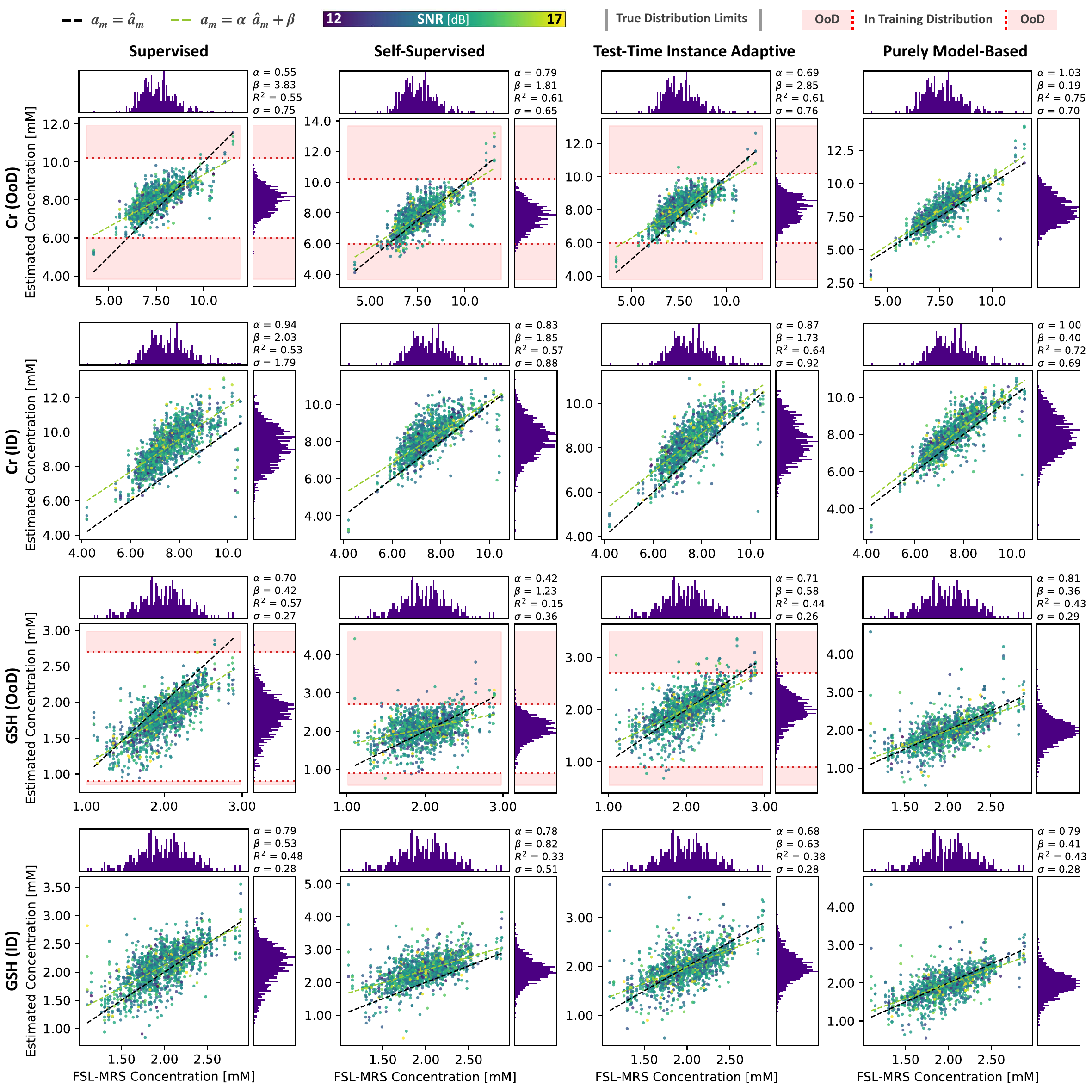}
    \caption{\enspace Scatter plots with marginal histograms comparing optimally scaled predicted versus pseudo-true (FSL-MRS estimates) concentrations of \ac{glu} and \ac{gaba} across 1,710 in-vivo spectra. Models are evaluated under two scenarios for the full concentration range: trained on mid-range concentrations (\ac{ood}) or trained on the full range (\ac{id}). Data-driven methods include supervised, self-supervised, and test-time instance adaptive approaches, compared with purely model-based fitting. Points are colored by \ac{snr}, and regression lines with corresponding statistics (slope $\alpha$, intercept $\beta$, R$^2$, and \ac{rmse} $\sigma$) are shown.
    \vspace{3cm}
    }
    \label{fig:invivo_fsl_64_dist_comb_1_mosae}
\end{figure*}

\begin{figure*}
    \vspace{3cm}
    \centering
    \includegraphics[width=2.0\columnwidth]{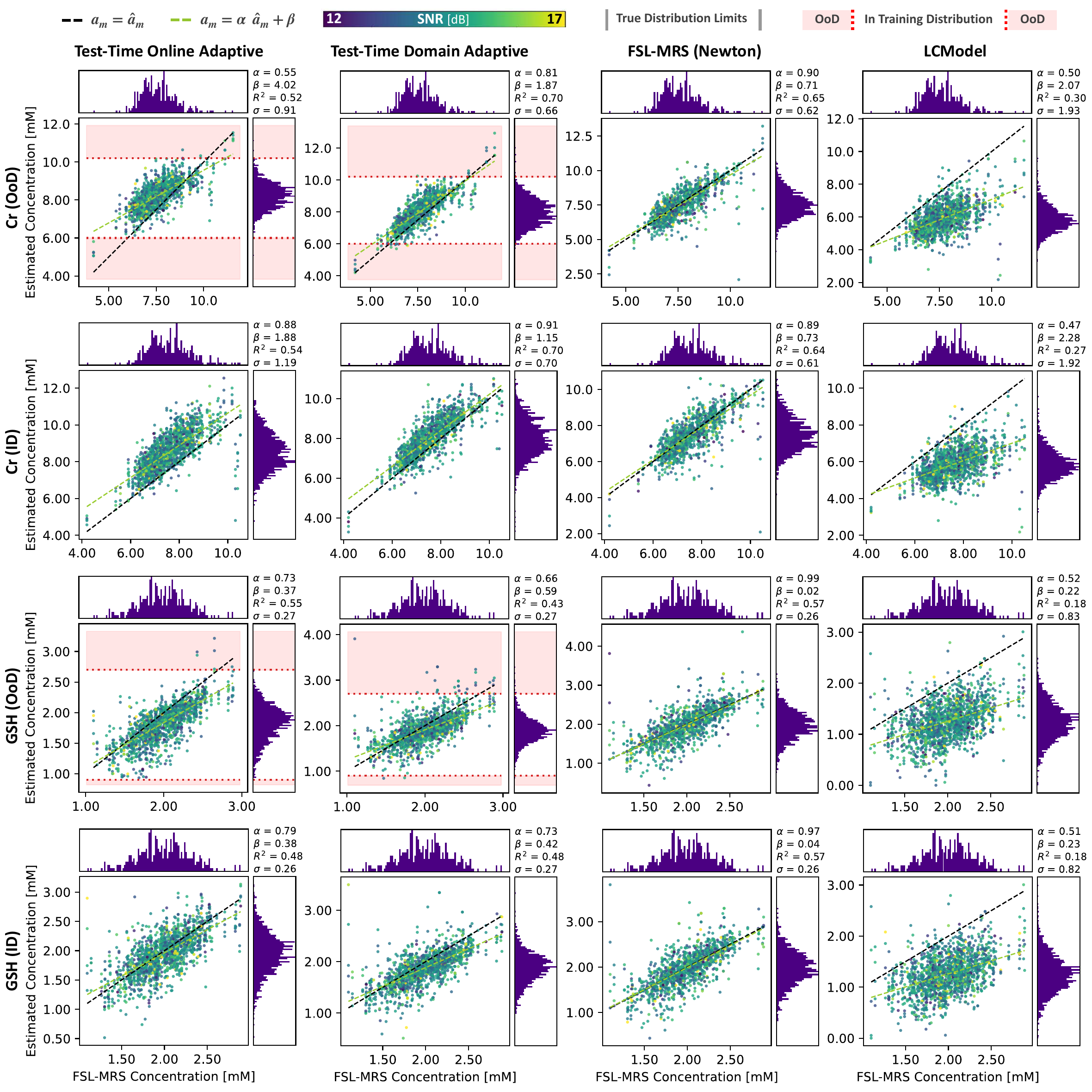}
    \caption{\enspace Scatter plots with marginal histograms comparing optimally scaled predicted versus pseudo-true (FSL-MRS estimates) concentrations of \ac{glu} and \ac{gaba} across 1,710 in-vivo spectra. Models are evaluated under two scenarios for the full concentration range: trained on mid-range concentrations (\ac{ood}) or trained on the full range (\ac{id}). Methods include test-time online adaptive approaches, test-time domain adaptive approaches compared with FSL-MRS (Newton) and LCModel. Points are colored by \ac{snr}, and regression lines with corresponding statistics (slope $\alpha$, intercept $\beta$, R$^2$, and \ac{rmse} $\sigma$) are shown.
    \vspace{3cm}
    }
    \label{fig:invivo_fsl_64_dist_comb_2_mosae}
\end{figure*}

\begin{figure*}
    \centering
    \includegraphics[width=2.0\columnwidth]{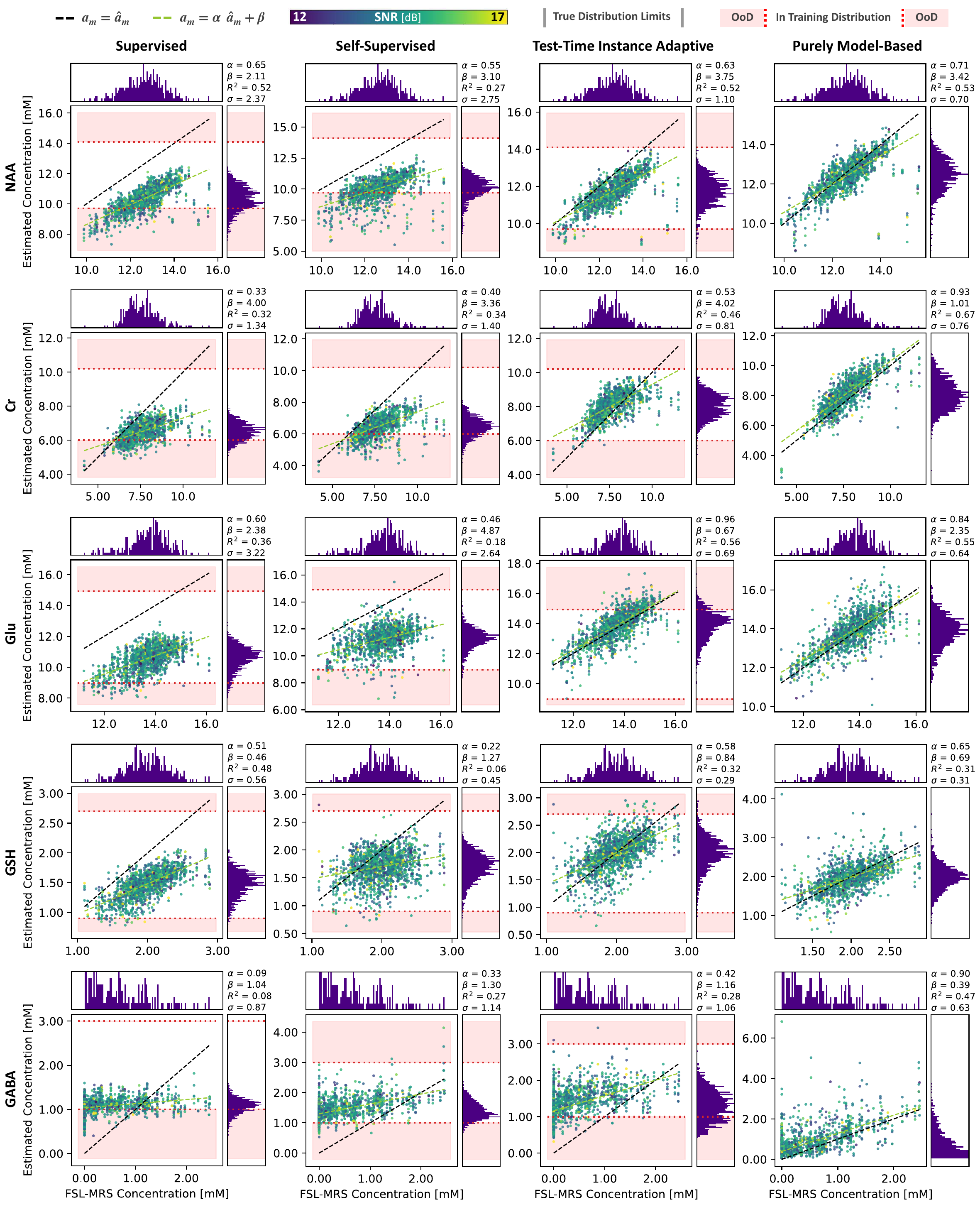}
    \caption{\enspace Scatter plots with marginal histograms comparing predicted versus pseudo-true (FSL-MRS estimates) concentrations for \ac{naa}, \ac{cr}, \ac{glu}, \ac{gsh}, and \ac{gaba} across 1,710 in-vivo spectra. Models were trained on mid-range concentrations and evaluated across the full concentration range to assess extrapolation performance. This figure shows data-driven methods: supervised, self-supervised, and test-time instance adaptive against purely model-based. Points are colored by \ac{snr}, and regression lines with corresponding statistics (R$^2$, slope, intercept, \ac{rmse}) are included.}
    \label{fig:invivo_fsl_64_dist_ood_1_sel_mae}
\end{figure*}

\begin{figure*}
    \centering
    \includegraphics[width=2.0\columnwidth]{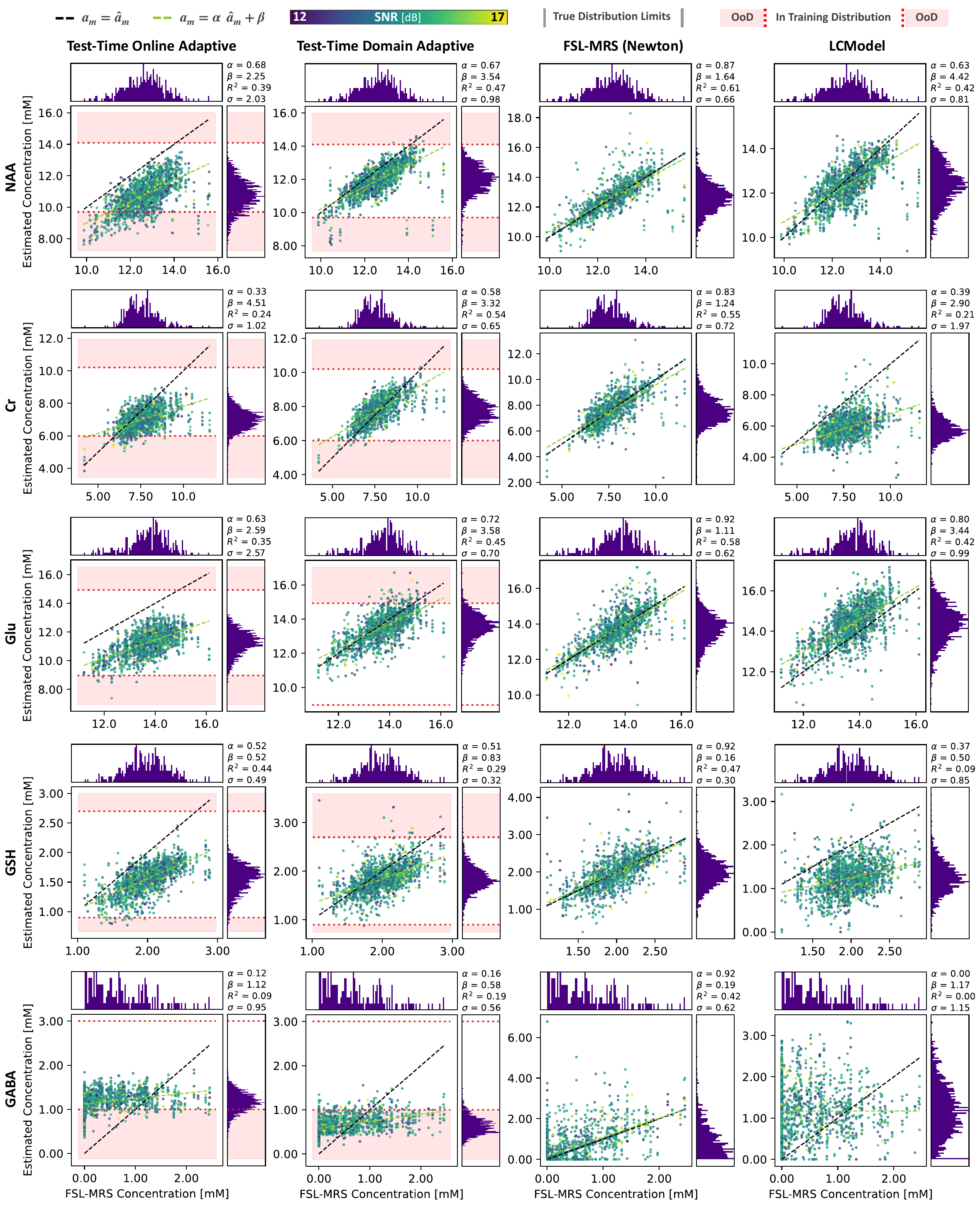}
    \caption{\enspace Scatter plots with marginal histograms comparing predicted versus pseudo-true (FSL-MRS estimates) concentrations for \ac{naa}, \ac{cr}, \ac{glu}, \ac{gsh}, and \ac{gaba} across 1,710 in-vivo spectra. Models were trained on mid-range concentrations and evaluated across the full concentration range. This figure shows adaptive and classical methods: test-time online adaptive, test-time domain adaptive, FSL-MRS (Newton), and LCModel. Points are colored by \ac{snr}, and regression lines with corresponding statistics (R$^2$, slope, intercept, \ac{rmse}) are included.}
    \label{fig:invivo_fsl_64_dist_ood_2_sel_mae}
\end{figure*}

\begin{figure*}
    \centering
    \includegraphics[width=2.0\columnwidth]{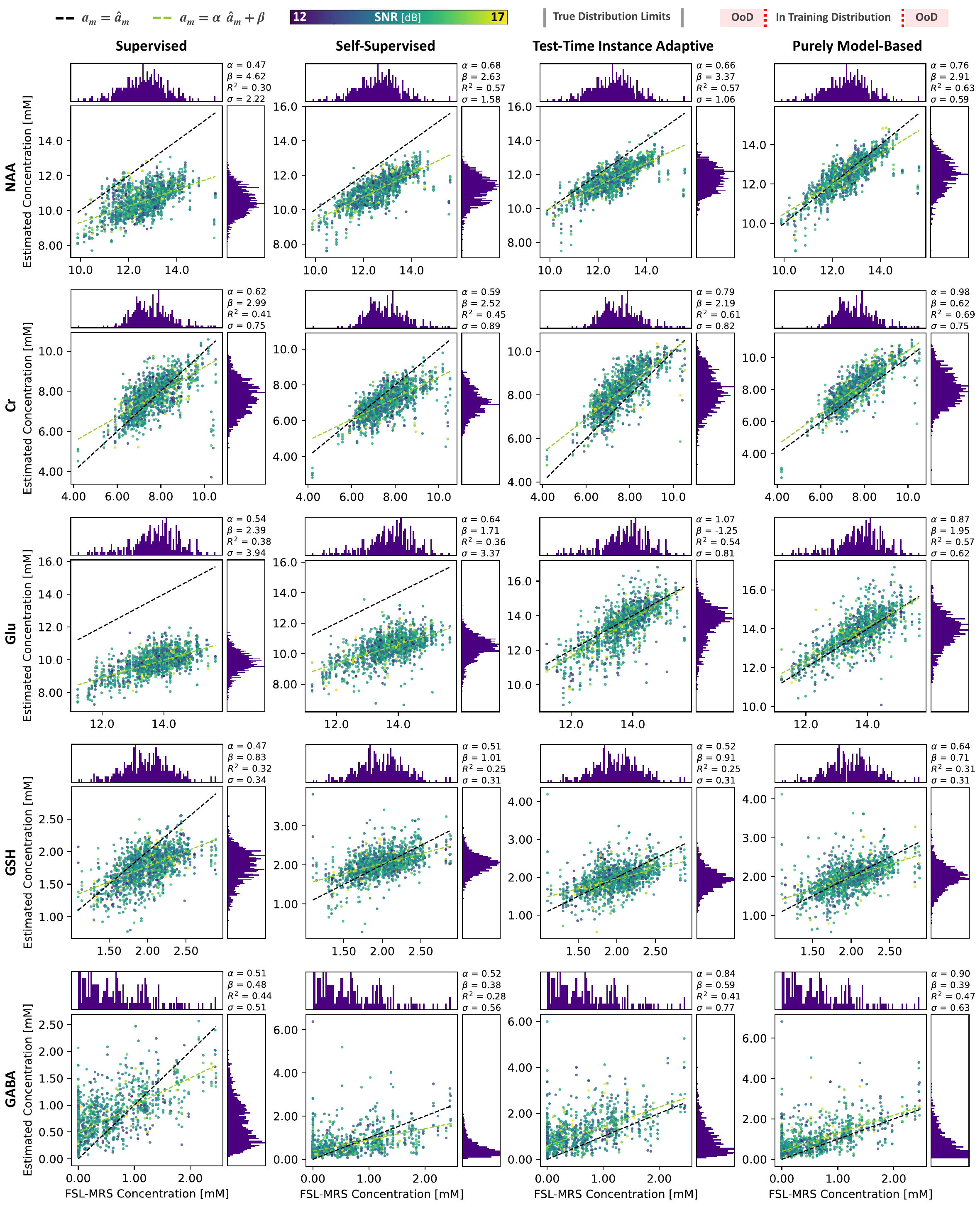}
    \caption{\enspace Scatter plots with marginal histograms comparing predicted versus pseudo-true (FSL-MRS estimates) concentrations for \ac{naa}, \ac{cr}, \ac{glu}, \ac{gsh}, and \ac{gaba} across 1,710 in-vivo spectra. Models were trained and tested across the full concentration range. This figure shows data-driven methods: supervised, self-supervised, and test-time instance adaptive against purely model-based. Points are colored by \ac{snr}, and regression lines with corresponding statistics (R$^2$, slope, intercept, \ac{rmse}) are included.}
    \label{fig:invivo_fsl_64_dist_idft_1_sel_mae}
\end{figure*}

\begin{figure*}
    \centering
    \includegraphics[width=2.0\columnwidth]{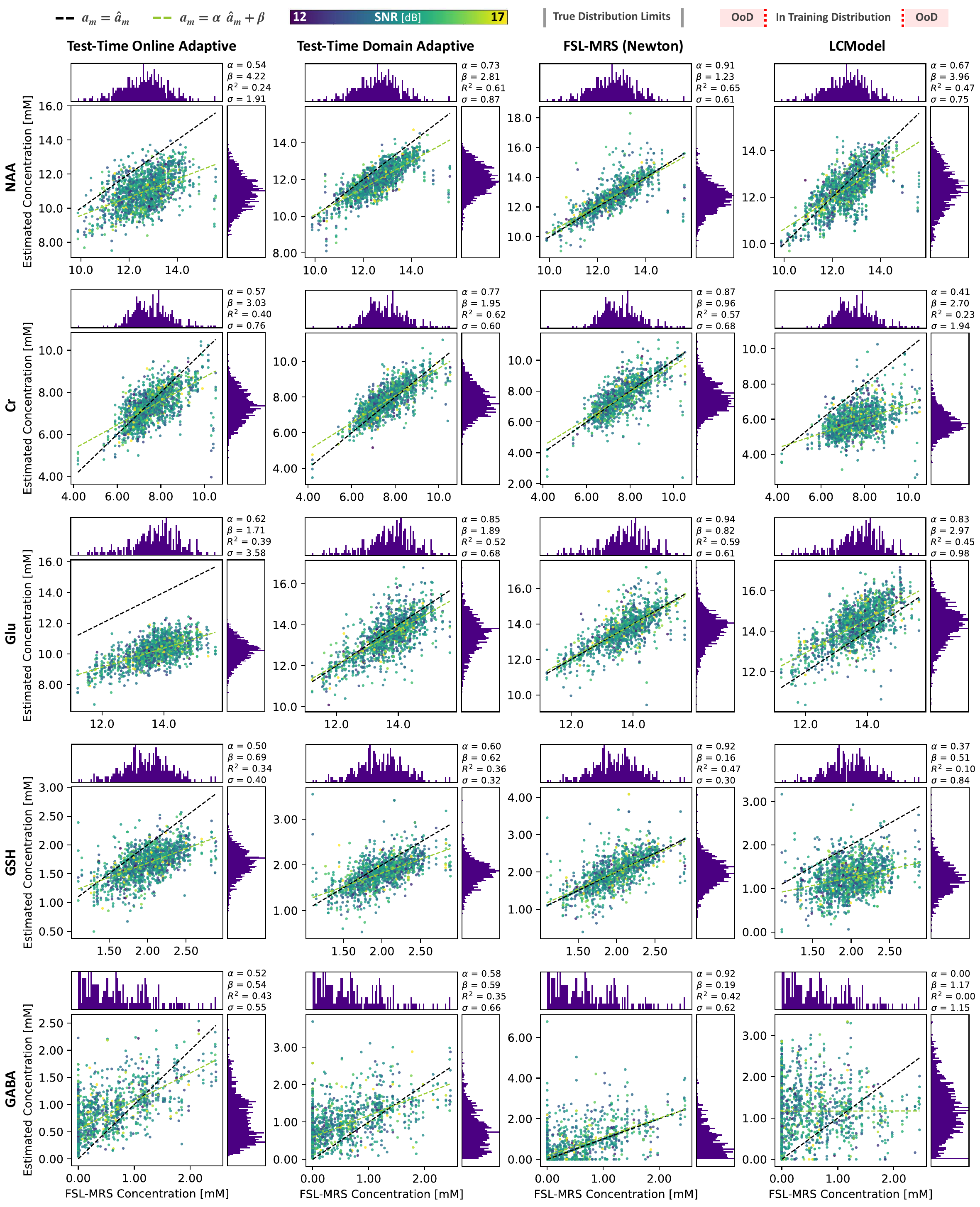}
    \caption{\enspace Scatter plots with marginal histograms comparing predicted versus pseudo-true (FSL-MRS estimates) concentrations for \ac{naa}, \ac{cr}, \ac{glu}, \ac{gsh}, and \ac{gaba} across 1,710 in-vivo spectra. Models were trained and tested across the full concentration range. This figure shows adaptive and classical methods: test-time online adaptive, test-time domain adaptive, FSL-MRS (Newton), and LCModel. Points are colored by \ac{snr}, and regression lines with corresponding statistics (R$^2$, slope, intercept, \ac{rmse}) are included.}
    \label{fig:invivo_fsl_64_dist_idft_2_sel_mae}
\end{figure*}

\begin{figure*}
    \centering
    \includegraphics[width=2.0\columnwidth]{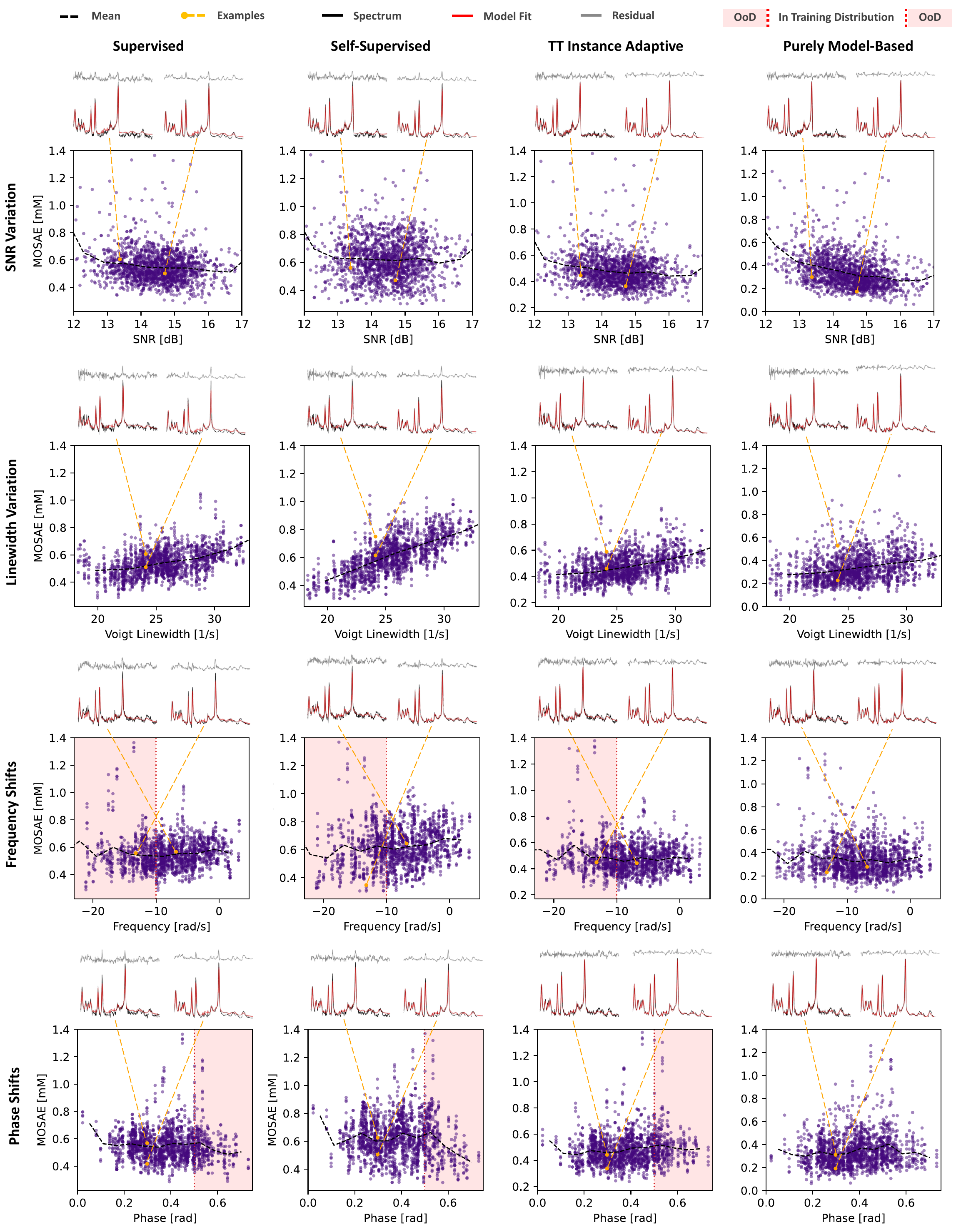}
    \caption{\enspace Scatter plots showing quantification accuracy (\ac{mosae}) across 1,710 in-vivo spectra as a function of estimated \ac{snr}, linewidth, zeroth-order phase shift, and frequency offset. Data-driven methods include supervised, self-supervised, and test-time instance adaptive compared against purely model-based fitting. Each point represents one spectrum, illustrating method-specific sensitivity to core signal parameter variations.}
    \label{fig:invivo_fsl_64_rang_1_1_mosae}
\end{figure*}

\begin{figure*}
    \vspace{5cm}
    \centering
    \includegraphics[width=2.0\columnwidth]{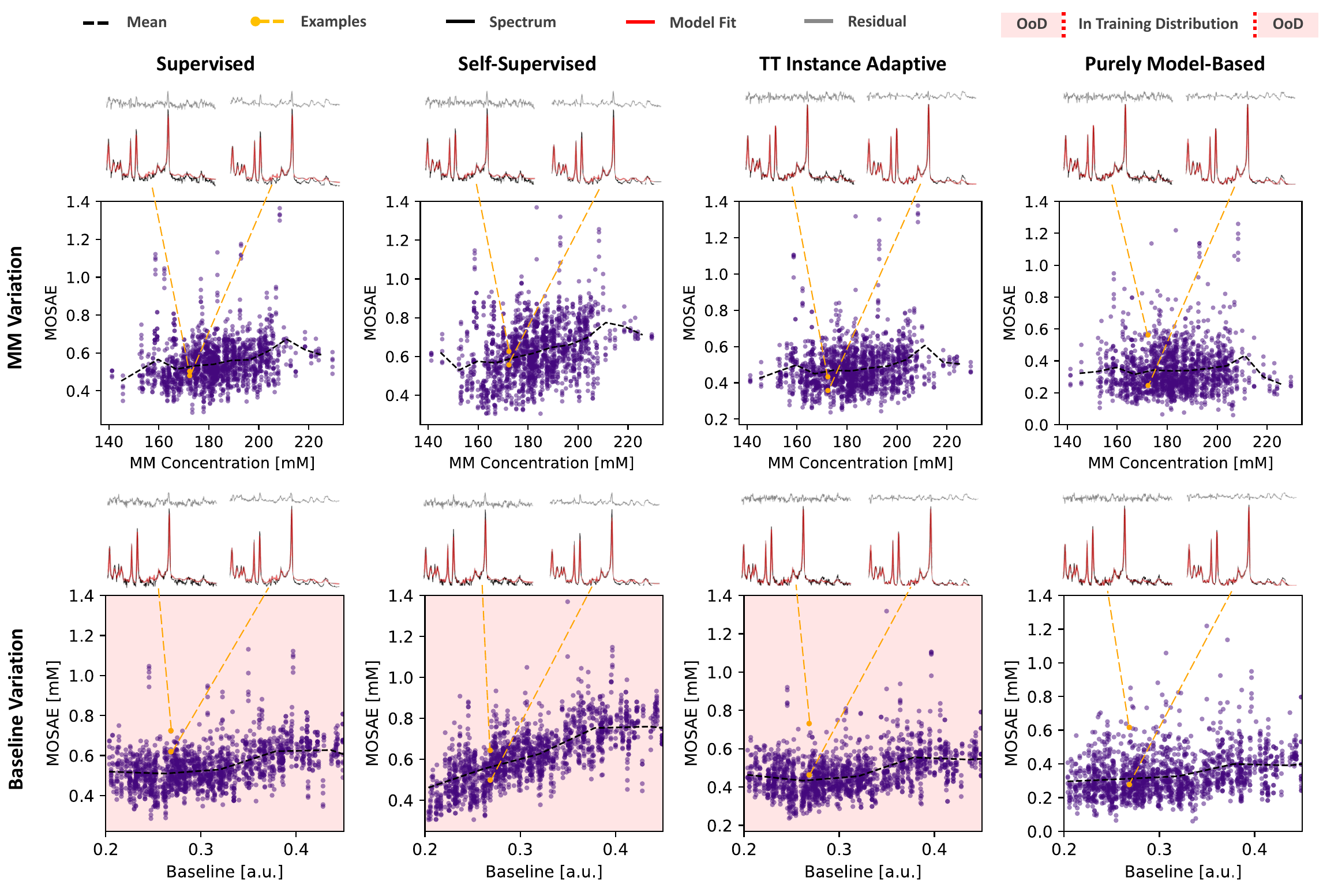}
    \caption{\enspace Scatter plots showing quantification accuracy (\ac{mosae}) across 1,710 in-vivo spectra under estimated macromolecular baseline (\ac{mm}), baseline variation, and random signal corruptions. Data-driven methods include supervised, self-supervised, and test-time instance adaptive compared against purely model-based fitting. Each point represents one spectrum, illustrating method-specific robustness to unmodeled spectral deviations.
    \vspace{5cm}
    }
    \label{fig:invivo_fsl_64_rang_1_2_mosae}
\end{figure*}

\begin{figure*}
    \centering
    \includegraphics[width=2.0\columnwidth]{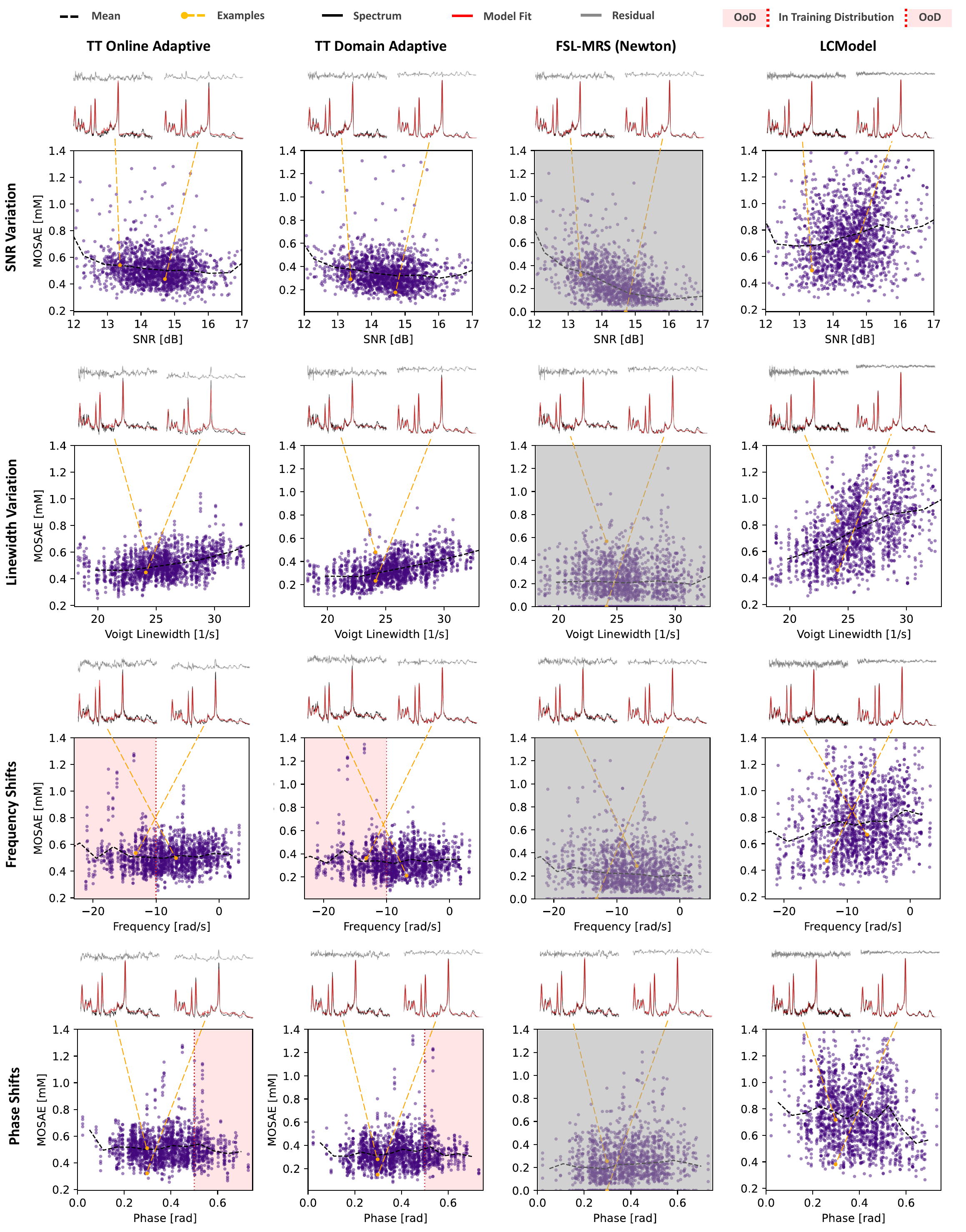}
    \caption{\enspace Scatter plots showing quantification accuracy (\ac{mosae}) across 1,710 in-vivo spectra as a function of estimated \ac{snr}, linewidth, zeroth-order phase shift, and frequency offset. Adaptive and classical methods include test-time online adaptive, test-time domain adaptive, FSL-MRS (Newton), and LCModel. Each point represents one spectrum, illustrating method-specific sensitivity to core signal parameter variations.}
    \label{fig:invivo_fsl_64_rang_2_1_mosae}
\end{figure*}

\begin{figure*}
    \vspace{6cm}
    \centering
    \includegraphics[width=2.0\columnwidth]{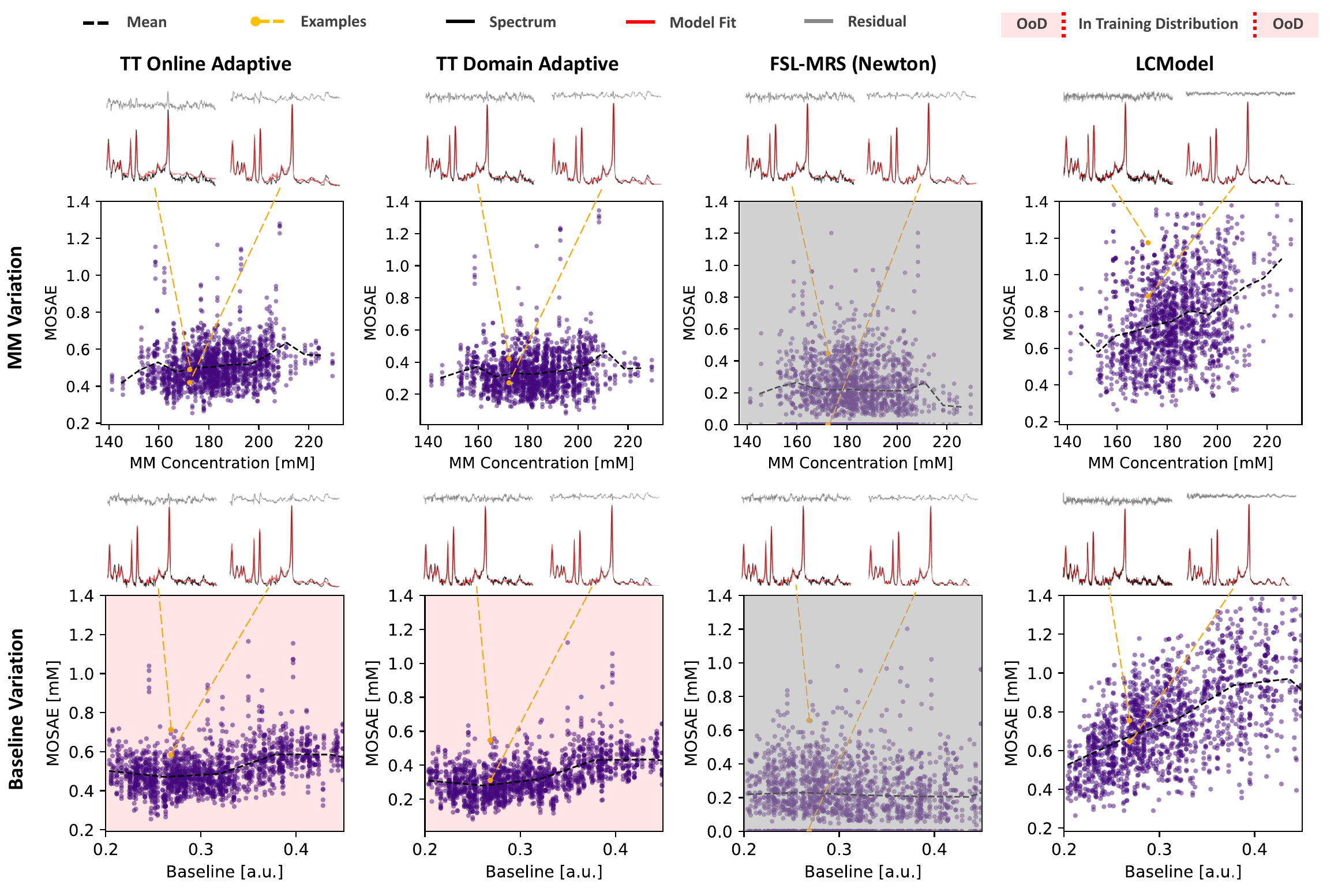}
    \caption{\enspace Scatter plots showing quantification accuracy (\ac{mosae}) across 1,710 in-vivo spectra under macromolecular baseline (\ac{mm}), baseline variation, and random signal corruptions. Adaptive and classical methods include test-time online adaptive, test-time domain adaptive, FSL-MRS (Newton), and LCModel. Each point represents one spectrum, illustrating method-specific robustness to unmodeled spectral deviations.
    \vspace{6cm}
    }
    \label{fig:invivo_fsl_64_rang_2_2_mosae}
\end{figure*}

\begin{figure*}
    \vspace{1cm}
    \centering
    \includegraphics[width=2.0\columnwidth]{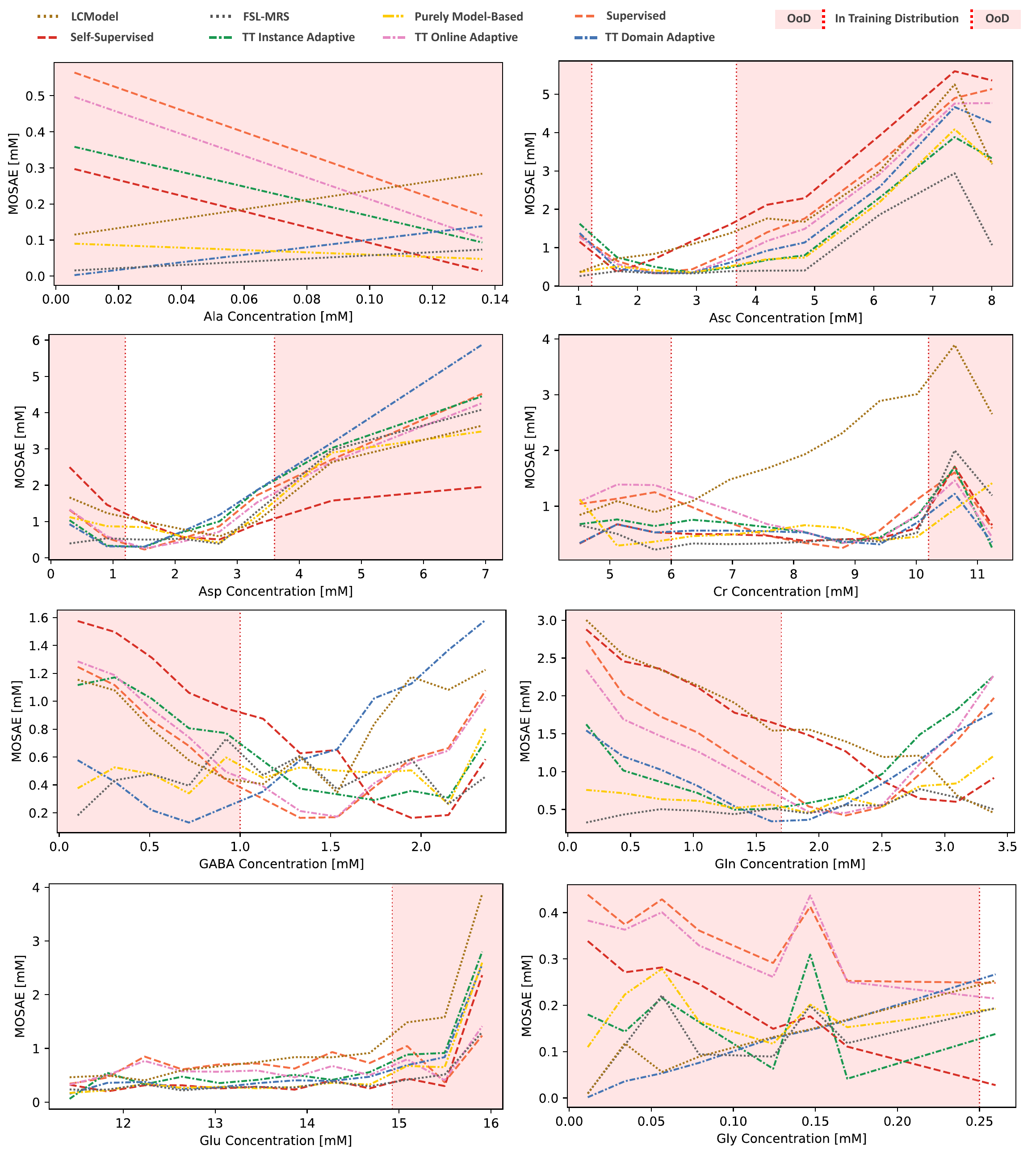}
    \caption{\enspace Summary of quantification performance across 1,710 in-vivo spectra for eight metabolites (Ala, Asc, Asp, Cr, GABA, Gln, Glu, Gly). Each subplot corresponds to one metabolite, showing the \ac{mosae} for all methods: purely model-based gradient descent, FSL-MRS (Newton), LCModel, supervised, self-supervised, and \ac{tta} strategies. This visualization allows comparison of method performance across metabolites under in-vivo conditions.
    \vspace{1cm}
    }
    \label{fig:invivo_fsl_64_perf_metabs_1_1_mosae}
\end{figure*}

\begin{figure*}
    \vspace{1cm}
    \centering
    \includegraphics[width=2.0\columnwidth]{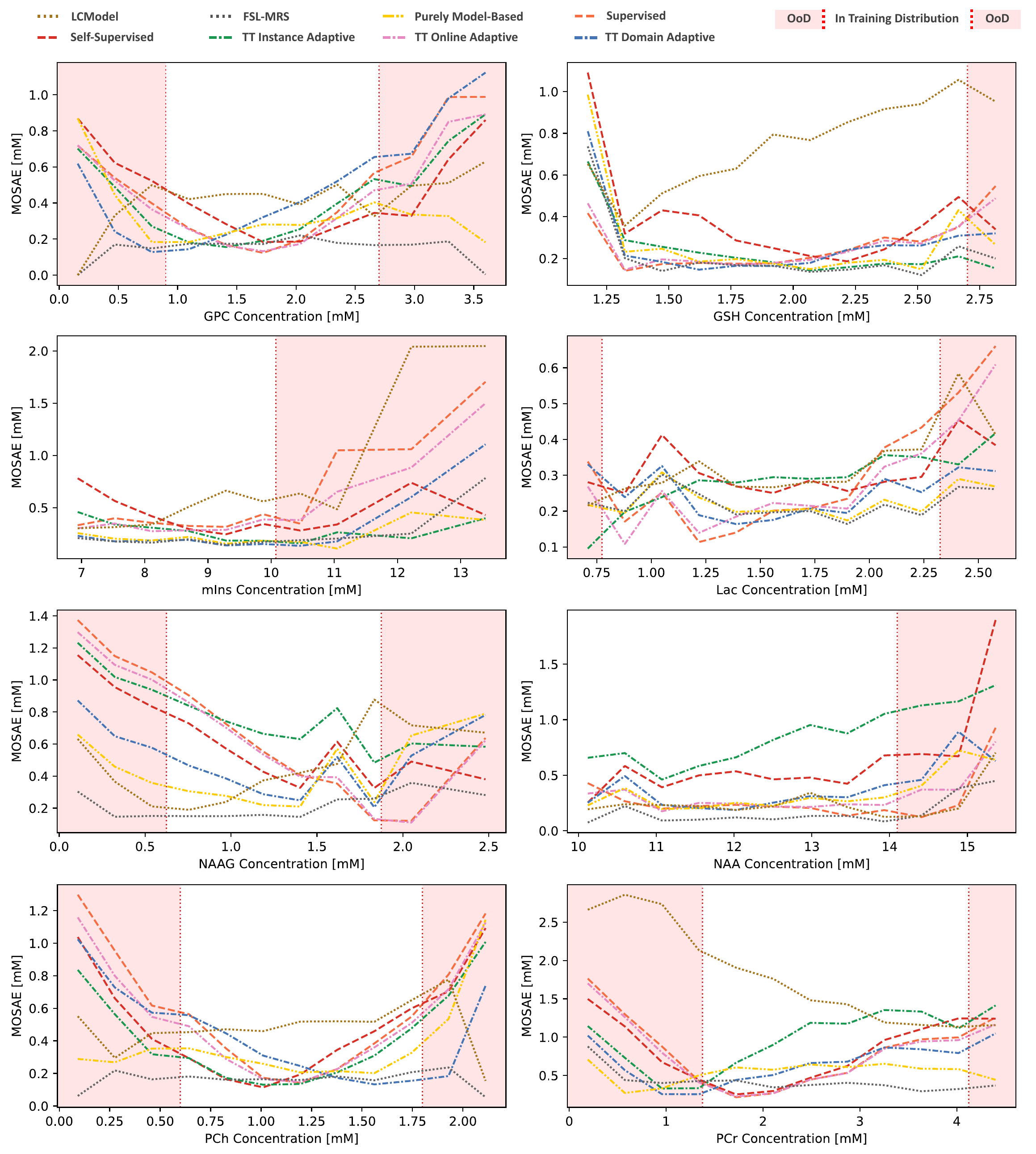}
    \caption{\enspace Summary of quantification performance across 1,710 in-vivo spectra for eight metabolites (GPC, GSH, mIns, Lac, NAAG, NAA, PCh, PCr). Each subplot corresponds to one metabolite, showing the \ac{mosae} for all methods: purely model-based gradient descent, FSL-MRS (Newton), LCModel, supervised, self-supervised, and \ac{tta} strategies. This visualization allows comparison of method performance across metabolites under in-vivo conditions.
    \vspace{1cm}
    }
    \label{fig:invivo_fsl_64_perf_metabs_1_2_mosae}
\end{figure*}

\begin{figure*}
    \vspace{6cm}
    \centering
    \includegraphics[width=2.0\columnwidth]{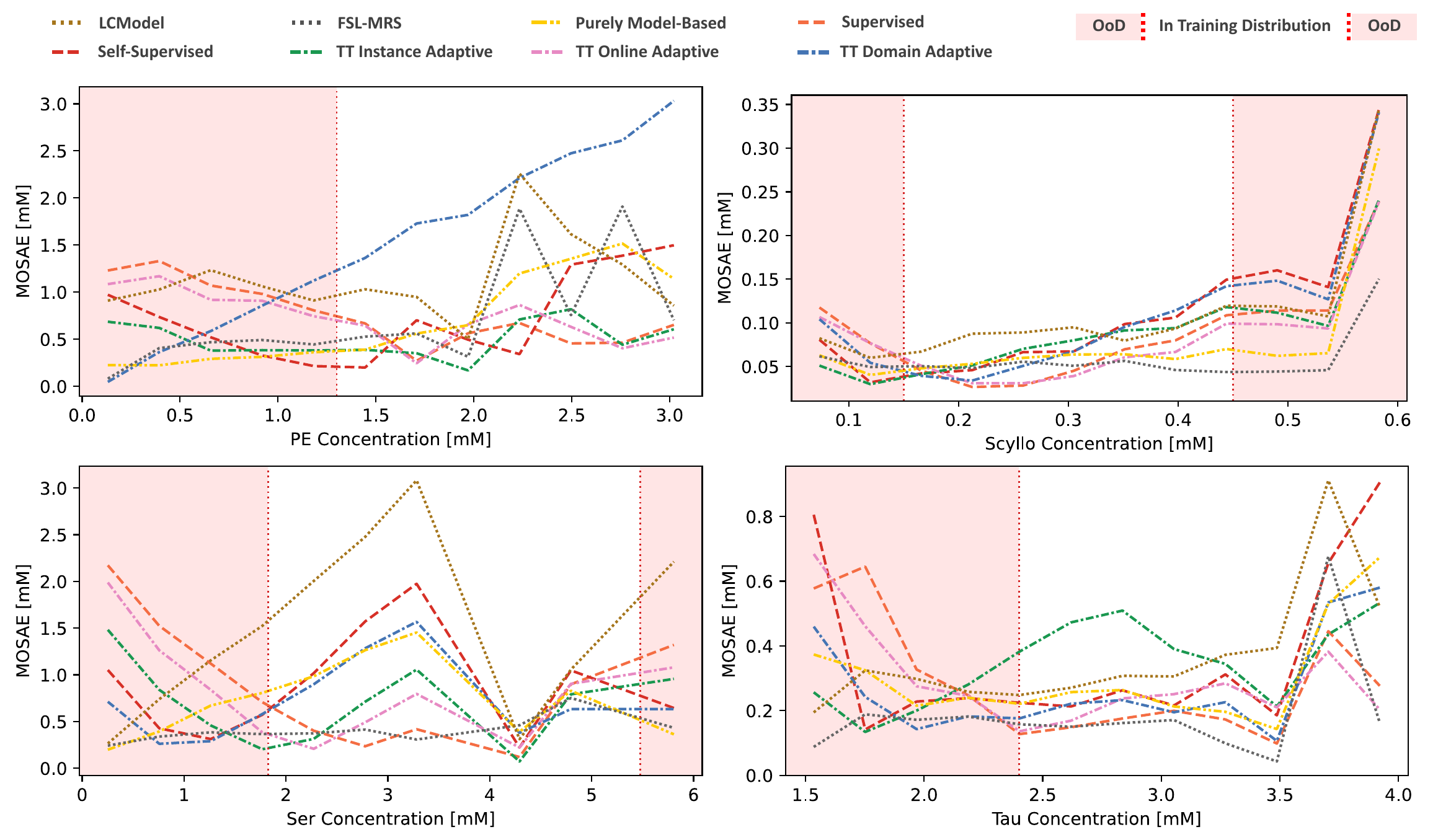}
    \caption{\enspace Summary of quantification performance across 1,710 in-vivo spectra for four metabolites (PE, Scyllo, Ser, Tau). Each subplot corresponds to one metabolite, showing the \ac{mosae} for all methods: purely model-based gradient descent, FSL-MRS (Newton), LCModel, supervised, self-supervised, and \ac{tta} strategies. This visualization allows comparison of method performance across metabolites under in-vivo conditions.
    \vspace{6cm}
    }
    \label{fig:invivo_fsl_64_perf_metabs_1_3_mosae}
\end{figure*}

\begin{figure*}
    \vspace{4cm}
    \centering
    \includegraphics[width=2.0\columnwidth]{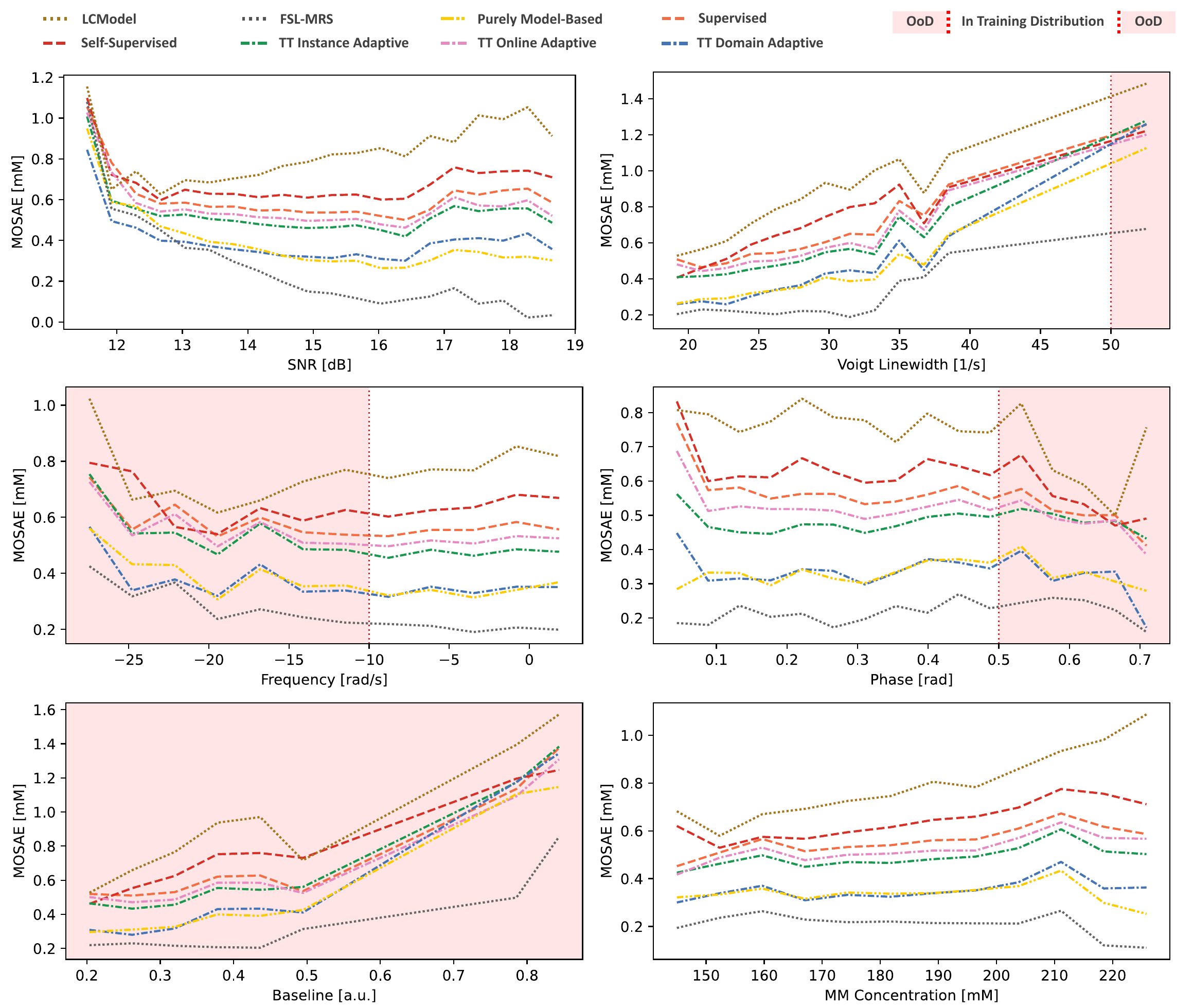}
    \caption{\enspace Summary of quantification performance across 1,710 in-vivo spectra. For each signal parameter, \ac{snr}, linewidth, frequency offset, phase shift, \ac{mm} baseline, and polynomial baseline, the \ac{mosae} is averaged within parameter bins. All methods are overlaid in each subplot, including purely model-based gradient descent, FSL-MRS (Newton), LCModel, supervised, self-supervised, and \ac{tta} strategies. Each point represents one spectrum, illustrating method-specific robustness to unmodeled spectral deviations in real in-vivo data.
    \vspace{4cm}
    }
    \label{fig:invivo_fsl_64_perf_other_1_mosae}
\end{figure*}

This section presents further results from the in-vivo data evaluation, including:
\begin{itemize}
    \item \textbf{Tables \ref{tab:invivo_fslmrs_mae}, \ref{tab:invivo_fslmrs_mosae}, \ref{tab:invivo_mix_mae}, \ref{tab:invivo_mix_mosae}, \ref{tab:invivo_lcmodel_mae}, \ref{tab:invivo_lcmodel_mosae}}: These tables extend the in-vivo quantification performance analysis using different pseudo ground truths (FSL-MRS, mean of FSL-MRS and LCModel, and LCModel), providing both \ac{mae} and \ac{mosae} across the \ac{id} and \ac{ood} scenarios. They also include results for CNN-based models and various \ac{tta} initialization and iteration settings.

    \item \textbf{Figures \ref{fig:invivo_fsl_64_dist_comb_1_mosae}, \ref{fig:invivo_fsl_64_dist_comb_2_mosae}}: These scatter plots with marginal histograms compare optimally scaled predicted versus pseudo-true concentrations for \ac{glu} and \ac{gaba} across 1,710 in-vivo spectra, under \ac{id} and \ac{ood} conditions.

    \item \textbf{Figures \ref{fig:invivo_fsl_64_dist_ood_1_sel_mae}, \ref{fig:invivo_fsl_64_dist_ood_2_sel_mae}, \ref{fig:invivo_fsl_64_dist_idft_1_sel_mae}, \ref{fig:invivo_fsl_64_dist_idft_2_sel_mae}}: These figures show predicted versus pseudo-true concentrations for \ac{naa}, \ac{cr}, \ac{glu}, \ac{gsh}, and \ac{gaba} in-vivo, under both \ac{ood} and \ac{id} full-range scenarios.

    \item \textbf{Figures \ref{fig:invivo_fsl_64_rang_1_1_mosae}, \ref{fig:invivo_fsl_64_rang_1_2_mosae}, \ref{fig:invivo_fsl_64_rang_2_1_mosae}, \ref{fig:invivo_fsl_64_rang_2_2_mosae}}: These scatter plots illustrate \ac{mosae} across in-vivo spectra as a function of estimated \ac{snr}, linewidth, zeroth-order phase shift, frequency offset, \ac{mm}, and baseline variation.

    \item \textbf{Figures \ref{fig:invivo_fsl_64_perf_metabs_1_1_mosae}, \ref{fig:invivo_fsl_64_perf_metabs_1_2_mosae}, \ref{fig:invivo_fsl_64_perf_metabs_1_3_mosae}, \ref{fig:invivo_fsl_64_perf_other_1_mosae}}: These figures provide a summary of quantification performance for various metabolites and signal parameters in in-vivo spectra, showing mean \ac{mosae} within parameter bins.
\end{itemize}

\begin{table*}
\centering
\caption{\enspace 
Comparison of quantification methods across test scenarios of 10,000 spectra.
\textbf{ID (Mid-Range)}: Models trained and tested on the central 50\% metabolite concentration range.
\textbf{OoD (Full-Range)}: Models trained on the central 50\% concentration range, but tested across the entire range of concentrations to assess extrapolation.
\textbf{ID (Full-Trained)}: Models trained and tested on the full concentration range.
}
\label{tab:sim_performance_mae}
\vspace{2mm}
\begin{adjustbox}{width=1.0\textwidth}
\begin{tabular}{lccc c}
\toprule
\multirow{2}{*}{\textbf{Method}} 
  & \multicolumn{3}{c}{\textbf{MAE ↓ ($\pm$ SE)}} 
  & \multirow{2}{*}{\textbf{Time ↓ (ms/sample)}} \\
\cmidrule(lr){2-4}
  & \textbf{ID (Mid-Range)} & \textbf{OoD (Full-Range)} & \textbf{ID (Full-Trained)} & \\
\midrule
Supervised            & \bftab 0.2962 (± 0.0033) & 0.5970 (± 0.0059) & \bftab 0.4121 (± 0.0047) & \bftab 0.1565 \\
Self-Supervised       & 0.4037 (± 0.0046) & 0.5846 (± 0.0060) & 0.4863 (± 0.0060) & 0.1574 \\
TT Instance Adaptive  & 0.4071 (± 0.0051) & \bftab 0.4767 (± 0.0054) & 0.4699 (± 0.0061) & 30.4637 \\
TT Online Adaptive    & 0.3025 (± 0.0033) & 0.5917 (± 0.0059) & 0.4182 (± 0.0048) & 0.2454 \\
TT Domain Adaptive    & 0.3977 (± 0.0046) & 0.5166 (± 0.0054) & 0.4888 (± 0.0056) & 104.5304 \\
Purely Model-Based    & 0.5881 (± 0.0071) & 0.5819 (± 0.0072) & 0.5822 (± 0.0072) & 424.9928 \\
& & & & \\
Supervised (CNN)                             & 0.3212 (± 0.0034) & 0.6483 (± 0.0063) & 0.4326 (± 0.0049) & 0.1642 \\
Self-Supervised (CNN)                        & 0.4075 (± 0.0046) & 0.5780 (± 0.0060) & 0.4866 (± 0.0059) & 0.1643 \\
TT Instance Adaptive (CNN)                     & 0.4649 (± 0.0051) & 0.6364 (± 0.0066) & 0.4479 (± 0.0056) & 40.7961 \\
TT Instance Adaptive (CNN, Self-Sup. Init.)    & 0.4676 (± 0.0058) & 0.5412 (± 0.0063) & 0.5247 (± 0.0072)  & 40.9133 \\
& & & & \\
TT Instance Adaptive (Self-Sup. Init.)         & 0.4375 (± 0.0053) & 0.5362 (± 0.0060) & 0.5152 (± 0.0070) & 30.5271 \\
TT Instance Adaptive (From Scratch Init.)      & 0.5140 (± 0.0059) & 0.6119 (± 0.0069) & 0.6038 (± 0.0069) & 4567.4828 \\
TT Online Adaptive (Self-Sup. Init.)  & 0.4107 (± 0.0046) & 0.5876 (± 0.0060) & 0.4967 (± 0.0061) & 0.2362 \\
TT Domain Adaptive (Self-Sup. Init.)  & 0.4677 (± 0.0050) & 0.6299 (± 0.0064) & 0.5632 (± 0.0065) & 105.9836 \\
& & & & \\
TT Instance Adaptive (10 Iter.)   & 0.3170 (± 0.0036) & 0.5285 (± 0.0053) & 0.4245 (± 0.0050) & 12.7126 \\
TT Instance Adaptive (50 Iter.)   & 0.4071 (± 0.0051) & 0.4767 (± 0.0054) & 0.4378 (± 0.0060) & 30.5217 \\
TT Instance Adaptive (100 Iter.)  & 0.4660 (± 0.0061) & 0.4873 (± 0.0060) & 0.5071 (± 0.0069) & 59.3287 \\
TT Instance Adaptive (500 Iter.)  & 0.5312 (± 0.0070) & 0.5138 (± 0.0070) & 0.5363 (± 0.0074) & 285.3065 \\
\bottomrule
\end{tabular}
\end{adjustbox}
\begin{tablenotes}
\end{tablenotes}
\end{table*}

\begin{table*}
\centering
\caption{\enspace 
Comparison of quantification methods across test scenarios of 10,000 spectra.
\textbf{ID (Mid-Range)}: Models trained and tested on the central 50\% metabolite concentration range.
\textbf{OoD (Full-Range)}: Models trained on the central 50\% concentration range, but tested across the entire range of concentrations to assess extrapolation.
\textbf{ID (Full-Trained)}: Models trained and tested on the full concentration range.
}
\label{tab:sim_performance_mosae} 
\vspace{2mm}
\begin{adjustbox}{width=1.0\textwidth}
\begin{tabular}{lccc c}
\toprule
\multirow{2}{*}{\textbf{Method}} 
  & \multicolumn{3}{c}{\textbf{MOSAE ↓ ($\pm$ SE)}} 
  & \multirow{2}{*}{\textbf{Time ↓ (ms/sample)}} \\
\cmidrule(lr){2-4}
  & \textbf{ID (Mid-Range)} & \textbf{OoD (Full-Range)} & \textbf{ID (Full-Trained)} & \\
\midrule
Supervised            & \bftab 0.2764 (± 0.0032) & 0.5537 (± 0.0058) & \bftab 0.3896 (± 0.0047) & \bftab 0.1565 \\
Self-Supervised       & 0.3734 (± 0.0044) & 0.5401 (± 0.0059) & 0.4548 (± 0.0059) & 0.1574 \\
TT Instance Adaptive  & 0.3767 (± 0.0049) & \bftab 0.4420 (± 0.0052) & 0.4378 (± 0.0060) & 30.4637 \\
TT Online Adaptive    & 0.2788 (± 0.0032) & 0.5520 (± 0.0058) & 0.3923 (± 0.0048) & 0.2454 \\
TT Domain Adaptive    & 0.3621 (± 0.0044) & 0.4717 (± 0.0053) & 0.4478 (± 0.0055) & 104.5304 \\
Purely Model-Based    & 0.5299 (± 0.0070) & 0.5238 (± 0.0071) & 0.5229 (± 0.0071) & 424.9928  \\
FSL-MRS               & 0.5369 (± 0.0073) & 0.5210 (± 0.0073) & 0.5201 (± 0.0074) & 962.7825 \\
LCModel               & 0.5704 (± 0.0067) & 0.5843 (± 0.0072) & 0.5832 (± 0.0072) & 78.0310 \\
& & & & \\
Supervised (CNN)                             & 0.3017 (± 0.0033) & 0.6073 (± 0.0062) & 0.4084 (± 0.0049) & 0.1642 \\
Self-Supervised (CNN)                        & 0.3767 (± 0.0045) & 0.5335 (± 0.0059) & 0.4553 (± 0.0058) & 0.1643 \\
TT Instance Adaptive (CNN)                     & 0.4229 (± 0.0048) & 0.5911 (± 0.0063) & 0.4169 (± 0.0055) & 40.7961 \\
TT Instance Adaptive (CNN, Self-Sup. Init.)    & 0.4316 (± 0.0056) & 0.5027 (± 0.0061) & 0.4843 (± 0.0068) & 40.9133 \\
& & & & \\
TT Instance Adaptive (Self-Sup. Init.)         & 0.4004 (± 0.0051) & 0.4985 (± 0.0059) & 0.4768 (± 0.0067) & 30.5271 \\
TT Instance Adaptive (From Scratch Init.)      & 0.4492 (± 0.0055) & 0.5497 (± 0.0066) & 0.5395 (± 0.0066) & 4567.4828 \\
TT Online Adaptive (Self-Sup. Init.)  & 0.4004 (± 0.0051) & 0.4985 (± 0.0059) & 0.4594 (± 0.0060) & 0.2362 \\
TT Domain Adaptive (Self-Sup. Init.)  & 0.4007 (± 0.0045) & 0.5571 (± 0.0061) & 0.5081 (± 0.0063) & 105.9836 \\
& & & & \\
TT Instance Adaptive (10 Iter.)   & 0.2932 (± 0.0035) & 0.4979 (± 0.0053) & 0.3961 (± 0.0049) & 12.7126 \\
TT Instance Adaptive (50 Iter.)   & 0.3767 (± 0.0049) & 0.4420 (± 0.0052) & 0.4378 (± 0.0060) & 30.5217 \\
TT Instance Adaptive (100 Iter.)  & 0.4283 (± 0.0058) & 0.4496 (± 0.0058) & 0.4684 (± 0.0066) & 59.3287 \\
TT Instance Adaptive (500 Iter.)  & 0.4859 (± 0.0066) & 0.4708 (± 0.0066) & 0.4913 (± 0.0070) & 285.3065 \\
\bottomrule
\end{tabular}
\end{adjustbox}
\begin{tablenotes}
\end{tablenotes}
\end{table*}


\begin{table*}
\caption{\enspace 
Comparison of quantification methods on in-vivo data using pseudo ground truth: \textbf{FSL-MRS}. The spectra are filtered to create equivalent scenarios to the simulated test scenarios: \textbf{ID (Mid-Range)}, \textbf{OoD (Full-Range)}, and \textbf{ID (Full-Trained)}.
}
\label{tab:invivo_fslmrs_mae}
\vspace{2mm}
\begin{adjustbox}{width=1.0\textwidth}
\begin{tabular}{lccc c}
\toprule
\multirow{2}{*}{\textbf{Method}} 
  & \multicolumn{3}{c}{\textbf{MAE ↓ ($\pm$ SE)}}
  & \multirow{2}{*}{\textbf{Time ↓ (ms/sample)}} \\
\cmidrule(lr){2-4}
  & \textbf{ID (Mid-Range)} & \textbf{OoD (Full-Range)} & \textbf{ID (Full-Trained)}\\
\midrule
Supervised            & 0.7953 (± 0.0265) & 0.8633 (± 0.0211) & 0.7623 (± 0.0233) & \bftab  0.1145 \\
Self-Supervised       & 0.8173 (± 0.0267) & 0.8942 (± 0.0218) & 0.7073 (± 0.0215) & 0.1347 \\
TT Instance Adaptive  & 0.4674 (± 0.0140) & 0.5063 (± 0.0116) & 0.4770 (± 0.0120) & 212.2983 \\
TT Online Adaptive    & 0.6751 (± 0.0220) & 0.7440 (± 0.0179) & 0.7055 (± 0.0212) & 0.3400 \\
TT Domain Adaptive    & 0.3403 (± 0.0122) & 0.3922 (± 0.0111) & 0.3950 (± 0.0111) & 312.6671 \\
Purely Model-Based    & 0.3500 (± 0.0134) & 0.3802 (± 0.0112) & 0.3802 (± 0.0112) & 3154.4558 \\
FSL-MRS               & \ittab 0.2390 (± 0.0130) & \ittab 0.2560 (± 0.0109) & \ittab 0.2560 (± 0.0109) & 475.4806 \\
LCModel               & 0.7288 (± 0.0259) & 0.7935 (± 0.0205) & 0.7935 (± 0.0205) & 274.6317 \\
& & & \\
Supervised (CNN)                             & 0.7484 (± 0.0221) & 0.8245 (± 0.0180) & 0.7757 (± 0.0190) & 0.7374 \\
Self-Supervised (CNN)                        & 0.8474 (± 0.0327) & 0.8963 (± 0.0248) & 0.5071 (± 0.0127) & 0.1551 \\
TT Instance Adaptive (CNN)                     & 0.6163 (± 0.0179) & 0.6662 (± 0.0143) & 0.5326 (± 0.0124) & 256.5153 \\
TT Instance Adaptive (CNN, Self-Sup. Init.)    & 0.5948 (± 0.0197) & 0.6474 (± 0.0156) & 0.4592 (± 0.0121) & 270.3693 \\
& & & \\
TT Instance Adaptive (Self-Sup. Init.)         & 0.5419 (± 0.0178) & 0.6040 (± 0.0148) & 0.4559 (± 0.0119) & 205.7727 \\
TT Instance Adaptive (From Scratch Init.)      & \bftab  0.3272 (± 0.0121) & \bftab  0.3739 (± 0.0109) & \bftab  0.3761 (± 0.0109) & 29970.4098 \\
TT Online Adaptive (Self-Sup. Init.)  & 0.6353 (± 0.0217) & 0.7001 (± 0.0176) & 0.5357 (± 0.0150) & 0.3331 \\
TT Domain Adaptive (Self-Sup. Init.)  & 0.3752 (± 0.0140) & 0.4260 (± 0.0122) & 0.4219 (± 0.0123) & 312.5201 \\
& & & \\
TT Instance Adaptive (10 Iter.)   & 0.6069 (± 0.0189) & 0.6744 (± 0.0156) & 0.6571 (± 0.0192) & 44.8087 \\
TT Instance Adaptive (50 Iter.)   & 0.4674 (± 0.0140) & 0.5063 (± 0.0116) & 0.4770 (± 0.0120) & 212.2983 \\
TT Instance Adaptive (100 Iter.)  & 0.4220 (± 0.0134) & 0.4657 (± 0.0114) & 0.4404 (± 0.0118) & 414.0273 \\
TT Instance Adaptive (500 Iter.)  & 0.4171 (± 0.0141) & 0.4565 (± 0.0121) & 0.4574 (± 0.0127) & 2061.0019 \\
\bottomrule
\end{tabular}
\end{adjustbox}
\end{table*}

\begin{table*}
\caption{\enspace 
Comparison of quantification methods on in-vivo data using pseudo ground truth: \textbf{FSL-MRS}. The spectra are filtered to create equivalent scenarios to the simulated test scenarios: \textbf{ID (Mid-Range)}, \textbf{OoD (Full-Range)}, and \textbf{ID (Full-Trained)}.
}
\label{tab:invivo_fslmrs_mosae} 
\vspace{2mm}
\begin{adjustbox}{width=1.0\textwidth}
\begin{tabular}{lccc c}
\toprule
\multirow{2}{*}{\textbf{Method}} 
  & \multicolumn{3}{c}{\textbf{MOSAE ↓ ($\pm$ SE)}} 
  & \multirow{2}{*}{\textbf{Time ↓ (ms/sample)}} \\
\cmidrule(lr){2-4}
  & \textbf{ID (Mid-Range)} & \textbf{OoD (Full-Range)} & \textbf{ID (Full-Trained)} \\
\midrule
Supervised            & 0.5164 (± 0.0170) & 0.5533 (± 0.0136) & 0.5562 (± 0.0159) & \bftab 0.1145 \\
Self-Supervised       & 0.5773 (± 0.0217) & 0.6249 (± 0.0171) & 0.5037 (± 0.0138) & 0.1347 \\
TT Instance Adaptive  & 0.4498 (± 0.0144) & 0.4818 (± 0.0115) & 0.4489 (± 0.0118) & 212.2983 \\
TT Online Adaptive    & 0.4793 (± 0.0158) & 0.5156 (± 0.0127) & 0.5101 (± 0.0143) & 0.3400 \\
TT Domain Adaptive    & 0.3052 (± 0.0120) & 0.3425 (± 0.0102) & 0.3490 (± 0.0104) & 312.6671 \\
Purely Model-Based    & 0.3168 (± 0.0130) & 0.3431 (± 0.0106) & 0.3431 (± 0.0106) & 3154.4558 \\
FSL-MRS               & \ittab 0.2104 (± 0.0119) & \ittab 0.2210 (± 0.0097) & \ittab 0.2210 (± 0.0097) & 475.4806 \\
LCModel               & 0.7027 (± 0.0262) & 0.7611 (± 0.0207) & 0.7611 (± 0.0207) & 274.6317 \\
& & & \\
Supervised (CNN)                             & 0.5543 (± 0.0184) & 0.5923 (± 0.0141) & 0.5302 (± 0.0131) & 0.7374 \\
Self-Supervised (CNN)                        & 0.7274 (± 0.0278) & 0.7538 (± 0.0211) & 0.4423 (± 0.0125) & 0.1551 \\
TT Instance Adaptive (CNN)                     & 0.5834 (± 0.0184) & 0.6199 (± 0.0141) & 0.4991 (± 0.0123) & 256.5153 \\
TT Instance Adaptive (CNN, Self-Sup. Init.)    & 0.5509 (± 0.0206) & 0.5902 (± 0.0161) & 0.4254 (± 0.0118) & 270.3693 \\
& & & \\
TT Instance Adaptive (Self-Sup. Init.)         & 0.5029 (± 0.0185) & 0.5518 (± 0.0150) & 0.4178 (± 0.0116) & 205.7727 \\
TT Instance Adaptive (From Scratch Init.)      & \bftab 0.2982 (± 0.0120) & \bftab 0.3307 (± 0.0102) & \bftab 0.3328 (± 0.0102) & 29970.4098 \\
TT Online Adaptive (Self-Sup. Init.)  & 0.5339 (± 0.0202) & 0.5842 (± 0.0160) & 0.4080 (± 0.0110) & 0.3331 \\
TT Domain Adaptive (Self-Sup. Init.)  & 0.3169 (± 0.0123) & 0.3528 (± 0.0104) & 0.3419 (± 0.0107) & 312.5201 \\
& & & \\
TT Instance Adaptive (10 Iter.)   & 0.4654 (± 0.0150) & 0.5018 (± 0.0122) & 0.5179 (± 0.0147) & 44.8087 \\
TT Instance Adaptive (50 Iter.)   & 0.4498 (± 0.0144) & 0.4818 (± 0.0115) & 0.4489 (± 0.0118) & 212.2983 \\
TT Instance Adaptive (100 Iter.)  & 0.4028 (± 0.0135) & 0.4361 (± 0.0110) & 0.4100 (± 0.0114) & 414.0273 \\
TT Instance Adaptive (500 Iter.)  & 0.3932 (± 0.0143) & 0.4221 (± 0.0117) & 0.4253 (± 0.0124) & 2061.0019 \\
\bottomrule
\end{tabular}
\end{adjustbox} 
\end{table*}

\begin{table*}
\caption{\enspace 
Comparison of quantification methods on in-vivo data using pseudo ground truth: \textbf{Mean of FSL-MRS and LCModel}. The spectra are filtered to create equivalent scenarios to the simulated test scenarios: \textbf{ID (Mid-Range)}, \textbf{OoD (Full-Range)}, and \textbf{ID (Full-Trained)}.
}
\label{tab:invivo_mix_mae} 
\vspace{2mm}
\begin{adjustbox}{width=1.0\textwidth}
\begin{tabular}{lccc c}
\toprule
\multirow{2}{*}{\textbf{Method}} 
  & \multicolumn{3}{c}{\textbf{MAE ↓ ($\pm$ SE)}}
  & \multirow{2}{*}{\textbf{Time ↓ (ms/sample)}} \\
\cmidrule(lr){2-4}
  & \textbf{ID (Mid-Range)} & \textbf{OoD (Full-Range)} & \textbf{ID (Full-Trained)}\\
\midrule
Supervised            & 0.7251 (± 0.0386) & 0.7498 (± 0.0219) & 0.8030 (± 0.0245) & \bftab 0.1145 \\
Self-Supervised       & 0.6487 (± 0.0353) & 0.6902 (± 0.0206) & 0.7501 (± 0.0220) & 0.1347 \\
TT Instance Adaptive  & 0.5806 (± 0.0250) & 0.5671 (± 0.0133) & 0.6576 (± 0.0154) & 212.2983 \\
TT Online Adaptive    & 0.6354 (± 0.0319) & 0.6617 (± 0.0185) & 0.7498 (± 0.0228) & 0.3400 \\
TT Domain Adaptive    & 0.4586 (± 0.0217) & 0.4723 (± 0.0122) & 0.4832 (± 0.0125) & 312.6671 \\
Purely Model-Based    & 0.4646 (± 0.0233) & 0.4868 (± 0.0130) & 0.4868 (± 0.0130) & 3154.4558 \\
FSL-MRS               & \ittab 0.5018 (± 0.0242) & \ittab 0.5256 (± 0.0138) & \ittab 0.5256 (± 0.0138) & 475.4806 \\
LCModel               & \ittab 0.4296 (± 0.0200) & \ittab 0.4534 (± 0.0117) & \ittab 0.4534 (± 0.0117) & 274.6317 \\
& & & & \\
Supervised (CNN)                             & 0.6885 (± 0.0322) & 0.7062 (± 0.0182) & 0.8248 (± 0.0217) & 0.7374 \\
Self-Supervised (CNN)                        & 0.8069 (± 0.0436) & 0.8180 (± 0.0242) & 0.5382 (± 0.0137) & 0.1551 \\
TT Instance Adaptive (CNN)                     & 0.5977 (± 0.0233) & 0.5977 (± 0.0130) & 0.7218 (± 0.0168) & 256.5153 \\
TT Instance Adaptive (CNN, Self-Sup. Init.)    & 0.4798 (± 0.0219) & 0.5139 (± 0.0125) & 0.4984 (± 0.0129) & 270.3693 \\
& & & & \\
TT Instance Adaptive (Self-Sup. Init.)         & \bftab 0.4250 (± 0.0183) & \bftab 0.4605 (± 0.0109) & 0.5286 (± 0.0132) & 205.7727 \\
TT Instance Adaptive (From Scratch Init.)      & 0.4541 (± 0.0218) & 0.4790 (± 0.0124) & \bftab 0.4768 (± 0.0124) & 29970.4098 \\
TT Online Adaptive (Self-Sup. Init.)  & 0.5103 (± 0.0238) & 0.5410 (± 0.0142) & 0.6028 (± 0.0160) & 0.3331 \\
TT Domain Adaptive (Self-Sup. Init.)  & 0.5091 (± 0.0244) & 0.5301 (± 0.0138) & 0.5094 (± 0.0135) & 312.5201 \\
& & & & \\
TT Instance Adaptive (10 Iter.)   & 0.6057 (± 0.0285) & 0.6258 (± 0.0162) & 0.7280 (± 0.0212) & 44.8087 \\
TT Instance Adaptive (50 Iter.)   & 0.5806 (± 0.0250) & 0.5671 (± 0.0133) & 0.6576 (± 0.0154) & 212.2983 \\
TT Instance Adaptive (100 Iter.)  & 0.5328 (± 0.0251) & 0.5315 (± 0.0135) & 0.6031 (± 0.0153) & 414.0273 \\
TT Instance Adaptive (500 Iter.)  & 0.4819 (± 0.0238) & 0.5045 (± 0.0133) & 0.5380 (± 0.0141) & 2061.0019 \\
\bottomrule
\end{tabular}
\end{adjustbox}
\end{table*}

\begin{table*}
\caption{\enspace 
Comparison of quantification methods on in-vivo data using pseudo ground truth: \textbf{Mean of FSL-MRS and LCModel}. The spectra are filtered to create equivalent scenarios to the simulated test scenarios: \textbf{ID (Mid-Range)}, \textbf{OoD (Full-Range)}, and \textbf{ID (Full-Trained)}.
}
\label{tab:invivo_mix_mosae} 
\vspace{2mm}
\begin{adjustbox}{width=1.0\textwidth}
\begin{tabular}{lccc c}
\toprule
\multirow{2}{*}{\textbf{Method}} 
  & \multicolumn{3}{c}{\textbf{MOSAE ↓ ($\pm$ SE)}}
  & \multirow{2}{*}{\textbf{Time ↓ (ms/sample)}} \\
\cmidrule(lr){2-4}
  & \textbf{ID (Mid-Range)} & \textbf{OoD (Full-Range)} & \textbf{ID (Full-Trained)}\\
\midrule
Supervised            & 0.5115 (± 0.0242) & 0.5372 (± 0.0134) & 0.6572 (± 0.0193) & \bftab 0.1145 \\
Self-Supervised       & 0.4936 (± 0.0210) & 0.5190 (± 0.0124) & 0.6098 (± 0.0163) & 0.1347 \\
TT Instance Adaptive  & 0.5641 (± 0.0258) & 0.5506 (± 0.0137) & 0.6402 (± 0.0160) & 212.2983 \\
TT Online Adaptive    & 0.4975 (± 0.0235) & 0.5156 (± 0.0129) & 0.6351 (± 0.0179) & 0.3400 \\
TT Domain Adaptive    & 0.4360 (± 0.0220) & 0.4441 (± 0.0122) & 0.4566 (± 0.0126) & 312.6671 \\
Purely Model-Based    & 0.4390 (± 0.0230) & 0.4641 (± 0.0129) & 0.4641 (± 0.0129) & 3154.4558 \\
FSL-MRS               & \ittab 0.4658 (± 0.0235) & \ittab 0.4845 (± 0.0133) & \ittab 0.4845 (± 0.0133) & 475.4806 \\
LCModel               & \ittab 0.4111 (± 0.0201) & \ittab 0.4336 (± 0.0117) & \ittab 0.4336 (± 0.0117) & 274.6317 \\
& & & & \\
Supervised (CNN)                             & 0.5713 (± 0.0240) & 0.5647 (± 0.0130) & 0.6846 (± 0.0173) & 0.7374 \\
Self-Supervised (CNN)                        & 0.7343 (± 0.0360) & 0.7452 (± 0.0199) & 0.5085 (± 0.0136) & 0.1551 \\
TT Instance Adaptive (CNN)                     & 0.5843 (± 0.0241) & 0.5798 (± 0.0130) & 0.6969 (± 0.0175) & 256.5153 \\
TT Instance Adaptive (CNN, Self-Sup. Init.)    & 0.4553 (± 0.0225) & 0.4831 (± 0.0127) & 0.4697 (± 0.0132) & 270.3693 \\
& & & & \\
TT Instance Adaptive (Self-Sup. Init.)         & \bftab 0.4034 (± 0.0186) & \bftab 0.4335 (± 0.0109) & 0.4945 (± 0.0136) & 205.7727 \\
TT Instance Adaptive (From Scratch Init.)      & 0.4357 (± 0.0222) & 0.4565 (± 0.0125) & \bftab 0.4542 (± 0.0125) & 29970.4098 \\
TT Online Adaptive (Self-Sup. Init.)  & 0.4495 (± 0.0194) & 0.4723 (± 0.0114) & 0.5258 (± 0.0135) & 0.3331 \\
TT Domain Adaptive (Self-Sup. Init.)  & 0.4652 (± 0.0236) & 0.4797 (± 0.0132) & 0.4601 (± 0.0129) & 312.5201 \\
& & & & \\
TT Instance Adaptive (10 Iter.)   & 0.5017 (± 0.0239) & 0.5137 (± 0.0129) & 0.6444 (± 0.0183) & 44.8087 \\
TT Instance Adaptive (50 Iter.)   & 0.5641 (± 0.0258) & 0.5506 (± 0.0137) & 0.6402 (± 0.0160) & 212.2983 \\
TT Instance Adaptive (100 Iter.)  & 0.5186 (± 0.0258) & 0.5134 (± 0.0139) & 0.5831 (± 0.0157) & 414.0273 \\
TT Instance Adaptive (500 Iter.)  & 0.4583 (± 0.0245) & 0.4766 (± 0.0136) & 0.5091 (± 0.0144) & 2061.0019 \\
\bottomrule
\end{tabular}
\end{adjustbox}
\end{table*}

\begin{table*}
\caption{\enspace 
Comparison of quantification methods on in-vivo data using pseudo ground truth: \textbf{LCModel}. The spectra are filtered to create equivalent scenarios to the simulated test scenarios: \textbf{ID (Mid-Range)}, \textbf{OoD (Full-Range)}, and \textbf{ID (Full-Trained)}.
}
\label{tab:invivo_lcmodel_mae} 
\vspace{2mm}
\begin{adjustbox}{width=1.0\textwidth}
\begin{tabular}{lccc c}
\toprule
\multirow{2}{*}{\textbf{Method}} 
  & \multicolumn{3}{c}{\textbf{MAE ↓ ($\pm$ SE)}} 
  & \multirow{2}{*}{\textbf{Time ↓ (ms/sample)}} \\
\cmidrule(lr){2-4}
  & \textbf{ID (Mid-Range)} & \textbf{OoD (Full-Range)} & \textbf{ID (Full-Trained)} \\
\midrule
Supervised            & 0.9662 (± 0.0470) & 0.9896 (± 0.0257) & 1.1211 (± 0.0290) & \bftab 0.1145 \\
Self-Supervised       & 0.8086 (± 0.0409) & 0.8429 (± 0.0233) & 1.0475 (± 0.0270) & 0.1347 \\
TT Instance Adaptive  & 0.8734 (± 0.0383) & 0.9130 (± 0.0214) & 1.0334 (± 0.0240) & 212.2983 \\
TT Online Adaptive    & 0.9033 (± 0.0417) & 0.9356 (± 0.0232) & 1.0788 (± 0.0278) & 0.3400 \\
TT Domain Adaptive    & 0.7955 (± 0.0356) & 0.8500 (± 0.0205) & 0.8558 (± 0.0210) & 312.6671 \\
Purely Model-Based    & 0.7648 (± 0.0360) & 0.8562 (± 0.0218) & 0.8562 (± 0.0218) & 3154.4558 \\
FSL-MRS               & 0.8368 (± 0.0383) & 0.9293 (± 0.0233) & 0.9293 (± 0.0233) & 475.4806 \\
LCModel               & \ittab 0.3248 (± 0.0211) & \ittab 0.3405 (± 0.0125) & \ittab 0.3405 (± 0.0125) & 274.6317 \\
& & & & \\
Supervised (CNN)                             & 0.8757 (± 0.0414) & 0.9077 (± 0.0232) & 1.1076 (± 0.0284) & 0.7374 \\
Self-Supervised (CNN)                        & 1.0287 (± 0.0492) & 1.0251 (± 0.0274) & 0.8611 (± 0.0215) & 0.1551 \\
TT Instance Adaptive (CNN)                     & 0.8242 (± 0.0339) & 0.8470 (± 0.0192) & 1.0844 (± 0.0258) & 256.5153 \\
TT Instance Adaptive (CNN, Self-Sup. Init.)    & 0.6809 (± 0.0300) & 0.7303 (± 0.0177) & 0.8428 (± 0.0209) & 270.3693 \\
& & & & \\
TT Instance Adaptive (Self-Sup. Init.)         & \bftab 0.6556 (± 0.0279) & \bftab 0.6977 (± 0.0163) & 0.8760 (± 0.0215) & 205.7727 \\
TT Instance Adaptive (From Scratch Init.)      & 0.7909 (± 0.0358) & 0.8580 (± 0.0210) & \bftab 0.8553 (± 0.0209) & 29970.4098 \\
TT Online Adaptive (Self-Sup. Init.)  & 0.7032 (± 0.0307) & 0.7430 (± 0.0180) & 0.9387 (± 0.0229) & 0.3331 \\
TT Domain Adaptive (Self-Sup. Init.)  & 0.8379 (± 0.0381) & 0.8977 (± 0.0220) & 0.8702 (± 0.0217) & 312.5201 \\
& & & & \\
TT Instance Adaptive (10 Iter.)   & 0.8840 (± 0.0400) & 0.9137 (± 0.0220) & 1.0636 (± 0.0270) & 44.8087 \\
TT Instance Adaptive (50 Iter.)   & 0.8734 (± 0.0383) & 0.9130 (± 0.0214) & 1.0334 (± 0.0240) & 212.2983 \\
TT Instance Adaptive (100 Iter.)  & 0.8432 (± 0.0383) & 0.8931 (± 0.0217) & 0.9775 (± 0.0240) & 414.0273 \\
TT Instance Adaptive (500 Iter.)  & 0.7846 (± 0.0360) & 0.8544 (± 0.0213) & 0.8819 (± 0.0222) & 2061.0019 \\
\bottomrule
\end{tabular}
\end{adjustbox}
\end{table*}

\begin{table*}
\caption{\enspace 
Comparison of quantification methods on in-vivo data using pseudo ground truth: \textbf{LCModel}. The spectra are filtered to create equivalent scenarios to the simulated test scenarios: \textbf{ID (Mid-Range)}, \textbf{OoD (Full-Range)}, and \textbf{ID (Full-Trained)}.
}
\label{tab:invivo_lcmodel_mosae} 
\vspace{2mm}
\begin{adjustbox}{width=1.0\textwidth}
\begin{tabular}{lccc c}
\toprule
\multirow{2}{*}{\textbf{Method}} 
  & \multicolumn{3}{c}{\textbf{MOSAE ↓ ($\pm$ SE)}} 
  & \multirow{2}{*}{\textbf{Time ↓ (ms/sample)}} \\
\cmidrule(lr){2-4}
  & \textbf{ID (Mid-Range)} & \textbf{OoD (Full-Range)} & \textbf{ID (Full-Trained)} \\
\midrule
Supervised            & 0.7992 (± 0.0364) & 0.8315 (± 0.0202) & 1.0168 (± 0.0262) & \bftab 0.1145 \\
Self-Supervised       & 0.6666 (± 0.0297) & 0.7009 (± 0.0172) & 0.9477 (± 0.0241) & 0.1347 \\
TT Instance Adaptive  & 0.8577 (± 0.0393) & 0.8972 (± 0.0221) & 1.0184 (± 0.0249) & 212.2983 \\
TT Online Adaptive    & 0.7959 (± 0.0364) & 0.8297 (± 0.0201) & 1.0029 (± 0.0252) & 0.3400 \\
TT Domain Adaptive    & 0.7749 (± 0.0360) & 0.8302 (± 0.0209) & 0.8349 (± 0.0214) & 312.6671 \\
Purely Model-Based    & 0.7426 (± 0.0361) & 0.8292 (± 0.0219) & 0.8292 (± 0.0219) & 3154.4558 \\
FSL-MRS               & 0.7905 (± 0.0382) & 0.8743 (± 0.0230) & 0.8743 (± 0.0230) & 475.4806 \\
LCModel               & \ittab 0.3129 (± 0.0211) & \ittab 0.3288 (± 0.0124) & \ittab 0.3288 (± 0.0124) & 274.6317 \\
& & & & \\
Supervised (CNN)                             & 0.7817 (± 0.0365) & 0.7990 (± 0.0202) & 1.0111 (± 0.0261) & 0.7374 \\
Self-Supervised (CNN)                        & 0.9606 (± 0.0437) & 0.9652 (± 0.0246) & 0.8377 (± 0.0219) & 0.1551 \\
TT Instance Adaptive (CNN)                     & 0.8104 (± 0.0349) & 0.8309 (± 0.0198) & 1.0695 (± 0.0266) & 256.5153 \\
TT Instance Adaptive (CNN, Self-Sup. Init.)    & 0.6585 (± 0.0306) & 0.7057 (± 0.0182) & \bftab 0.8200 (± 0.0214) & 270.3693 \\
& & & & \\
TT Instance Adaptive (Self-Sup. Init.)         & \bftab 0.6330 (± 0.0283) & \bftab 0.6751 (± 0.0167) & 0.8496 (± 0.0221) & 205.7727 \\
TT Instance Adaptive (From Scratch Init.)      & 0.7748 (± 0.0365) & 0.8427 (± 0.0214) & 0.8398 (± 0.0214) & 29970.4098 \\
TT Online Adaptive (Self-Sup. Init.)  & 0.6548 (± 0.0281) & 0.6901 (± 0.0163) & 0.8871 (± 0.0220) & 0.3331 \\
TT Domain Adaptive (Self-Sup. Init.)  & 0.8035 (± 0.0382) & 0.8621 (± 0.0220) & 0.8327 (± 0.0217) & 312.5201 \\
& & & & \\
TT Instance Adaptive (10 Iter.)   & 0.8000 (± 0.0369) & 0.8357 (± 0.0205) & 1.0080 (± 0.0257) & 44.8087 \\
TT Instance Adaptive (50 Iter.)   & 0.8577 (± 0.0393) & 0.8972 (± 0.0221) & 1.0184 (± 0.0249) & 212.2983 \\
TT Instance Adaptive (100 Iter.)  & 0.8272 (± 0.0393) & 0.8764 (± 0.0223) & 0.9604 (± 0.0247) & 414.0273 \\
TT Instance Adaptive (500 Iter.)  & 0.7606 (± 0.0370) & 0.8307 (± 0.0219) & 0.8567 (± 0.0228) & 2061.0019 \\
\bottomrule
\end{tabular}
\end{adjustbox}
\end{table*}




\section{Implementation Details} \label{app:impl_details}
This section offers additional information about the specific implementation choices made throughout this work.

\subsection{Metabolite Concentration Ranges}
The origins of the metabolite concentration ranges used for simulation are detailed in the following tables. Bounds were derived from a combination of literature values reported by De Graaf 2019 \cite{de_graaf_vivo_2019} and empirical distributions obtained by fitting all in-vivo spectra using LCModel \cite{provencher_estimation_1993} and FSL-MRS \cite{clarke_fslmrs_2021} (Tables \ref{tab:sim_params_deGraaf}, \ref{tab:sim_params_lcmodel}, and \ref{tab:sim_params_fsl}).

\begin{table*}
\vspace{2cm}
\caption{\enspace Overview of the metabolite concentration ranges [mM] taken from De Graaf 2019 \cite{de_graaf_vivo_2019}. \label{tab:sim_params_deGraaf}}%
\vspace{2mm}
\begin{adjustbox}{width=0.5\textwidth}
\begin{tabular*}{0.6\textwidth}{@{\extracolsep\fill\hspace{1mm}}lcc@{\extracolsep\fill\hspace{2mm}}}
\toprule
\textbf{Parameter}                     & \textbf{Notation} & \textbf{Range}                   \\
\midrule
Alanine (Ala)                          & \( a_1 \)         & \( \mathcal{U}[0.1, 1.6] \)   \\
Ascorbate (Asc)                        & \( a_2 \)         & \( \mathcal{U}[0.4, 1.7] \)   \\
Aspartate (Asp)                        & \( a_3 \)         & \( \mathcal{U}[1.0, 2.0] \)   \\
Creatine (Cr)                          & \( a_4 \)         & \( \mathcal{U}[4.5, 10.5] \)   \\
Gamma-Aminobutyric Acid (GABA)         & \( a_5 \)         & \( \mathcal{U}[1.0, 2.0] \)   \\
Glutamine (Gln)                        & \( a_{6} \)       & \( \mathcal{U}[3.0, 6.0] \)  \\
Glutamate (Glu)                        & \( a_{7} \)       & \( \mathcal{U}[6.0, 12.5] \)  \\
Glycine (Gly)                          & \( a_{8} \)       & \( \mathcal{U}[0.2, 1.0] \)  \\
Glycerophosphocholine (GPC)            & \( a_{9} \)       & \( \mathcal{U}[0.4, 1.7] \)   \\
Glutathione (GSH)                      & \( a_{10} \)      & \( \mathcal{U}[1.7, 3.0] \)   \\ 
Myo-Inositol (mIns)                    & \( a_{11} \)      & \( \mathcal{U}[4.0, 9.0] \)  \\
\bottomrule
\end{tabular*}
\end{adjustbox}
\begin{adjustbox}{width=0.5\textwidth}
\begin{tabular*}{0.6\textwidth}{@{\extracolsep\fill\hspace{1mm}}lcc@{\extracolsep\fill\hspace{2mm}}}
\toprule
\textbf{Parameter}                     & \textbf{Notation} & \textbf{Range}                   \\
\midrule
Lactate (Lac)                          & \( a_{12} \)      & \( \mathcal{U}[0.2, 1.0] \)  \\
N-Acetylaspartylglutamate (NAAG)       & \( a_{13} \)      & \( \mathcal{U}[0.5, 2.5] \)   \\
N-Acetylaspartate (NAA)                & \( a_{14} \)      & \( \mathcal{U}[7.5, 12.0] \)  \\
Phosphocholine (PCh)                   & \( a_{15} \)      & \( \mathcal{U}[0.2, 1.0] \)   \\
Phosphocreatine (PCr)                  & \( a_{16} \)      & \( \mathcal{U}[3.0, 5.5] \)   \\
Phosphoethanolamine (PE)               & \( a_{17} \)      & \( \mathcal{U}[1.0, 2.0] \)   \\
Scyllo-Inositol (Scyllo)               & \( a_{18} \)      & \( \mathcal{U}[0.2, 0.5] \)   \\
Serine (Ser)                           & \( a_{18} \)      & \( \mathcal{U}[0.2, 2.0] \)   \\  
Taurine (Tau)                          & \( a_{20} \)      & \( \mathcal{U}[3.0, 6.0] \)   \\
Macromolecules (MMs)$^{*}$             & \( a_{21} \)      & \( \mathcal{U}[0.0, 400.0] \)  \\
& & \\
\bottomrule
\end{tabular*}
\end{adjustbox}
\begin{tablenotes}
\item[$^{*}$] Not taken from De Graaf 2019.
\end{tablenotes}

\caption{\enspace The metabolite concentration ranges  [mM] obtained by fitting all in-vivo
spectra of Section \ref{ssec:in-vivo_data} using LCModel \cite{provencher_estimation_1993}. \label{tab:sim_params_lcmodel}}%
\vspace{2mm}
\begin{adjustbox}{width=0.5\textwidth}
\begin{tabular*}{0.6\textwidth}{@{\extracolsep\fill\hspace{1mm}}lcc@{\extracolsep\fill\hspace{2mm}}}
\toprule
\textbf{Parameter}                     & \textbf{Notation} & \textbf{Range}                   \\
\midrule
Alanine (Ala)                          & \( a_1 \)         & \( \mathcal{U}[0.0, 1.0] \)   \\
Ascorbate (Asc)                        & \( a_2 \)         & \( \mathcal{U}[0.0, 3.7] \)   \\
Aspartate (Asp)                        & \( a_3 \)         & \( \mathcal{U}[0.8, 4.0] \)   \\
Creatine (Cr)                          & \( a_4 \)         & \( \mathcal{U}[3.9, 8.0] \)   \\
Gamma-Aminobutyric Acid (GABA)         & \( a_5 \)         & \( \mathcal{U}[0.0, 4.0] \)   \\
Glutamine (Gln)                        & \( a_{6} \)       & \( \mathcal{U}[1.9, 6.8] \)  \\
Glutamate (Glu)                        & \( a_{7} \)       & \( \mathcal{U}[11.7, 16.9] \)  \\
Glycine (Gly)                          & \( a_{8} \)       & \( \mathcal{U}[0.0, 0.5] \)  \\
Glycerophosphocholine (GPC)            & \( a_{9} \)       & \( \mathcal{U}[0.0, 3.2] \)   \\
Glutathione (GSH)                      & \( a_{10} \)      & \( \mathcal{U}[0.0, 1.7] \)   \\ 
Myo-Inositol (mIns)                    & \( a_{11} \)      & \( \mathcal{U}[5.9, 10.0] \)  \\
\bottomrule
\end{tabular*}
\end{adjustbox}
\begin{adjustbox}{width=0.5\textwidth}
\begin{tabular*}{0.6\textwidth}{@{\extracolsep\fill\hspace{1mm}}lcc@{\extracolsep\fill\hspace{2mm}}}
\toprule
\textbf{Parameter}                     & \textbf{Notation} & \textbf{Range}                   \\
\midrule
Lactate (Lac)                          & \( a_{12} \)      & \( \mathcal{U}[0.0, 2.8] \)  \\
N-Acetylaspartylglutamate (NAAG)       & \( a_{13} \)      & \( \mathcal{U}[0.2, 2.0] \)   \\
N-Acetylaspartate (NAA)                & \( a_{14} \)      & \( \mathcal{U}[9.8, 14.0] \)  \\
Phosphocholine (PCh)                   & \( a_{15} \)      & \( \mathcal{U}[0.0, 2.4] \)   \\
Phosphocreatine (PCr)                  & \( a_{16} \)      & \( \mathcal{U}[0.9, 5.2] \)   \\
Phosphoethanolamine (PE)               & \( a_{17} \)      & \( \mathcal{U}[0.0, 5.2] \)   \\
Scyllo-Inositol (Scyllo)               & \( a_{18} \)      & \( \mathcal{U}[0.0, 0.4] \)   \\
Serine (Ser)                           & \( a_{18} \)      & \( \mathcal{U}[0.0, 3.6] \)   \\  
Taurine (Tau)                          & \( a_{20} \)      & \( \mathcal{U}[1.2, 3.1] \)   \\
Macromolecules (MMs)                   & \( a_{21} \)      & \( \mathcal{U}[175.7, 318.9] \)  \\
& & \\
\bottomrule
\end{tabular*}
\end{adjustbox}
\begin{tablenotes}
\end{tablenotes}

\caption{\enspace The metabolite concentration ranges  [mM] obtained by fitting all in-vivo
spectra of Section \ref{ssec:in-vivo_data} using FSL-MRS \cite{clarke_fslmrs_2021}. \label{tab:sim_params_fsl}}%
\vspace{2mm}
\begin{adjustbox}{width=0.5\textwidth}
\begin{tabular*}{0.6\textwidth}{@{\extracolsep\fill\hspace{1mm}}lcc@{\extracolsep\fill\hspace{2mm}}}
\toprule
\textbf{Parameter}                     & \textbf{Notation} & \textbf{Range}                   \\
\midrule
Alanine (Ala)                          & \( a_1 \)         & \( \mathcal{U}[0.0, 0.5] \)   \\
Ascorbate (Asc)                        & \( a_2 \)         & \( \mathcal{U}[0.9, 4.9] \)   \\
Aspartate (Asp)                        & \( a_3 \)         & \( \mathcal{U}[0.0, 4.8] \)   \\
Creatine (Cr)                          & \( a_4 \)         & \( \mathcal{U}[5.0, 12.3] \)   \\
Gamma-Aminobutyric Acid (GABA)         & \( a_5 \)         & \( \mathcal{U}[0.0, 2.2] \)   \\
Glutamine (Gln)                        & \( a_{6} \)       & \( \mathcal{U}[0.0, 3.6] \)  \\
Glutamate (Glu)                        & \( a_{7} \)       & \( \mathcal{U}[11.1, 17.9] \)  \\
Glycine (Gly)                          & \( a_{8} \)       & \( \mathcal{U}[0.0, 0.2] \)  \\
Glycerophosphocholine (GPC)            & \( a_{9} \)       & \( \mathcal{U}[0.2, 3.6] \)   \\
Glutathione (GSH)                      & \( a_{10} \)      & \( \mathcal{U}[1.0, 3.6] \)   \\ 
Myo-Inositol (mIns)                    & \( a_{11} \)      & \( \mathcal{U}[6.5, 12.1] \)  \\
\bottomrule
\end{tabular*}
\end{adjustbox}
\begin{adjustbox}{width=0.5\textwidth}
\begin{tabular*}{0.6\textwidth}{@{\extracolsep\fill\hspace{1mm}}lcc@{\extracolsep\fill\hspace{2mm}}}
\toprule
\textbf{Parameter}                     & \textbf{Notation} & \textbf{Range}                   \\
\midrule
Lactate (Lac)                          & \( a_{12} \)      & \( \mathcal{U}[0.1, 3.1] \)  \\
N-Acetylaspartylglutamate (NAAG)       & \( a_{13} \)      & \( \mathcal{U}[0.0, 2.3] \)   \\
N-Acetylaspartate (NAA)                & \( a_{14} \)      & \( \mathcal{U}[9.7, 16.3] \)  \\
Phosphocholine (PCh)                   & \( a_{15} \)      & \( \mathcal{U}[0.0, 2.2] \)   \\
Phosphocreatine (PCr)                  & \( a_{16} \)      & \( \mathcal{U}[0.0, 5.1] \)   \\
Phosphoethanolamine (PE)               & \( a_{17} \)      & \( \mathcal{U}[0.0, 3.5] \)   \\
Scyllo-Inositol (Scyllo)               & \( a_{18} \)      & \( \mathcal{U}[0.0, 0.6] \)   \\
Serine (Ser)                           & \( a_{18} \)      & \( \mathcal{U}[0.0, 7.3] \)   \\  
Taurine (Tau)                          & \( a_{20} \)      & \( \mathcal{U}[1.5, 3.4] \)   \\
Macromolecules (MMs)                   & \( a_{21} \)      & \( \mathcal{U}[133.8, 242.1] \)  \\
& & \\
\bottomrule
\end{tabular*}
\end{adjustbox}
\begin{tablenotes}
\end{tablenotes}
\vspace{5cm}
\end{table*}

\subsection{Model Architecture \& Setup Configuration} \label{app:modelArch}
Layer-wise architectures and configuration parameters used for training and testing the models are provided in Tables \ref{tab:arch_compare} and \ref{tab:ml_config}.

\begin{table*}[b]
    \vspace{0.5cm}
    \caption{\enspace  Layer-wise architecture of the MLP and CNN models, implemented in PyTorch. Convolutional parameters shown as (kernel, stride, padding). ELU activations follow each hidden layer. Parameter counts: MLP = 532K, CNN = 3.0M.}
    \label{tab:arch_compare}
    \vspace{2mm}
    \centering
    \begin{adjustbox}{width=1.0\textwidth}
    \begin{tabular*}{2.53\columnwidth}{@{\extracolsep{\fill}}lllccc}
    \toprule
    \textbf{Arch} & \textbf{Layer} & \textbf{Operation} & \textbf{Params (k,s,p)} & \textbf{Input Dim} & \textbf{Output Dim} \\
    \midrule
    \multirow{6}{*}{\textbf{MLP}} 
        & Input         & BatchNorm1d        & -          & 355           & 355 \\
        & Flatten       & Flatten            & -          & 2 × 355       & -   \\
        & FC0           & Linear + ELU       & -          & 710           & 512 \\
        & FC1           & Linear + ELU       & -          & 512           & 256 \\
        & FC2           & Linear + ELU       & -          & 256           & 128 \\
        & Output$^{*}$  & Linear             & -          & 128           & 32  \\
    \midrule
    \multirow{9}{*}{\textbf{CNN}} 
        & Input         & BatchNorm1d        & -          & 355           & 355 \\
        & Conv0         & Conv1d + ELU       & (3, 1, 0)  & 4             & 8   \\
        & Conv1         & Conv1d + ELU       & (3, 1, 0)  & 8             & 16  \\
        & Conv2         & Conv1d + ELU       & (3, 1, 0)  & 16            & 32  \\
        & Flatten       & Flatten            & -          & -             & -   \\
        & FC0           & Linear + ELU       & -          & flattened     & 512 \\
        & FC1           & Linear + ELU       & -          & 512           & 256 \\
        & FC2           & Linear + ELU       & -          & 256           & 128 \\
        & Output$^{*}$  & Linear             & -          & 128           & 32  \\
    \bottomrule
    \end{tabular*}
    \end{adjustbox}
    \vspace{-3mm}
    \begin{tablenotes}
    \item[\textbf{*}] Output vector uses component-wise activations: \texttt{softplus} for metabolite amplitudes and linewidths, with a \texttt{+1} offset for Gaussian and Lorentzian broadening to ensure values $> 1$. First-order phase uses a scaled \texttt{tanh} activation (\(\tanh(x) \times 10^{-4}\)) to keep values in the stable range \(\mathcal{U}[-10^{-5}, 10^{-5}]\). Remaining parameters (zeroth-order phase, frequency shift, and baseline) are linear.
    \end{tablenotes}
    \vspace{0.5cm}
    \caption{\enspace Overview of the configuration and training parameters for the MLP and CNN models. \hspace{12cm}} \label{tab:ml_config}%
    \vspace{2mm}
    \begin{adjustbox}{width=1.0\textwidth}
    \begin{tabular*}{2.53\columnwidth}{@{\extracolsep{\fill}\hspace{1mm}}lll@{\extracolsep{\fill}\hspace{1mm}}}
    \toprule
    \textbf{Parameter} & \textbf{Value} & \textbf{Description}\\
    \midrule
    \textbf{Data Settings} \\
    dataType & aumc2\_ms & Dataset used for training and evaluation. \\
    basisFmt & 7tslaser & Format of the MRS basis set. \\
    path2basis & .../7T\_sLASER\_OIT\_TE34.basis & Path to the basis set. \\
    specType & auto & Automatically selects ppm region. \\
    ppmlim & (0.5, 4.0) & ppm limits of the spectra. \\
    test\_size & 10000 & Number of test samples. \\
    \midrule
    \textbf{Architecture Settings} \\
    arch & mlp / cnn & Architecture type: MLP or CNN. \\
    activation & elu & Nonlinearity used in hidden layers. \\
    dropout & 0.0 & Dropout probability. \\
    width & 512 & Width of first fully connected layer. \\
    depth & 3 & Number of FC layers (after input / conv layers). \\
    conv\_depth & 3 & Number of Conv1D layers (CNN only). \\
    kernel\_size & 3 & Kernel size for Conv1D layers. \\
    stride & 1 & Stride for Conv1D layers. \\
    \midrule
    \textbf{Optimization} \\
    loss & mse\_specs / mae\_all\_scale & Loss function for training. \\
    optimizer & Adam~\cite{kingma2015adam} & Optimizer used for training. \\
    batch & 16 & The batch size. \\
    trueBatch & 16 & Accumulates the gradients over trueBatch/batch. \\
    check\_val\_every\_n\_epoch & None & None, if trained with generator, otherwise the number of epochs between validations. \\
    learning & 0.0001 & Learning rate. \\
    max\_epochs & -1 & Maximum number of epochs. \\
    max\_steps & -1 & Maximum number of steps/iterations. \\
    val\_check\_interval & 256 & The number of iterations per between validations. \\
    val\_size & 1,024 & Validation size (in samples). \\
    \midrule
    \textbf{Adaptation / Inner Loop} \\
    adaptMode & per\_spec\_adapt & Mode of inner-loop adaptation (instance/online/domain adaptation). \\
    innerEpochs & 50 & Number of adaptation epochs. \\
    innerBatch & 1 & Batch size for inner loop. \\
    innerLr & 1e-4 & Learning rate for inner loop. \\
    innerLoss & mse\_specs & Loss function for inner loop. \\
    bnState & train & BatchNorm mode in inner loop. \\
    \bottomrule
    \end{tabular*}
    \end{adjustbox}
    \vspace{0.5cm}
\end{table*}

\section{MRS in MRS} \label{app:mrs_in_mrs}
This section presents tables that detail the acquisitions, experimental setup, processing, and data analysis methods employed in the study, adhering to the \ac{mrsinmrs} guidelines \cite{lin_minimum_2021}, made easier by REMY~\cite{Susnjar2025REMY}.

\begin{table*}%
    \caption{\enspace MRSinMRS for the data of Section \ref{ssec:results_invivo}. \hspace{12cm}} \label{tab:mrsinmrs_invivo}%
    \vspace{2mm}
    \begin{adjustbox}{width=1.0\textwidth}
    \begin{tabular*}{2.53\columnwidth}{@{\extracolsep{\fill}\hspace{1mm}}ll@{\extracolsep{\fill}\hspace{1mm}}}
    \toprule
    \textbf{Site (name or number)} & Amsterdam UMC \\
    \midrule
    \textbf{1. Hardware} & \\
    a. Field strength [T] & 7 T (298030131 MHz) \\
    b. Manufacturer & Philips \\
    c. Model (software version if available) & 5.1.7; .1.7; \\
    d. RF coils: nuclei (transmit/receive), number of channels, type, body part & 1H, 32 channel, head coil \\
    e. Additional hardware & - \\
    \midrule
    \textbf{2. Acquisition} & \\
    a. Pulse sequence & Semi-LASER \\
    b. Volume of interest (VOI) locations & Anterior cingulate cortex \\
    c. Nominal VOI size [cm$^3$, mm$^3$] & 25 × 18 × 18 mm$^3$ \\
    d. Echo time (TE) / repetition time (TR) [ms, s] & 36 ms / 5000 ms \\
    e. Total number of excitations or acquisitions per spectrum & 64 averages \\
    f. Additional sequence parameters & 3000 Hz bandwidth, 1024 sample points, \\
    g. Water suppression method & VAPOR \\
    h. Shimming method, reference peak, and thresholds for "acceptance of shim" chosen & HOS-DLT \cite{Boer2020HOS_DLT} \\
    i. Triggering or motion correction method (respiratory, peripheral, cardiac triggering) & - \\
    \midrule
    \textbf{3. Data Analysis Methods and Outputs} & \\
    a. Analysis software & In-house Python scripts, \\
    & FSL-MRS \cite{clarke_fslmrs_2021} (version 2.1.20), \\
    & LCModel \cite{provencher_estimation_1993} (version 6.3-1L) \\
    b. Processing steps (deviating from quoted reference or product) & NIfTI-MRS Header (ProcessingApplied): \\
    & Method: "Custom coil combination (adaptive)", \\
    & Details: \textit{own\_nifti\_coil\_combination\_adaptive, data, reference} \\
    & \textit{fsl\_mrs\_preproc} \\
    & \hspace{1cm}\textit{--data \{save\_path\}/\{item\}/block\{i + 1\}/metab.nii.gz} \\
    & \hspace{1cm}\textit{--reference \{save\_path\}/\{item\}/block\{i + 1\}/wref.nii.gz} \\
    & \hspace{1cm}\textit{--output \{save\_path\}/\{item\}/block\{i + 1\}/\{sub\_folder\}} \\
    & \hspace{1cm}\textit{--hlsvd --conjugate --overwrite --report} \\
    c. Output measure (e.g. absolute concentration, institutional units, ratio) & Absolute concentrations [mM] \\
    d. Quantification references and assumptions, fitting model assumptions & 7T Semi-LASER OIT basis set with TE 34ms \\
    & (metabolite list seen in Table \ref{tab:sim_params}, \ac{mm} \cite{cudalbu_contribution_2021}) \\
    & LCModel control:\\
    & \hspace{1cm}\textit{\$LCMODL, nunfil=1024, deltat=3.333e-04,} \\
    & \hspace{1cm}\textit{hzpppm=hzpppm=2.9803e+02, ppmst=4.0,} \\
    & \hspace{1cm}\textit{ppmend=0.5, dows=T, doecc=F, neach=50,} \\
    & \hspace{1cm}\textit{filbas='example.basis', filraw='example.raw',} \\
    & \hspace{1cm}\textit{filh2o='example.h20', filps='example.ps',} \\
    & \hspace{1cm}\textit{filcoo='example.coord', filtab='example.table',} \\
    & \hspace{1cm}\textit{ltable=7, lcoord=9, lps=8, nsimul=0, echot=36,} \\
    & \hspace{1cm}\textit{dkntmn=0.5, nuse1=3, chcomb(1)='Glu+Gln',} \\
    & \hspace{1cm}\textit{hcomb(2)='Cr+PCr', chcomb(3)='NAA+NAAG',} \\
    & \hspace{1cm}\textit{chcomb(4)='GPC+PCh'atth2o=0.7, wconc=59297,} \\
    & \hspace{1cm}\textit{\$END} \\
    & FSL-MRS: \textit{from fsl\_mrs.utils import fitting} \\
    & \textit{fitting.fit\_FSLModel, method='Newton' ppmlim=(0.5, 4.0),} \\
    & \hspace{1cm}\textit{baseline\_order=2} \\
    \midrule
    \textbf{4. Data Quality} & \\
    a. Reported variables (SNR, linewidth) & S/N = 20.0 - 54.0, FWHM = 0.029 - 0.079 ppm \\
    & (LCModel estimates)\\
    b. Data exclusion criteria & 4 participants excluded based on visual inspection \\
    c. Quality measures of postprocessing model fitting & - \\
    d. Sample spectra (and mean) & 
    \begin{minipage}{\textwidth}
    \includegraphics[width=0.4\textwidth]{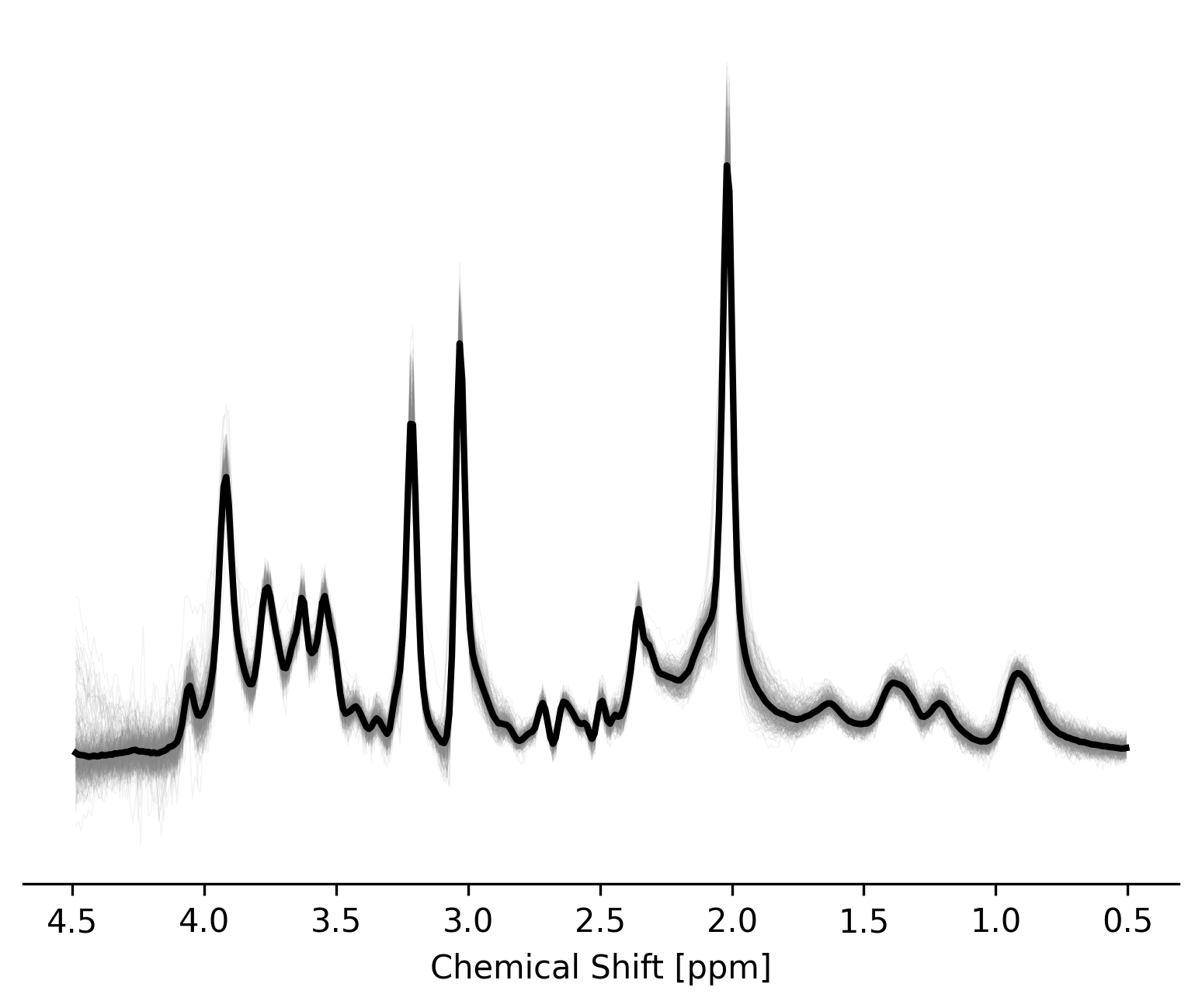}
    \end{minipage} \\
    \bottomrule
    \end{tabular*}
    \end{adjustbox}
\end{table*}

\section{Hardware \& Software Environment}

\subsection{Simulation Environment}
All simulated experiments and runtime benchmarks were executed on a workstation with the following specifications:
\begin{itemize}
  \item \textbf{CPU}: AMD Ryzen 9 7950X 
  \item \textbf{GPU}: NVIDIA RTX 6000 Ada
  \item \textbf{Memory}: 128 GB DDR5-6000
  \item \textbf{OS}: Ubuntu 24.04.2 LTS
  \item \textbf{Software}: Python~3.10.16, PyTorch~2.6.0, CUDA~12.4
\end{itemize}
Runtimes reported in tables \ref{tab:sim_performance_sel}, \ref{tab:sim_performance_mae} and \ref{tab:sim_performance_mosae} reflect per-sample inference times measured on simulated test data, using the most efficient configuration for each method (e.g., GPU where applicable, multiprocessing, etc.).

\subsection{In-Vivo Environment}
All in-vivo experiments were executed on a high-performance computing cluster managed by SLURM. Jobs were scheduled on dual-socket AMD EPYC 7662 nodes (128 cores, 256 threads) with the following resource allocation:
\begin{itemize}
  \item \textbf{CPU}: 8~CPUs per task
  \item \textbf{GPU}: NVIDIA A100 with 10--40~GB memory (depending on job configuration)
    \item \textbf{Memory}: 128~GB system memory
  \item \textbf{OS}: Red Hat Enterprise Linux 8.10 (Ootpa)
  \item \textbf{Software}: Python~3.11.13, PyTorch~2.7.1, CUDA~12.8
\end{itemize}
Runtimes reported for in-vivo experiments in tables \ref{tab:invivo_fslmrs_sel_mosae}, \ref{tab:invivo_fslmrs_mae}, \ref{tab:invivo_fslmrs_mosae}, \ref{tab:invivo_mix_mae}, \ref{tab:invivo_mix_mosae}, \ref{tab:invivo_lcmodel_mae}, and \ref{tab:invivo_lcmodel_mosae} correspond to these allocated resources, and were measured within SLURM-managed jobs on dedicated compute nodes.

\end{document}